\documentclass[preprint]{elsarticle}

\usepackage{amsmath,amssymb,amsfonts}
\usepackage{mathrsfs}
\usepackage{epsfig}
\usepackage{array}
\usepackage{subfigure}
\usepackage{IEEEtrantools}
\usepackage{exscale}
\usepackage{float}
\usepackage{bbm}
\usepackage{hyperref}
\usepackage{mathtools}
\usepackage{lineno}

\newcommand{\N}{{\mathbb{N}}}
\newcommand{\Z}{{\mathbb{Z}}}

\newcommand{\C}{{\mathbb{C}}}
\newcommand{\1}{{\mathbbm{1}}}
\renewcommand{\Re}{\operatorname{Re}}

\newcommand{\Tr}{\operatorname{Tr}}
\newcommand{\Id}{I}
\newcommand{\ord}{\operatorname{ord}}

\newcolumntype{L}[1]{>{\raggedright\let\newline\\\arraybackslash\hspace{0pt}}m{#1}}
\newcolumntype{C}[1]{>{\centering\let\newline\\\arraybackslash\hspace{0pt}}m{#1}}
\newcolumntype{R}[1]{>{\raggedleft\let\newline\\\arraybackslash\hspace{0pt}}m{#1}}

\journal{Annals of Physics}
\begin{document}

\begin{frontmatter}
  
\title{Doubled Lattice Chern-Simons-Yang-Mills Theories\\with Discrete Gauge Group}

\author{S.\ Caspar}
\author{D.\ Mesterh\'azy\corref{1}}
\cortext[1]{Corresponding author}
\ead{mesterh@itp.unibe.ch}
\author{T.\ Z.\ Olesen\corref{2}}
\author{N.\ D.\ Vlasii}
\author{and U.-J.\ Wiese}
\address{Albert Einstein Center for Fundamental Physics, 
  Institute for Theoretical Physics,
  University of Bern, Sidlerstrasse 5, 3012 Bern, Switzerland}

\begin{abstract}
  We construct doubled lattice Chern-Simons-Yang-Mills theories with discrete gauge group $G$ in the Hamiltonian formulation. Here, these theories are considered on a square spatial lattice and the fundamental degrees of freedom are defined on pairs of links from the direct lattice and its dual, respectively. This provides a natural lattice construction for topologically-massive gauge theories, which are invariant under parity and time-reversal symmetry. After defining the building blocks of the doubled theories, paying special attention to the realization of gauge transformations on quantum states, we examine the dynamics in the group space of a single cross, which is spanned by a single link and its dual. The dynamics is governed by the single-cross electric Hamiltonian and admits a simple quantum mechanical analogy to the problem of a charged particle moving on a discrete space affected by an abstract electromagnetic potential. Such a particle might accumulate a phase shift equivalent to an Aharonov-Bohm phase, which is manifested in the doubled theory in terms of a nontrivial ground-state degeneracy on a single cross. We discuss several examples of these doubled theories with different gauge groups including the cyclic group $\Z(k)\subset U(1)$, the symmetric group $S_3\subset O(2)$, the binary dihedral (or quaternion) group $\bar{D}_2\subset SU(2)$, and the finite group $\Delta(27)\subset SU(3)$. In each case the spectrum of the single-cross electric Hamiltonian is determined exactly. We examine the nature of the low-lying excited states in the full Hilbert space, and emphasize the role of the center symmetry for the confinement of charges. Whether the investigated doubled models admit a non-Abelian topological state which allows for fault-tolerant quantum computation will be addressed in a future publication.
\end{abstract}


\end{frontmatter}

%\linenumbers

\section{Introduction}
\label{Sec:Introduction}

With recent experimental progress the design and control of macroscopic quantum many-body systems has become a reality \cite{Bloch:2008,Lewenstein:2012}. For the first time, this enables us to study intriguing many-body phenomena that are otherwise difficult to realize in solid-state materials or to model with classical computers. The vision of a universal quantum simulator \cite{Feynman:1981tf} has become the primary motivation for an impressive effort, both from the side of theory and experiment, to bring us closer to a device that transcends the boundaries of classical computation. One of the fundamental challenges towards reliable quantum computing is to defeat decoherence and other unavoidable errors that arise during computation. Topological quantum computing \cite{Kitaev:1997wr} is a particularly intriguing proposal, which allows for the fault-tolerant encoding and processing of quantum information \cite{Shor:1996qc,Gottesman:1997qd,Preskill:1997uk}; while information is stored nonlocally in topological states of a given many-body system, the braiding of quasiparticles with fractional statistics (anyons) \cite{Wilczek:1983cy,Frohlich:1988di,Frohlich:1988qh,Wilczek:1990ik} implements unitary quantum operations \cite{Kitaev:1997wr,Kitaev:2006lla,Nayak:2008zza}. For non-Abelian anyons whose representation lies dense in the unitary group, the representation of these braiding operations provide a universal set of quantum gates.

Kitaev's toric code \cite{Kitaev:1997wr}, which corresponds to a $\Z(2)$ Ising gauge theory in the low-energy limit, is a prime example of a topological quantum memory \cite{Dennis:2001nw}, but it does not allow for the implementation of a fault-tolerant universal set of quantum gates. Therefore suitable generalizations thereof need to be considered if one desires a model with these qualities. Accordingly, the toric code has been generalized to the non-Abelian variants \cite{Bombin:2007qv} and the comprehensive class of Levin-Wen models \cite{Levin:2004mi}. These and other theoretical developments have significantly contributed to our understanding of the nature of different topological phases of matter, as well as their quasiparticle excitations, and have helped us to advance the development of quantum algorithms \cite{Aharonov:2005,Aharonov:2006,Wocjan:2006}. However, the practical realization of such models, allowing for braiding operations on non-Abelian anyons, remains elusive to date. From the side of experiment, the most promising evidence in favor of non-Abelian anyonic quasiparticles has been presented for the $\nu = 5/2$ fractional quantum Hall state \cite{Willett:2010,An:2011,Willett:2013}, but it is fair to say that, so far, the interpretation of these experiments remains inconclusive. This raises the question whether one can design viable quantum systems that realize topological phases with the sought-after non-Abelian anyonic quasiparticle excitations \cite{Aguado:2008,Brennen:2009}.

In fact, there are two issues here. Of course, one needs to ask whether the given model allows for universal quantum computing. For some of the proposed models, this can be answered affirmatively. But we should also inquire whether the model lends itself to a practical implementation, i.e., is it stable at nonvanishing temperature \cite{Dennis:2001nw,Alicki:2007,Castelnovo:2007,Chesi:2010,Brell:2013wsa,Brown:2014,Freeman:2014} and most importantly, is there an experimental platform which can quantum simulate the model? Although there has been some progress in this direction \cite{Lu:2009zzb,Passante:2009zz,Muller:2011} both of these questions have not been answered satisfactorily so far and therefore it is desirable to consider a wider class of models. Motivated by a number of recent works \cite{Banerjee:2012pg,Zohar:2012xf,Banerjee:2012xg,Marcos:2013aya,Zohar:2013zla,Zohar:2015hwa} we inquire whether it is possible to directly engineer lattice gauge theories \cite{Wilson:1974sk} in the Hamiltonian formulation \cite{Kogut:1974ag,Kogut:1979wt,Kogut:1982ds} that feature topological phases with the desired quasiparticle excitations. The proper theoretical framework is that of a Chern-Simons theory \cite{Chern:1974ft} with finite (discrete) gauge group \cite{Freed:1991bn,Bais:1991pe,Bais:1992ca,Bais:2008xi}. Their lattice variants allow for a finite Hilbert space, which can be mapped onto an atomic (gauge) simulator. The benefit of such a bottom-up approach is that the behavior of the system is strictly controlled, both from the point of view of theory and experiment.

Pure Chern-Simons gauge theory has a vanishing Hamiltonian -- the properties of the physical state space are dictated by topological properties of the manifold supporting the gauge degrees of freedom \cite{Witten:1988hf,Atiyah:1989vu}. This makes it challenging (at best) from the perspective of experiments to engineer physical states that are governed by such a theory. A standard way to address this problem is to impose that the topologically nontrivial states of interest should span the physical state space only in the low-energy limit. Instead of considering a pure non-Abelian Chern-Simons theory, one might add a Chern-Simons term to the Yang-Mills action, which endows the gauge fields with a mass \cite{Schonfeld:1980kb,Jackiw:1980kv,Deser:1981wh,Deser:1982vy}. This allows the target topological field theory to be recovered in the limit when the mass is sent to infinity (which corresponds to an infinite separation between distinct energy levels). Such a theory can be solved exactly when the theory is projected onto the lowest state in a given charge sector by examining the spectrum of the single-link electric Hamiltonian \cite{Dunne:1998qy}. This proves to be useful for the construction of Hamiltonians in the framework of lattice gauge theories with discrete gauge group that might eventually be realized in experiment.

A natural way to construct lattice gauge theories with fractional statistics is to consider so-called doubled models \cite{Kantor:1991ty,Eliezer:1991cv,Eliezer:1991qh,Eliezer:1992sq,Adams:1997eb}. In addition to assigning gauge degrees of freedom to the links of the (direct) lattice, these theories allow for independent degrees of freedom defined on the links of its dual. A noncommuting algebra of group multiplication and projection operators is imposed on a single link and its dual. In the Hamiltonian formulation this algebra, as well as its representation on the state space, can be fully characterized in terms of a one-cocycle. The dynamics governed by the corresponding electric Hamiltonian admits a simple quantum mechanical analogy in terms of the motion of an abstract charged particle in the presence of an artificial $U(1)\times U(1)$ gauge field, which implements a parallel transport between nearest neighbors in group space \cite{Olesen:2015baa}. By the successive hopping in the discrete group space this abstract particle might accumulate an Aharonov-Bohm phase \cite{Aharonov:1959fk}, which determines the degeneracies of the Hilbert space of a single pair of links. Here, we consider a doubled lattice Chern-Simons-type theory for particular examples of finite Abelian and non-Abelian gauge groups, which include the cyclic group $\Z(k)\subset U(1)$, the symmetric group $S_3\subset O(2)$ (i.e., the permutation group of three objects), the binary dihedral (quaternion) group $\bar{D}_2\subset SU(2)$, and the $27$-element group $\Delta(27)\subset SU(3)$ (see, e.g., Refs.\ \cite{Fairbairn:1964sga,Fairbairn:1982jx,Luhn:2007uq,Ludl:2011gn}). The present work serves to extend the considerations of Ref.\ \cite{Olesen:2015baa} to discrete non-Abelian groups.

The resulting lattice theory can be seen as a discretized version of two Chern-Simons theories with opposite chirality. That is, the constructed doubled theory is invariant under parity and time-reversal symmetry. Such theories were examined in Ref.\ \cite{Freedman:2004lda} as low-energy effective models for interacting electrons and similarly in Ref.\ \cite{Fendley:2005} for lattice loop models, which have been proposed in the context of spin liquids and topological fluids \cite{Chen:1989xs,Wen:1990fv,Semenoff:1990vr}. Here, we inquire about the properties of the doubled lattice gauge theories and investigate under what circumstances these theories allow for topological order. The presence and nature of excitations with fractional statistics (anyons) will be addressed in a future publication. Finally, we mention that several proposals have been put forward over the years based on a construction that employs the degrees of freedom of the direct lattice without reference to its dual degrees of freedom; this includes pure Abelian Chern-Simons models \cite{Berruto:2000dp} in the Lagrangian formalism, as well as non-Abelian lattice Chern-Simons models in the Hamiltonian picture \cite{Doucot:2003mh,Doucot:2005vb}.

This article is organized as follows: In Sec.\ \ref{Sec:Hamiltonian Formulation of Standard Lattice Gauge Theories with Discrete Gauge Group} we introduce the basic building blocks for standard Wilson-type lattice gauge theories with discrete gauge group $G$ in the Hamiltonian formulation. This includes the link-based operator algebra, gauge symmetry and Gauss's law, as well as the electric and magnetic contribution to the Hamiltonian. In \mbox{Sec.\ \ref{Sec:Hamiltonian Formulation of Doubled Lattice Gauge Theories with Discrete Gauge Group}} we present the doubled lattice gauge theories that are defined in terms of a one-cocycle, which specifies the nature of the noncommuting algebra of group multiplication operators on the Hilbert space of paired links, composed of a link on the direct lattice and its dual. We classify all possible one-cocycles in the case of the finite Abelian group $\Z(k)$ and the finite discrete non-Abelian groups $S_3$, $\bar{D}_2$, and $\Delta(27)$ and address the role of the center symmetry for confinement. Our findings are summarized in Sec.\ \ref{Sec:Conclusions}.
  
\section{Hamiltonian Formulation of Standard Lattice Gauge Theories\\ with Discrete Gauge Group}
\label{Sec:Hamiltonian Formulation of Standard Lattice Gauge Theories with Discrete Gauge Group}

In this section, we consider standard lattice Yang-Mills (YM) theories with discrete gauge group in the Hamiltonian formulation for which the operator algebra is link-based. We discuss the realization of gauge transformations and Gauss's law and construct both the link-based electric and plaquette-based magnetic Hamiltonian. Some background on the theory of discrete groups is provided in \ref{Sec:Basics of the Theory of Discrete Groups}. This section serves to familiarize the reader with the basic properties of lattice YM theories with discrete gauge group before the doubled theories are introduced in \mbox{Sec.\ \ref{Sec:Hamiltonian Formulation of Doubled Lattice Gauge Theories with Discrete Gauge Group}}.

\subsection{Link-based operator algebra}
\label{SubSec:Link-based operator algebra}

Let us consider a square spatial lattice $\Lambda$ with periodic boundary conditions consisting of $N_{\Lambda}$ distinct sites separated by a lattice spacing $a$, as well as the set of links $\ell(\Lambda)$ connecting nearest-neighbor sites on the lattice. The generalization to other lattice geometries is straightforward but not very illuminating in the present context.

The operator algebra of a standard lattice gauge theory is link-based, i.e., all operators of the quantum theory are associated with a directed link $(x,i)\in\ell(\Lambda)$, $i = 1,2$, connecting two neighboring lattice sites from $x - (a/2) \hat i$ to $x + (a/2) \hat i$, where $x$ refers to the midpoint of the link and $\hat i$ is the unit vector in the $i$-direction. The total number of degrees of freedom is given by $N_{\ell(\Lambda)} = 2 N_{\Lambda}$ corresponding to the number of links on the lattice. Note that in the following we will often suppress link (site) indices when it is clear that only single-link (single-site) quantities are considered.

The basic dynamical variables of a lattice gauge theory are parallel transporters $U\equiv U_{x,i}$, associated with a single link, which take their values in the gauge group $G$. They define an orthonormal basis $\mathcal{B} = \{|U\rangle \hspace{1pt}|\hspace{1.5pt} U\in G\}$ for the single-link Hilbert space $\mathcal{H}$, which may be identified with the vector space of dimension $d_G = \ord(G)$ that is freely generated by the elements of $G$ (i.e., $\mathcal{H}\cong \C[G]$, the group algebra over the field of complex numbers). The orthonormal basis states $|U\rangle$ can be viewed as ``coordinate'' eigenstates in the group space.

Alternatively, we may consider the conjugate ``momentum'' states $|\Gamma_p,ab\rangle$, with $a,b = 1,2,\dots,$ $d_{\Gamma_p}$, where $d_{\Gamma_p}$ is the dimension of the unitary irreducible representation $\Gamma_p$ of $G$. They define the flux basis, which is obtained by a Fourier transformation on the group $G$, i.e.,
\begin{equation}
  \label{Eq:FluxBasis}
  |\Gamma_p,ab\rangle = \sum_{U\in G} |U\rangle \langle U|\Gamma_p,ab\rangle .
\end{equation}
Here, the matrix elements are defined as $\langle U|\Gamma_p,ab\rangle = \sqrt{d_{\Gamma_p} / d_G} \,\Gamma_p(U)_{ab}$. Note that the unitarity of the matrix elements, i.e., $\Gamma_p(U)_{ba}^{\ast} = \Gamma_p(U^{-1})_{ab}$, follows from the completeness relation
\begin{equation}
  \sum_{U\in G} \langle \Gamma_p,ab|U\rangle \langle U|\Gamma_q,cd\rangle = \frac{\sqrt{d_{\Gamma_p} d_{\Gamma_q}}}{d_G} \sum_{U\in G} \Gamma_p(U)_{ab}^{\ast} \Gamma_q(U)_{cd} = \delta_{p,q} \delta_{a,c} \delta_{b,d} ,
\end{equation}
where ${}^{\ast}$ denotes complex conjugation. The conjugate ``momentum'' variables can be associated with the electric field degrees of freedom. While for a continuous gauge group electric fields correspond to infinitesimal Hermitian generators of group multiplications that induce an infinitesimal group transformation on a single link, for a discrete gauge group it is most natural to introduce two distinct  unitary operators $L$ and $R$ that represent multiplications by group elements from the left and from the right, respectively\footnote{In the case of an Abelian gauge group there is no distinction between left and right multiplication and we may define a single operator $T(\Omega) \equiv L(\Omega) = R(\Omega^{-1})$, so that $T(\Omega)|U\rangle = |\Omega U \rangle = |U \Omega \rangle$.}
\begin{equation}
  \label{Eq:LeftRightMultiplication}
  L(\Omega) R(\Omega')|U\rangle = |\Omega U \Omega'^{-1}\rangle ,
\end{equation}
where $\Omega, \Omega'\in G$. Using Eq.\ \eqref{Eq:FluxBasis} and \eqref{Eq:LeftRightMultiplication}, we obtain the representation of the left and right operators in the flux basis
\begin{equation}
  \label{Eq:LeftRightMultiplicationFluxRepresentation}
  L(\Omega) R(\Omega') |\Gamma_p,ab\rangle = \sum_{cd} \Gamma_p (\Omega)_{ca}^{\ast} \hspace{1.5pt}|\Gamma_p, cd\rangle\hspace{0.5pt}\Gamma_p (\Omega')_{db} . 
\end{equation}
Note that according to our conventions $\Gamma_p (\1)_{ab} = \delta_{a,b}$.

We proceed by specifying the full operator algebra on a single link. We require that left- and right-multiplication operators obey the group multiplication law
\begin{equation}
  \label{Eq:GroupMultiplicationLawLeftRightOperators}
  L(\Omega) L(\Omega') = L(\Omega\Omega') , \quad
  R(\Omega) R(\Omega') = R(\Omega\Omega') ,
\end{equation}
and from the associativity of the gauge group, i.e., $(\Omega U) {\Omega'}^{-1} = \Omega (U {\Omega'}^{-1})$, it follows that the $L$ and $R$ operators commute with each other
\begin{equation}
  \label{Eq:CommutingLeftRightOperators}
  [L(\Omega), R(\Omega')] = 0 .
\end{equation}
Apart from the left- and right-multiplication operators, we define the operator $P$, which implements a projection on a single link
\begin{equation}
  \label{Eq:ProjectionOperatorSingleLink}
  P(\Omega) |U\rangle = \delta_{\Omega,U} |U\rangle ,
\end{equation}
and satisfies the algebra
\begin{equation}
  \label{Eq:ProjectionAlgebra}
  P(\Omega)P(\Omega') = \delta_{\Omega,\Omega'} P(\Omega) .
\end{equation}
Note that the projection operator does not commute with left and right operators, i.e.,
\begin{equation}
  \label{Eq:LeftRightMultiplicationProjector}
  L(\Omega)R(\Omega')P(\Omega'') = P(\Omega\Omega''\Omega'^{-1}) L(\Omega)R(\Omega') .
\end{equation}

\subsection{Gauge symmetry and Gauss's law}
\label{SubSec:Gauge symmetry and Gauss's law}

Under gauge transformations $\Omega_x\in G$, associated with lattice sites $x\in\Lambda$, the parallel transport variables transform in the following way
\begin{equation}
  U_{x,i} \rightarrow \Omega_{x - \frac{a}{2}\hat{i}} \hspace{1pt}U_{x,i}\hspace{1.5pt} \Omega_{x + \frac{a}{2}\hat{i}}^{-1} .
\end{equation}
The corresponding gauge transformation on the full Hilbert space of the standard YM theory $\mathcal{H}_{\ell(\Lambda)} = \mathcal{H}^{\otimes N_{\ell(\Lambda)}}$ is represented by the unitary operator $V = \bigotimes_{(x,i)\in\ell(\Lambda)} V_{x,i}$, where the single-link operator $V_{x,i}$ is expressed in terms of a combination of left and right multiplications induced by the local gauge transformations on adjacent sites
\begin{equation}
  \label{Eq:GaugeTransformationSingleLink}
  V_{x,i} = L_{x,i}\big(\Omega_{x-\frac{a}{2}\hat{i}}\big) R_{x,i}\big(\Omega_{x+\frac{a}{2}\hat{i}}\big) .
\end{equation}
Any physical (i.e., gauge invariant) state $|\Psi\rangle\in\mathcal{H}_{\ell(\Lambda)}$ must obey Gauss's law
\begin{equation}
  V |\Psi\rangle = |\Psi\rangle.
\end{equation}

\subsection{Hamiltonian formulation}
\label{SubSec:Hamiltonian formulation}

The Hamiltonian of a standard Wilson-type lattice gauge theory is given by the sum of an electric and magnetic part
\begin{equation}
  H = H_E + H_B .
\end{equation}
For the following discussion it is useful to define orthonormal basis states on the set of links $\ell(\Lambda)$ of the spatial lattice $\Lambda$, i.e., $|\Psi_U\rangle\equiv \bigotimes_{(x,i)\in\ell(\Lambda)} |U_{x,i}\rangle$ in the link basis, as well as $|\Psi_{\Gamma}\rangle\equiv \bigotimes_{(x,i)\in\ell(\Lambda)} |\Gamma_{p;\hspace{1pt}x,i}, ab\rangle$ in the flux basis. $\Psi_U$ (and $\Psi_{\Gamma}$) denote a single configuration of parallel transporters (fluxes) on the links of the spatial lattice. The magnetic contribution to the Hamiltonian $H_B$ is given by
\begin{equation}
  H_B = \sum_{\Psi_U} H_B\left(\Psi_U\right) |\Psi_U\rangle\langle\Psi_U| ,
\end{equation}
while the electric contribution $H_E$ reads
\begin{equation}
  H_E = \sum_{\Psi_{\Gamma}} H_E\left(\Psi_{\Gamma}\right) |\Psi_{\Gamma}\rangle\langle\Psi_{\Gamma}| .
\end{equation}
Diagonalizing the full Hamiltonian $H$ is a nontrivial task in general that cannot be achieved analytically.

In the following we provide the general form of $H_B(\Psi_U)$ and $H_E(\Psi_{\Gamma})$ for arbitrary discrete groups, while specific examples, i.e., the cyclic group \mbox{$\Z(k)\subset U(1)$}, the symmetric group $S_3\subset O(2)$, the binary dihedral group $\bar{D}_2\subset SU(2)$, as well as the finite group \mbox{$\Delta(27)\subset SU(3)$}, are considered in Secs.\ \ref{SubSec:Standard Zk lattice gauge theory} -- \ref{SubSec:Standard Delta27 lattice gauge theory}.

\subsubsection{Plaquette-based magnetic Hamiltonian}
\label{SubSubSec:Plaquette-based magnetic Hamiltonian}

The magnetic energy $H_B(\Psi_U)$ for a given link configuration $\Psi_U$ is naturally expressed in terms of elementary plaquette variables $U_{\Box,x}$, $x\in\Lambda$, i.e.,
\begin{equation}
  \label{Eq:MagneticSingleLinkHamiltonian}
  H_B\left(\Psi_U\right) = \frac{2}{(e a)^2} \sum_{x\in\Lambda} h_{B}(U_{\Box,x}) , 
\end{equation}
where $e$ is the unit of electric charge with engineering dimension $[e] = 1$ and $h_{B}(U_{\Box,x})$ is the magnetic energy of a single plaquette. The plaquette variables are constructed by the group multiplication of four link variables
\begin{equation}
  U_{\Box,x} \equiv U_{x + \frac{a}{2} \hat{1},1} U_{x + a \hat{1} + \frac{a}{2} \hat{2},2} U_{x + a \hat{2} + \frac{a}{2} \hat{1},1}^{-1} U_{x + \frac{a}{2} \hat{2},2}^{-1} ,
\end{equation}
where the initial and final link variables are attached to the site $x\in\Lambda$. Under gauge transformations $\Omega_x\in G$ they transform as
\begin{equation}
  \label{Eq:PlaquetteGaugeTransformation}
  U_{\Box,x} \rightarrow \Omega_x U_{\Box,x} \Omega_x^{-1} .
\end{equation}

Since the magnetic contribution to the Hamiltonian has to be invariant under \eqref{Eq:PlaquetteGaugeTransformation}, it is clear that $h_B$ should take constant values on the conjugacy classes of $G$ [cf.\ \ref{SubSec:Conjugacy classes}]. Thus, we may express the magnetic energy of an elementary plaquette in terms of the character $\chi_{{}_{\Gamma}}$ for a given representation $\Gamma$ of the gauge group, i.e.,
\begin{equation}
  \chi_{{}_{\Gamma}}(U_{\Box,x}) = \Tr \Gamma(U_{\Box,x}) .
\end{equation}
It is natural to use a ``fundamental'' (irreducible) representation $\Gamma$ from which one can generate other representations by tensor product reduction [cf.\ \ref{SubSec:Irreducible representations and characters}]. Furthermore, if the considered group $G$ has a nontrivial center $Z(G)\subset G$, then $\Gamma$ should also provide a nontrivial representation if restricted to $Z(G)$.

In the following, we choose the single-plaquette energy to be of the following form
\begin{equation}
  a h_B(U_{\Box}) = d_{\Gamma} - \Re \chi_{{}_{\Gamma}}(U_{\Box}) ,
\end{equation}
where the additional constant term is introduced so that the magnetic energy is zero for vanishing magnetic flux, i.e., $a h_B(\1) = 0$.

\subsubsection{Link-based electric Hamiltonian}
\label{SubSubSec:Link-based electric Hamiltonian}

While the magnetic energy is expressed in terms of elementary plaquette variables, the electric energy $H_E(\Psi_{\Gamma})$ for a given flux configuration $\Psi_{\Gamma}$ is given by
\begin{equation}
  \label{Eq:ElectricSingleLinkHamiltonian}
  H_E(\Psi_{\Gamma}) = \frac{(e a)^2}{2} \sum_{(x,i)\in\ell(\Lambda)}\hspace{1pt} h_E\hspace{0.5pt}\big(\Gamma_{p;\hspace{1pt}x,i}\big) ,
\end{equation}
where $h_E\hspace{0.5pt}\big(\Gamma_{p;\hspace{1pt}x,i}\big)$ is the electric energy of a single link. The latter is defined such that the Hamiltonian 
\begin{equation}
  \label{Eq:SingleLinkHamiltonianFluxBasis}
  H_E^{\textrm{single-link}} = \frac{(e a)^2}{2} \sum_{p, ab} h_E(\Gamma_p) |\Gamma_p, ab \rangle \langle \Gamma_p, ab | ,
\end{equation}
describes a symmetric hopping between nearest neighbors in the group space $G$. If we change from the flux to the link basis  
\begin{equation}
  \label{Eq:SingleLinkElectricHamiltonianLinkBasis}
  \frac{2}{(e a)^2} \hspace{1pt}\langle U'| H_E^{\textrm{single-link}} |U\rangle = \frac{1}{d_G} \sum_p d_{\Gamma_p} h_E\left( \Gamma_p \right) \chi_{{}_{\Gamma_p}}(U' U^{-1}) , 
\end{equation}
and require that the single-link Hamiltonian takes the form of a (negative) discrete Laplacian on group space
\begin{equation}
  \frac{2}{e^2 a}\langle U'| H_E^{\textrm{single-link}} |U\rangle = - \delta_{\langle U, U'\rangle} + N_U \delta_{U,U'} , 
\end{equation}
then we obtain a condition that can be solved for the single-link energies. Here, $\delta_{\langle U, U'\rangle} = 1$ if the group elements $U$ and $U'$ are nearest neighbors, and $\delta_{\langle U, U'\rangle} = 0$ otherwise; $N_U$ is the number of nearest neighbors of $U$ in $G$. Whether or not $U$ and $U'$ are nearest neighbors is determined by the (pseudo)metric $\mu(U,U')$, which we define with respect to some unitary irreducible representation $\Gamma$ [cf.\ \ref{SubSec:Distance between group elements in G}]. While there is no unique choice for $\Gamma$, it is natural to employ the same representation that we have used to define the elementary-plaquette energy, i.e., a higher-dimensional representation, which reflects the possibly nontrivial center properties of the gauge group.

Note that the single-link Hamiltonian Eq.\ \eqref{Eq:SingleLinkHamiltonianFluxBasis} is manifestly gauge invariant, i.e.,
\begin{equation}
  [L(\Omega) R(\Omega'), H_E^{\textrm{single-link}}] = 0 ,
\end{equation}
where the left and right operators correspond to the action of a gauge transformation on both ends of the link. This follows immediately from the representation of the left- and right-multiplication operators in the flux basis [cf.\ Eq.\ \eqref{Eq:LeftRightMultiplicationFluxRepresentation}].

\subsection{Standard $\Z(k)$ lattice gauge theory}
\label{SubSec:Standard Zk lattice gauge theory}

The group elements of the cyclic group $\Z(k)$ can be represented in terms of the $k$-th roots of unity, i.e., $\Z(k) = \{ z_n = e^{2\pi i n/k} \hspace{1pt}|\hspace{1.5pt} n = 0, 1, \ldots, k-1 \}$, and every group element defines its own conjugacy class $\mathcal{C}_p$, with $p = 1,2,\ldots ,k$. The group $\Z(k)$ admits $k$ one-dimensional representations $\Gamma_p$, $p = 1,2,\ldots,k$, that describe single-particle excitations with $(p-1)$ units of charge, i.e., $\Gamma_p(z_n) = z_n^{p-1}$. Further properties of the cyclic groups are summarized in \ref{Sec:The Cyclic Group Zk}.

To define the magnetic energy of a single (elementary) plaquette we use the irreducible representation with unit charge, $\Gamma_2$,  whereby 
\begin{equation}
  a h_B(U_{\Box} = z_n) = 1 - \cos(2 \pi n/k) .
\end{equation}
The electric energy of a single link is defined such that the single-link Hamiltonian describes a nearest-neighbor hopping in group space. We refer to \ref{SubSec:Distance between group elements Zk} for a discussion of the distance between group elements in $\Z(k)$. By requiring that the single-link Hamiltonian should take the form of a discrete Laplacian\footnote{The case $k = 2$ is special, since there is only one nearest neighbor to each group element. In that case, we define \begin{equation}\frac{2}{e^2 a}\langle z_m|H_E^{\textrm{single-link}}|z_n\rangle = - \delta_{m,[n+1]_2} + \delta_{m,n} ,\end{equation} which yields the eigenvalues $a h_E(\Gamma_1) = 0$ and $a h_E(\Gamma_2) = 2$.}
\begin{equation}
  \label{Eq:DiscreteLaplacianCyclicGroup}
  \frac{2}{e^2 a}\langle z_m|H_E^{\textrm{single-link}}|z_n\rangle = - \delta_{m,[n+1]_k} - \delta_{m,[n-1]_k} + 2 \delta_{m,n} ,
\end{equation}
where $[n]_k \equiv n \,(\!\!\!\!\mod k)$, we obtain the spectrum of the single-link energies by diagonalization, i.e.,
\begin{equation}
  \label{Eq:SpectrkumZk}
  a h_E(\Gamma_p) = 2\!\left[ 1 - \cos(2 \pi (p-1)/k)\right] .
\end{equation}
Clearly, the spectrum is bounded from below, with lowest eigenvalue, $a h_E(\Gamma_{1}) = 0$.

\subsection{Standard $S_3$ lattice gauge theory}
\label{SubSec:Standard S3 lattice gauge theory}

The group elements of the symmetric group $S_3$ can be expressed in terms of their action on the set $\{ 1, 2, 3\}$. They can be divided into three conjugacy classes, which entail the identity element $\mathcal{C}_1 = \{ \1\}$, the set of pair permutations $\mathcal{C}_2 = \{ P_{12}, P_{23}, P_{31} \}$, and the set of cyclic permutations $\mathcal{C}_3 = \{ P_{231}, P_{321} \}$. This group admits two one-dimensional representations, $\Gamma_1$ and $\Gamma_2$, as well as a two-dimensional representation, $\Gamma_3$. Further properties of $S_3$ are provided in \ref{Sec:The Symmetric Group S3}.

Here, we define the magnetic energy of a single plaquette by using the character of the two-dimensional ``fundamental'' representation $\Gamma_3$, i.e.,
\begin{equation}
  a h_B(U_{\Box}) = 2 - \chi_{{}_{\Gamma_3}}(U_\Box) .
\end{equation}
Thus, we obtain 
\begin{subequations}
\begin{IEEEeqnarray}{RCl}
  a h_B(\mathcal{C}_1) &=& 0, \\
  a h_B(\mathcal{C}_2) &=& 2, \\
  a h_B(\mathcal{C}_3) &=& 3,
\end{IEEEeqnarray}
\end{subequations}
for the distinct conjugacy classes.

We determine the spectrum of the single-link Hamiltonian using the distance \eqref{Eq:Distance} on $S_3$ [cf.\ \ref{SubSec:Distance between group elements in S3}]. We obtain the following matrix representation
\begin{equation}
  \label{Eq:SingleLinkHoppingHamiltonianS3}
  \frac{2}{e^2 a} \langle U'|H_E^{\textrm{single-link}}|U\rangle = -\delta_{\langle U,U'\rangle} + 3 \delta_{U,U'} ,
\end{equation}
where the adjacency matrix $\delta_{\langle U,U'\rangle}$ is given by
\begin{equation}
  \label{Eq:NearestNeighborKroneckerDelta}
  \delta_{\langle U,U'\rangle} = (1 - \delta_{\mathcal{C},\mathcal{C}'}) \left[ 1 - \big(1 - \delta_{\ord(\mathcal{C}),d_{\mathcal{C}}^{\textrm{max}}}\big) \big(1 - \delta_{\ord(\mathcal{C}'),d_{\mathcal{C}}^{\textrm{max}}}\big) \right] .
\end{equation}
Here, we adopt the following short-hand notation: $\mathcal{C},\mathcal{C}'\in\{\mathcal{C}_1,\mathcal{C}_2,\mathcal{C}_3\}$ denote the conjugacy classes associated to the group elements $U$ and $U'$, i.e., $U\in\mathcal{C}$ and $U'\in\mathcal{C}'$, while the order (cardinality) of the largest conjugacy class is given by $d_{\mathcal{C}}^{\textrm{max}} = \max_p \ord(\mathcal{C}_p)$. We emphasize that Eq.\ \eqref{Eq:NearestNeighborKroneckerDelta} holds only for the two-dimensional irreducible representation $\Gamma_3$ of $S_3$, which enters the definition of the distance $\mu$ between group elements.

Eq.\ \eqref{Eq:NearestNeighborKroneckerDelta} effectively identifies group elements in the conjugacy classes $\mathcal{C}_1$ and $\mathcal{C}_3$ as far as the dynamics of the theory is concerned -- the Hamiltonian describes a hopping between group elements of the two distinct sets $\mathcal{C}_{13}\equiv \mathcal{C}_1\cup\hspace{1.5pt}\mathcal{C}_3$ and $\mathcal{C}_2$.\footnote{In slightly more mathematical terms: The definition of nearest neighbors given above implies that the single-link Hamiltonian of the $S_3$ gauge theory can be interpreted as a hopping between distinct group elements in the Abelianization of $S_3$, given by \mbox{$S_3\hspace{0.5pt}/\hspace{1pt}\Z(3)\cong \Z(2)$}. Recall that the Abelianization of the group $G$ is defined as $A_G = G / [G,G]$, where $[G,G]$ is the commutator subgroup of $G$, which is normal in $G$.} In fact, the same definition of nearest-neighbors, \mbox{Eq.\ \eqref{Eq:NearestNeighborKroneckerDelta}}, applies also for other finite discrete non-Abelian groups considered in this work, as long as the representation $\Gamma$ used to define the distance between group elements is chosen to be an irreducible representation of maximal degree for the considered gauge group. Therefore, the identification of different conjugacy classes with regard to the dynamics of the theory seems to be a general feature of these theories.

Upon diagonalization of Eq.\ \eqref{Eq:SingleLinkHoppingHamiltonianS3} we obtain the following spectrum
\begin{equation}
  a h_E(\Gamma_p) = 3 \big[ 1 - \chi_{{}_{\Gamma_p}}(\mathcal{C}_2) \big] ,
\end{equation}
which yields
\begin{subequations}
\begin{IEEEeqnarray}{RCl}
  a h_E(\Gamma_1) &=& 0 , \\
  a h_E(\Gamma_2) &=& 6 , \\
  a h_E(\Gamma_3) &=& 3 ,
\end{IEEEeqnarray}
\end{subequations}
for the different irreducible representations of $S_3$.

\subsection{Standard $\bar{D}_2$ lattice gauge theory}
\label{SubSec:Standard D2bar lattice gauge theory}

The group elements of the binary dihedral (i.e., quaternion) group $\bar{D}_2$ can be represented in terms of the $2\times 2$ matrices $\pm \1$ and $\pm i\sigma_{\alpha}$, $\alpha=1,2,3$, where $\sigma_{\alpha}$ denote the Pauli matrices. The elements of $\bar{D}_2$ can be organized into five distinct conjugacy classes \mbox{$\mathcal{C}_1 = \{ \1\}$}, $\mathcal{C}_2 = \{ -\1\}$, $\mathcal{C}_3 = \{ \pm i\sigma_1\}$, $\mathcal{C}_4 = \{ \pm i\sigma_2\}$, and $\mathcal{C}_5 = \{ \pm i\sigma_3\}$. We have five (inequivalent) irreducible representations $\Gamma_p$, four of which, $\Gamma_1$, $\Gamma_2$, $\Gamma_3$, and $\Gamma_4$, are one-dimensional while one of them, $\Gamma_5$, is two-dimensional. The theory of the binary dihedral group $\bar{D}_2$ is presented in \ref{Sec:D2bar}.

We define the single-plaquette magnetic energy by using the character of the two-dimensional fundamental representation $\Gamma_5$, i.e.,
\begin{equation}
  a h_B(U_\Box) = 2 - \Re \chi_{{}_{\Gamma_5}}(U_\Box) ,
\end{equation}
and
\begin{subequations}
\begin{IEEEeqnarray}{RCl}
  a h_B(\mathcal{C}_1) &=& 0 , \\
  a h_B(\mathcal{C}_2) &=& 4 , \\
  a h_B(\mathcal{C}_p) &=& 2 , \quad p = 3, 4, 5 .
\end{IEEEeqnarray}
\end{subequations}

The distances between $\bar{D}_2$ group elements are discussed in \ref{SubSec:Distance between group elements in D2bar}. The corresponding matrix elements of the single-link electric Hamiltonian take the following form
\begin{equation}
  \frac{2}{e^2 a}\langle U'|H_E^{\textrm{single-link}}|U\rangle = -\delta_{\langle U,U'\rangle} + 6 \delta_{U,U'} ,
\end{equation}
where $\delta_{\langle U,U'\rangle}$ is given by the same expression as in Eq.\ \eqref{Eq:NearestNeighborKroneckerDelta}. We find that the group elements in the conjugacy classes $\mathcal{C}_1$ and $\mathcal{C}_2$ are identified by the dynamics, such that a hopping in the group space of a single link is allowed only between elements of $\mathcal{C}_{12}\equiv\mathcal{C}_1\cup\hspace{1.5pt}\mathcal{C}_2$, $\mathcal{C}_3$, $\mathcal{C}_4$, and $\mathcal{C}_5$. The electric energy of a single link reads
\begin{equation}
  a h_E(\Gamma_p) = 6 - 2\!\left[\chi_{{}_{\Gamma_p}}(\mathcal{C}_3) + \chi_{{}_{\Gamma_p}}(\mathcal{C}_4) + \chi_{{}_{\Gamma_p}}(\mathcal{C}_5)\right] ,
\end{equation}
and for the various irreducible representations $\Gamma_p$, we have
\begin{subequations}
\begin{IEEEeqnarray}{RCl}
  a h_E(\Gamma_1) &=& 0 , \\
  a h_E(\Gamma_p) &=& 8 , \quad p = 2,3,4, \\
  a h_E(\Gamma_5) &=& 6 .
\end{IEEEeqnarray}
\end{subequations}

\subsection{Standard $\Delta(27)$ lattice gauge theory}
\label{SubSec:Standard Delta27 lattice gauge theory}

The group $\Delta(27)$ consists of $27$ elements that can be represented by $3\times 3$ matrices, $D(a,b)$, $U(a,b)$, and $L(a,b)$, with $a,b\in\{ 1, e^{2\pi i/3}, e^{-2\pi i/3}\}$, which are defined in \ref{Sec:Delta27}. The group admits 11 distinct conjugacy classes $\mathcal{C}_p$ and an equal number of irreducible representations $\Gamma_p$, $p = 1,2, \ldots, 11$. In our conventions, the representations $\Gamma_1$, $\Gamma_2$, \ldots, $\Gamma_9$ are one-dimensional, while $\Gamma_{10}$ and $\Gamma_{11}$ are both three-dimensional. Further properties of $\Delta(27)$ are summarized in the appendix.

We define the magnetic energy by using the character of the three-dimensional fundamental representation $\Gamma_{10}$ (or equivalently the anti-fundamental representation $\Gamma_{11} = \overline{\Gamma}_{10}$), i.e.,
\begin{equation}
  a h_B(U_\Box) = 3 - \Re \chi_{{}_{\Gamma_{10}}}(U_\Box) ,
\end{equation}
and
\begin{subequations}
\begin{IEEEeqnarray}{RCl}
  a h_B(\mathcal{C}_1) &=& 0 , \\
  a h_B(\mathcal{C}_p) &=& 9/2 , \quad p = 2, 3, \\
  a h_B(\mathcal{C}_q) &=& 3 , \qquad q = 4,5, \ldots, 11 .
\end{IEEEeqnarray}
\end{subequations}

The distances between $\Delta(27)$ group elements are discussed in \ref{SubSec:Distance between group elements in Delta27} and the corresponding matrix elements of the single-link electric Hamiltonian take the following form
\begin{equation}
  \frac{2}{e^2 a}\langle U'|H_E^{\textrm{single-link}}|U\rangle = -\delta_{\langle U,U'\rangle} + 24 \delta_{U,U'} ,
\end{equation}
where $\delta_{\langle U,U'\rangle}$ is given by Eq.\ \eqref{Eq:NearestNeighborKroneckerDelta}. The group elements in $\mathcal{C}_1$, $\mathcal{C}_2$, and $\mathcal{C}_3$ are combined into a single set $\mathcal{C}_{123}\equiv\mathcal{C}_1\cup\mathcal{C}_2\cup\mathcal{C}_3$, such that the dynamics can be viewed as a hopping between elements of distinct sets $\mathcal{C}_{123}$, $\mathcal{C}_4$, \ldots, $\mathcal{C}_{11}$ in group space. By the diagonalization of the single-link electric Hamiltonian we obtain the spectrum
\begin{equation}
  a h_E(\Gamma_p) = 24 - 3\!\left[\chi_{{}_{\Gamma_p}}(\mathcal{C}_4) + \dots + \chi_{{}_{\Gamma_p}}(\mathcal{C}_{11})\right] ,
\end{equation}
which evaluates to
\begin{subequations}
\begin{IEEEeqnarray}{RCl}
  a h_E(\Gamma_1) &=& 0 , \\
  a h_E(\Gamma_p) &=& 27, \quad p = 2,3, \dots,9 , \\ 
  a h_E(\Gamma_q) &=& 24, \quad q = 10, 11 . 
\end{IEEEeqnarray}
\end{subequations}

\section{Hamiltonian Formulation of Doubled Lattice Gauge Theories\\ with Discrete Gauge Group}
\label{Sec:Hamiltonian Formulation of Doubled Lattice Gauge Theories with Discrete Gauge Group}

In this section we introduce the doubled lattice gauge theories, which consist of two gauge fields associated with the links of the original lattice $\Lambda$ and its dual $\widetilde{\Lambda}$. We examine the realization of gauge transformations on quantum states when a noncommuting algebra of group multiplication operators is imposed on a pair of links, i.e., a direct link and its dual. Thus, in contrast to the standard lattice YM theories, the operator algebra is no longer link- but cross-based. The noncommuting algebra is characterized in terms of a one-cocycle, which we determine explicitly for the finite Abelian group $\Z(k)$ and the non-Abelian groups $S_3$, $\bar{D}_2$, and $\Delta(27)$. We refer to earlier work \cite{Olesen:2015baa} for an in depth discussion of the theory with gauge group $\Z(k)$. A full characterization of one-cocycles for arbitrary discrete groups will be provided elsewhere \cite{Mesterhazy:2016}. Here, we list all possible doubled theories for the finite non-Abelian gauge groups and investigate their spectra on a single cross, which allows us to construct physical states of the doubled lattice theory in the strong-coupling limit. Finally, we comment on the confinement of charges, emphasizing in particular the role of the center symmetry.

\subsection{Cross-based operator algebra}
\label{SubSec:Cross-based operator algebra}

\begin{figure}[!t]
  \centering
  \includegraphics[width=0.32\textwidth]{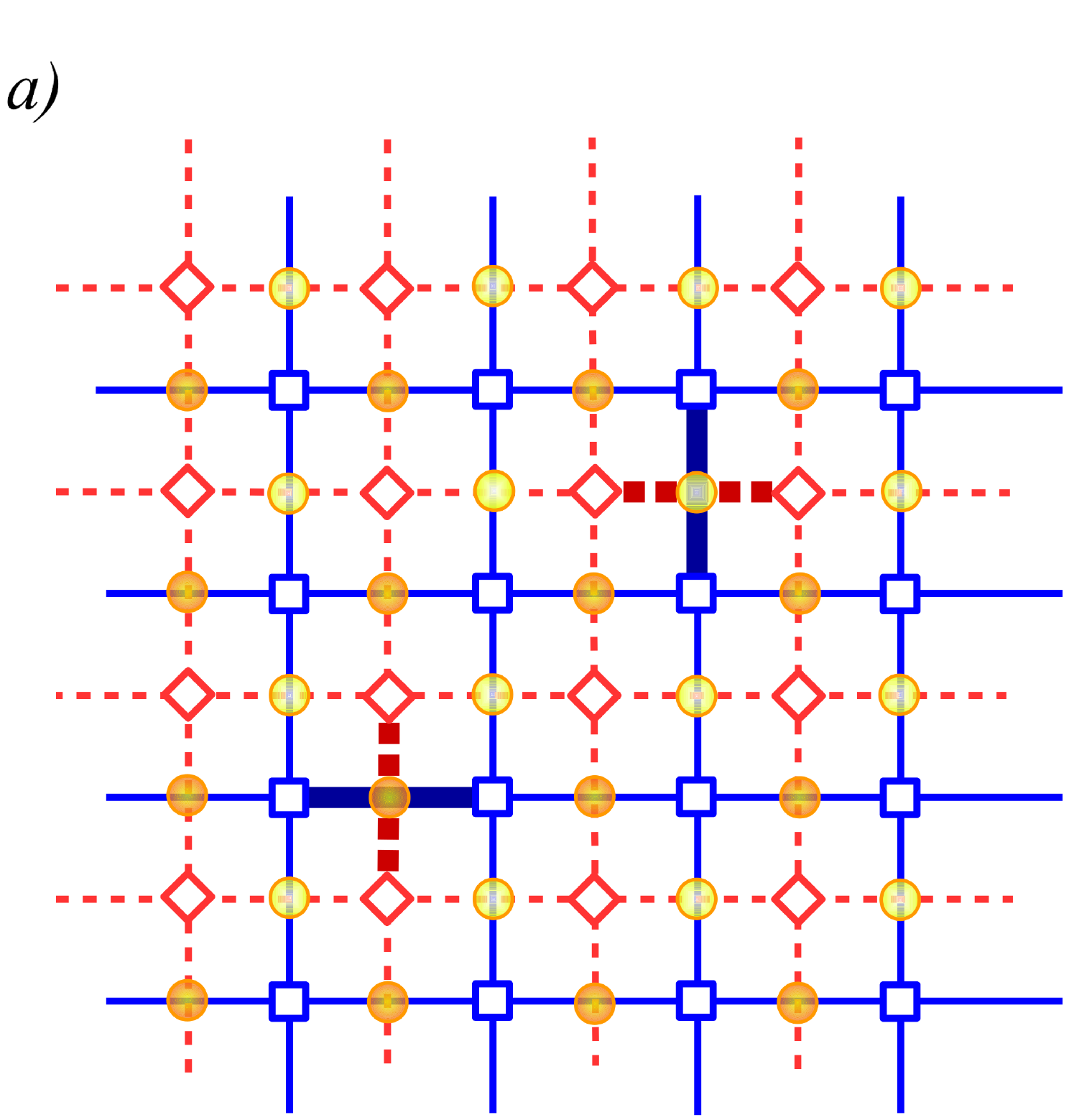}\hspace{20pt}
  \includegraphics[width=0.11\textwidth]{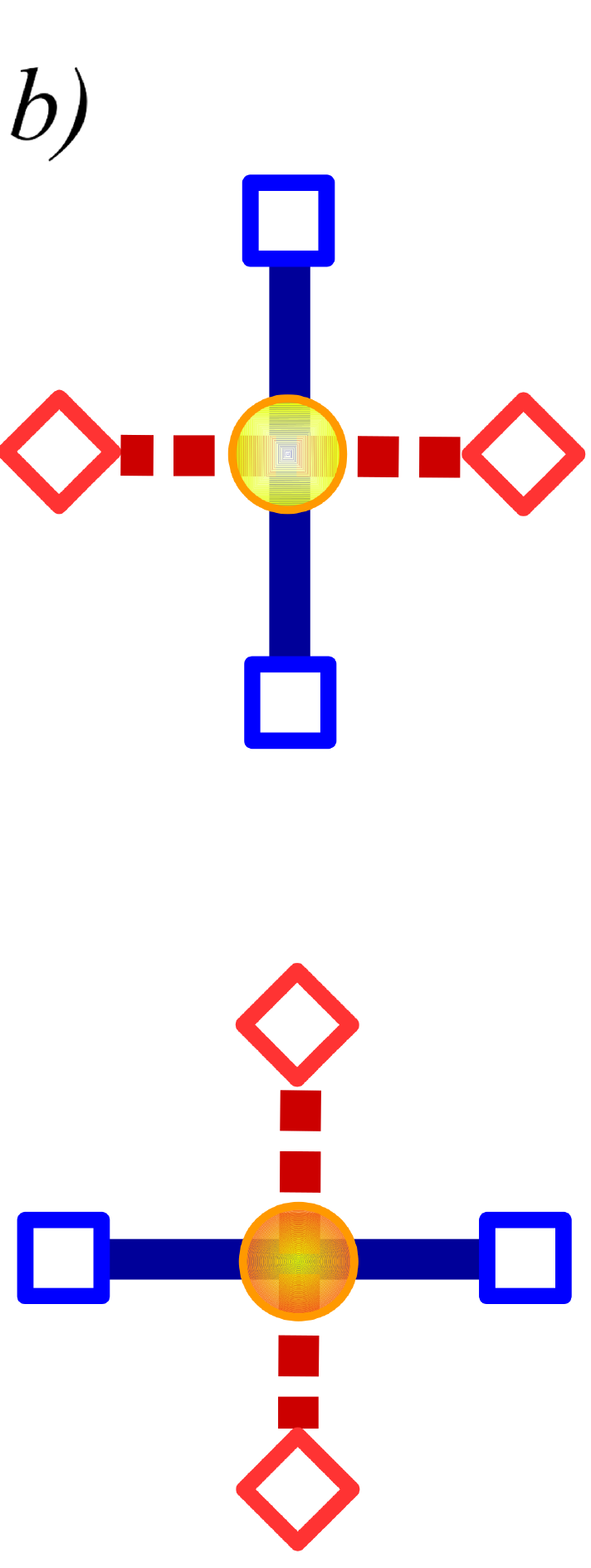}
  \caption{\label{Fig:DoubledLattice}a) Doubled lattice as a combination of the original square lattice $\Lambda$ and its dual $\widetilde{\Lambda}$. Circles denote the midpoints of pairs of links (crosses) on which the algebra of left- and right-multiplication operators is defined. b) The two distinct types of crosses on the doubled square lattice.}
\end{figure}

To construct the doubled lattice gauge theories we define the combined set of nearest-neighbor links $\ell(\Lambda)\cup\ell(\widetilde{\Lambda})$ from the square lattice $\Lambda$ and its dual $\widetilde{\Lambda}$. The basic dynamical variables in the doubled theory are pairs of parallel transporters, which are defined on a single link and its dual, i.e., $(U_{x,i}, U_{x,j})$, with $i\neq j$ and $i,j\in \{1,2\}$. We have two different types of crosses on the doubled lattice, e.g., $(U_{x,1},U_{x,2})$ or $(U_{x,2},U_{x,1})$ [cf.\ \mbox{Fig.\ \ref{Fig:DoubledLattice}}]. To avoid an unnecessary complication of notation we will drop the index $(x,i)$ in the following when single-link (single-cross) degrees of freedom are considered.

States on a single cross can be expressed in terms of linear combinations of orthonormal basis states $|U\rangle\otimes |U'\rangle\in\mathcal{H}\otimes\mathcal{H}$, where $\mathcal{H}$ corresponds to the Hilbert space of a single link. As in the standard lattice gauge theories that we discussed in the previous section, every link of the direct lattice carries two unitary operators $L$ and $R$ that realize left and right group multiplications on the associated link variable, as well as an operator $P$, which implements projections on a single link. They satisfy the operator algebra defined by \mbox{Eqs.\ \eqref{Eq:GroupMultiplicationLawLeftRightOperators} -- \eqref{Eq:LeftRightMultiplicationProjector}}. In addition, the operators $\widetilde{L}$, $\widetilde{R}$, and $\widetilde{P}$, are defined on the dual links and satisfy the same algebra. The projection operators $P$ and $\widetilde{P}$ commute, i.e., for $\Omega$, $\Omega'\in G$, we have
\begin{equation}
  \label{Eq:OperatorAlgebra1}
  [P(\Omega), \widetilde{P}(\Omega')] = 0 ,
\end{equation}
while the left- and right-multiplication operators associated with a direct link and its dual do not commute in general
\begin{equation}
  \label{Eq:OperatorAlgebra2}
  \widetilde{L}(\Omega') L(\Omega) = W_{L\widetilde{L}}(\Omega,\Omega') L(\Omega) \widetilde{L}(\Omega') .
\end{equation}
Clearly, $W_{\widetilde{L}L}(\Omega',\Omega) = W_{L\widetilde{L}}(\Omega,\Omega')^{-1}$ and similar relations apply for the other unitary group multiplication operators with an obvious adaption of notation. The set of $W$ factors $\{ W_{L\widetilde{L}}$, $W_{L\widetilde{R}}$, $W_{R\widetilde{L}}$, $W_{R\widetilde{R}}\}$ fully define the noncommuting algebra on a single cross. However, note that they are not necessarily independent.

Here, we employ the following operator representation
\begin{subequations}
  \begin{IEEEeqnarray}{RCl}
    L(\Omega)R(\Omega')\hspace{1pt}|U\rangle\otimes|U'\rangle &=& \omega_L(\Omega,U')\hspace{1pt}\omega_R(\Omega',U')|\Omega U\Omega'^{-1}\rangle\otimes|U'\rangle , \label{Eq:OneCocycle1} \\
    \widetilde{L}(\Omega)\hspace{1pt}\widetilde{R}(\Omega')\hspace{1pt}|U\rangle\otimes|U'\rangle &=& \omega_{\widetilde{L}}(\Omega,U)\hspace{1pt}\omega_{\widetilde{R}}(\Omega',U)|U\rangle\otimes|\Omega U'\Omega'^{-1}\rangle , \label{Eq:OneCocycle2}
  \end{IEEEeqnarray}
\end{subequations}
which is defined in terms of $U(1)$ phase factors $\omega_L$, $\omega_R$, $\omega_{\widetilde{L}}$, and $\omega_{\widetilde{R}}$. The latter depend both on the group element by which the state of the direct (dual) link is acted upon as well as the original state of the dual (direct) link. As we will show in the following section, \mbox{Eqs.\ \eqref{Eq:OneCocycle1} and \eqref{Eq:OneCocycle2}} may lead to nontrivial $W$ factors and therefore define possible realizations of a noncommuting operator algebra of group multiplication operators. Certainly, one might also consider more complicated representations of group multiplications (with a full dependence on the state of the cross, e.g., $\omega_L(\Omega, U, U')$, $\omega_{\widetilde{L}}(\Omega, U, U')$ etc.) that would similarly yield a noncommuting algebra. However, these considerations go beyond the scope of the present work and are left for future study \cite{Mesterhazy:2016}.

Apart from the link-dual-link basis given in terms of pairs of parallel transporters, i.e., \mbox{$|U\rangle\otimes |U'\rangle$}, we also introduce the flux-dual-link basis, which is defined by the set of orthonormal states $|\Gamma_p, ab\rangle\otimes |U\rangle$ and the link-dual-flux basis states, given by \mbox{$|U\rangle\otimes|\Gamma_p, ab\rangle$.} Note, however, that the noncommuting nature of group multiplication operators \eqref{Eq:OperatorAlgebra2} implies that we cannot (in general) define states on a single cross by simultaneously specifying the flux $\Gamma_p$ on both the direct link and its dual. This is only possible in the case of the naive (trivially) doubled YM theories. In addition to Eqs.\ \eqref{Eq:OneCocycle1} and \eqref{Eq:OneCocycle2}, we provide the action of the left and right operators in the flux-dual-link basis, which will be useful later on
\begin{subequations}
  \begin{IEEEeqnarray}{RCl}
    && \hspace{-20pt} L(\Omega)R(\Omega')\hspace{1pt}|\Gamma_p,ab\rangle\otimes |U\rangle \nonumber\\
    &=& \omega_L(\Omega,U)\hspace{0.5pt}\omega_R(\Omega',U) \left(\sum_{cd} \Gamma_p(\Omega)_{ca}^{\ast} \hspace{1pt}|\Gamma_p,cd\rangle\hspace{0.5pt}\Gamma_p(\Omega')_{db}\right)\!\otimes |U\rangle , \label{Eq:FluxDualLinkNontrivialOneCocycleA} \\
    && \hspace{-20pt} \widetilde{L}(\Omega)\widetilde{R}(\Omega')\hspace{1pt}|\Gamma_p,ab\rangle\otimes |U\rangle \nonumber\\
    &=& \frac{\sqrt{d_p}}{d_G} \left[ \sum_{q, cd} \sqrt{d_q} \hspace{1pt}\left(\hspace{1pt}\sum_{U'\in G} \omega_{\widetilde{L}}(\Omega,U')\hspace{0.5pt}\omega_{\widetilde{R}}(\Omega',U') \Gamma_p(U')_{ab} \Gamma_q(U')_{cd}^{\ast} \right) |\Gamma_q,cd\rangle\right]\!\otimes|\Omega U \Omega'^{-1}\rangle . \nonumber\\[-8pt] && \label{Eq:FluxDualLinkNontrivialOneCocycleB}
  \end{IEEEeqnarray}
\end{subequations}

\subsection{One-cocycles and noncommuting operator algebra}
\label{SubSec:One-cocycles and noncommuting operator algebra}

The phase factors introduced in Eqs.\ \eqref{Eq:OneCocycle1} and \eqref{Eq:OneCocycle2} essentially define the operator algebra of the lattice theory with discrete gauge group and here we establish their basic properties. Specific examples for the different discrete groups considered in this work are provided in the appendix. 
 
Let us apply a sequence of left-multiplication operators on the state of a single cross in the link-dual-link basis, e.g., $|U\rangle\otimes|U'\rangle$, whereby
\begin{equation}
  L(\Omega)L(\Omega')|U\rangle\otimes|U'\rangle = \omega_L(\Omega,U')\hspace{1pt}\omega_L(\Omega',U') |\Omega\Omega'U\rangle\otimes|U'\rangle .
\end{equation}
Using the group multiplication law \eqref{Eq:GroupMultiplicationLawLeftRightOperators} and 
\begin{equation}
  L(\Omega\Omega')|U\rangle\otimes|U'\rangle = \omega_L(\Omega\Omega',U') |(\Omega\Omega')\hspace{0.5pt}U\rangle\otimes|U'\rangle ,
\end{equation}
we obtain the following consistency condition
\begin{equation}
  \label{Eq:OneCocycleGroupMultiplication}
  \omega_L(\Omega\Omega',U) = \omega_L(\Omega,U)\hspace{1pt}\omega_L(\Omega',U) .
\end{equation}
Phase factors that satisfy this relation are called one-cocycles.\footnote{In fact, this defines a one-cocycle with trivial group action on the dual link.} Similar relations can be derived for $\omega_R$, $\omega_{\widetilde{L}}$, and $\omega_{\widetilde{R}}$. Thus, whenever possible, we will drop the $L$ index, which refers to the type of group multiplication. Eq.\ \eqref{Eq:OneCocycleGroupMultiplication} states that $\omega$ is a one-dimensional representation of $G$. That is, the one-cocycle $\omega(\Omega,U)$ assigns a one-dimensional representation $\Gamma_{p(U)}$ to every element $U\in G$, i.e.,
\begin{equation}
  \label{Eq:OneCocycleRepresentation}
  \omega(\Omega,U) = \Gamma_{p(U)}(\Omega) .
\end{equation}
Distinct one-cocycles $\omega_i$ are defined in terms of the functions $p_i$, $i = 1,2,\ldots$, which assign different one-dimensional representations to a given group element.\footnote{Since $\omega(\Omega\Omega', U) = \omega(\Omega'\Omega, U)$, the one-cocycle $\omega$ does not distinguish between noncommuting elements of $G$ and therefore descends to a representation of the Abelianization $A_G$ to itself. This observation is particularly useful in the classification of possible one-cocycles for arbitrary discrete groups \cite{Mesterhazy:2016}.}
  
Finally, we may derive an expression for the coefficient $W_{L\widetilde{L}}$. We proceed by applying Eq.\ \eqref{Eq:OperatorAlgebra2}, its defining equation, to a given state $|U\rangle\otimes|U'\rangle$ of a single cross. It is easy to check that this yields the following expression
\begin{equation}
  \label{Eq:WFactorOneCocyclesPreliminary}
  W_{L\widetilde{L}}(\Omega,\Omega') = \omega_{\widetilde{L}}(\Omega',\Omega U)\hspace{1pt}\omega_{\widetilde{L}}(\Omega',U)^{-1}\omega_L(\Omega,U')\hspace{1pt}\omega_L(\Omega,\Omega'U')^{-1} ,
\end{equation}
which apparently depends on the chosen state. However, additional consistency requirements remove this spurious state dependence. The group multiplication law for the left-multiplication operators $L$ and $\widetilde{L}$ implies that $W_{L\widetilde{L}}(\Omega,\Omega')$ should satisfy the group composition law with respect to both of its arguments. As an example, let us consider the group multiplication law in the second argument 
\begin{equation}
  W_{L\widetilde{L}}(\Omega,\Omega')W_{L\widetilde{L}}(\Omega,\Omega'') = W_{L\widetilde{L}}(\Omega,\Omega'\Omega'') ,
\end{equation}
and substitute \eqref{Eq:WFactorOneCocyclesPreliminary} into this relation. This imposes another constraint on the one-cocycles, which is of the following form
\begin{equation}
  \label{Eq:CocycleConsistencyCondition}
  \omega_L(\Omega,\Omega'U)\hspace{1pt}\omega_L(\Omega,\Omega''U) = \omega_L(\Omega,U)\hspace{1pt}\omega_L(\Omega,\Omega'\Omega''U) .
\end{equation}
Note that we may solve this equation by requiring
\begin{equation}
  \label{Eq:OneCocycleGroupMultiplication2}
  \omega_L(\Omega,U)\hspace{1pt}\omega_L(\Omega,U') = \omega_L(\Omega,U U') .
\end{equation}
If a solution to this equation can be found, then it clearly satisfies Eq.\ \eqref{Eq:CocycleConsistencyCondition}. In fact, solutions to Eq.\ \eqref{Eq:OneCocycleGroupMultiplication2} provide all possible cocycles up to an overall phase factor, which we may set to one (this corresponds to a particular choice of gauge; cf.\ Sec.\ \ref{SubSec:Hamiltonian formulation doubled lattice gauge theory}). They yield a $W$ factor, which is manifestly state independent
\begin{equation}
  \label{Eq:WFactorOneCocycles}
  W_{L\widetilde{L}}(\Omega,\Omega') = \omega_{\widetilde{L}}(\Omega',\Omega)\hspace{1pt}\omega_L(\Omega,\Omega')^{-1} .
\end{equation}
Similar expressions can be derived for the coefficients $W_{L\widetilde{R}}$, $W_{R\widetilde{L}}$, and $W_{R\widetilde{R}}$.

So far we have given only a representation of the $W$ factors in terms of the one-cocycles, but have not derived an explicit expression. From the representation of one-cocycles \eqref{Eq:OneCocycleRepresentation}, we may derive an alternative form of the consistency condition \eqref{Eq:OneCocycleGroupMultiplication2}:
\begin{equation}
  \Gamma_{p(U)}(\Omega) \otimes\Gamma_{p(U')}(\Omega) = \Gamma_{p(UU')}(\Omega) ,
\end{equation}
which is a constraint on the function $p$. Solving this equation is a simple exercise in group theory [cf.\ \ref{SubSec:Conjugacy classes and irreducible representations}]. Explicit $W$ factors for the different groups considered in this work are listed in the appendix.

\subsection{Gauge transformations and Gauss's law}

We have already introduced the unitary operator that realizes gauge transformations on the links of the direct lattice, Eq.\ \eqref{Eq:GaugeTransformationSingleLink}. Here, we provide the operator $\widetilde{V} = \bigotimes_{(x,i)\in\ell(\widetilde{\Lambda})} \widetilde{V}_{x,i}$ that acts on the set of dual links $\ell(\widetilde{\Lambda})$ and is induced by gauge transformation on the dual sites $x\in\widetilde{\Lambda}$. On a single (dual) link $(x,i)$, we define the operator
\begin{equation}
  \widetilde{V}_{x,i} = \widetilde{L}_{x,i}\big(\Omega_{x-\frac{a}{2}\hat{i}}\big) \widetilde{R}_{x,i}\big(\Omega_{x+\frac{a}{2} \hat{i}}\big) .
\end{equation}
Gauge transformations on the full Hilbert space of the doubled theory are built from the successive application of operators $(V\otimes\Id)$ and $(\Id\otimes\widetilde{V})$ on the direct and dual lattice, respectively, where $\Id$ denotes the unit operator on the Hilbert space of the direct and dual lattice. Both operators associated to the gauge transformation on the direct and dual lattice commute, which is a direct consequence of the imposed lattice symmetries.

States $|\Psi\rangle\in\mathcal{H}_{\ell(\Lambda)}\otimes\mathcal{H}_{\ell(\widetilde{\Lambda})}$ that satisfy
\begin{equation}
  (V\otimes\Id)|\Psi\rangle = (\Id\otimes\widetilde{V})|\Psi\rangle = |\Psi\rangle ,
\end{equation}
define the physical Hilbert space of gauge-invariant states.

\subsection{Hamiltonian formulation}
\label{SubSec:Hamiltonian formulation doubled lattice gauge theory}

As in the case of the standard Wilson-type lattice gauge theories, the full Hamiltonian of the doubled theory decomposes into an electric and a magnetic part $H = H_E + H_B$. In the following we first provide the full Hamiltonian for the trivially doubled YM theory. Then, we turn to the CS-type lattice gauge theories.

For the trivially doubled YM theory, the magnetic Hamiltonian is naturally expressed as a sum of different plaquette contributions, both from the direct and dual lattice, i.e.,
\begin{equation}
  \label{Eq:DoubledYMMagneticHamiltonian}
  H_B = \sum_{\Psi_U} \left\{ H_B(\Psi_U) \hspace{2pt}|\Psi_U\rangle \langle\Psi_U|\otimes \Id + \widetilde{H}_B(\Psi_U) \hspace{2pt}\Id\otimes|\Psi_U\rangle \langle\Psi_U| \right\} ,
\end{equation}
where $H_B (\Psi_U)$ is defined according to Eq.\ \eqref{Eq:MagneticSingleLinkHamiltonian} and
\begin{equation}
  \widetilde{H}_B (\Psi_U) =  \frac{2}{(e a)^2} \sum_{x\in\widetilde{\Lambda}} h_B\big(U_{\Box,x}\big) .
\end{equation}
Note that the same expression is used for the single-plaquette energy (and coupling) on the direct and dual lattice respectively. This is not a necessary assumption for the following construction, but we restrict ourselves to this scenario in the following. The electric contribution to the Hamiltonian reads
\begin{equation}
  H_E = \sum_{\Psi_{\Gamma}} \left\{ H_E(\Psi_{\Gamma}) \hspace{2pt}|\Psi_{\Gamma}\rangle \langle\Psi_{\Gamma}|\otimes\Id + \widetilde{H}_E(\Psi_{\Gamma}) \hspace{2pt}\Id\otimes|\Psi_{\Gamma}\rangle \langle\Psi_{\Gamma}| \right\} ,
\end{equation}
with $H_E (\Psi_{\Gamma})$ given by Eq.\ \eqref{Eq:ElectricSingleLinkHamiltonian} and
\begin{equation}
  \widetilde{H}_E (\Psi_{\Gamma}) = \frac{(e a)^2}{2} \sum_{(x,i)\in\ell(\widetilde{\Lambda})} h_E\big(\Gamma_{p;\hspace{1pt}x,i}\big) .
\end{equation}
Again, we choose the same functional form and coupling strength for the direct and dual lattice.

Within the deformed, CS-type theories the magnetic part of the Hamiltonian \eqref{Eq:DoubledYMMagneticHamiltonian} remains unaltered -- only the electric contribution is modified. To outline the essential steps in this construction, it is convenient to consider the electric Hamiltonian only on a single cross. It takes the following form
\begin{equation}
  H_E^{\textrm{single-cross}} = \frac{(e a)^2}{2} \sum_{p, ab} h_E(\Gamma_p) \left\{ |\Gamma_p, ab \rangle\langle \Gamma_p, ab|\otimes\Id + \Id\otimes|\Gamma_p, ab\rangle\langle\Gamma_p, ab| \right\} ,
\end{equation}
where, by a slight abuse of notation, $\Id$ denotes the identity operator on the single-link Hilbert space. If we change to the link-dual-link basis, we obtain
\begin{IEEEeqnarray}{RCl}
  && \hspace{-28pt}\frac{2}{e^2 a}\langle U_1'|\otimes \langle U_2'|\hspace{1.5pt} H_E^{\textrm{single-cross}}\hspace{2pt}| U_1\rangle\otimes | U_2\rangle \nonumber\\[5pt]
  &=& \frac{1}{d_G} \sum_p d_{\Gamma_p} a h_E\left( \Gamma_p \right) \left\{ \delta_{U_2,U_2'} \hspace{1pt}\chi_{{}_{\Gamma_p}}(U_1' U_1^{-1}) + \delta_{U_1,U_1'} \hspace{1pt}\chi_{{}_{\Gamma_p}}(U_2' U_2^{-1}) \right\} \\[0pt]
  &=& \left( -\delta_{\langle U_1,U_1'\rangle} + N_{U_1} \delta_{U_1,U_1'} \right) \delta_{U_2,U_2'} + \left(  -\delta_{\langle U_2,U_2'\rangle} + \widetilde{N}_{U_2} \delta_{U_2,U_2'} \right) \delta_{U_1,U_1'} ,
\end{IEEEeqnarray}
where $N_{U_1}$ and $\widetilde{N}_{U_2}$ denote the number of nearest-neighbor elements of $U_1$ and $U_2$ in the group space of the direct and dual link, respectively. We see that in the trivially doubled YM theory two hopping contributions are included -- one on the direct lattice and another one on its dual. To arrive at a doubled theory of CS type we need to account for the noncommutativity of the single-cross algebra [cf.\ Sec.\ \ref{SubSec:Cross-based operator algebra}]. Thus, we expect that the hopping on the link of the direct lattice will depend on the state of the dual link (and vice versa). Here, we consider the following class of Hamiltonians
\begin{IEEEeqnarray}{RCl}
  && \hspace{-38pt}\frac{2}{e^2 a}\langle U_1'|\otimes \langle U_2'|\hspace{1.5pt} H_E^{\textrm{single-cross}}\hspace{2pt}| U_1\rangle\otimes | U_2\rangle \nonumber\\ &=& \left( - e^{i\varphi(\Omega_1, U_2)} \delta_{\langle U_1,U_1'\rangle} + N_{U_1} \delta_{U_1,U_1'} \right) \delta_{U_2,U_2'} + \left(  - e^{i\widetilde{\varphi}(\Omega_2, U_1)} \delta_{\langle U_2,U_2'\rangle} + \widetilde{N}_{U_2} \delta_{U_2,U_2'} \right) \delta_{U_1,U_1'} ,
  \label{Eq:HamiltonianOneCocycle}
\end{IEEEeqnarray}
where $\Omega_1 = U_1' U_1^{-1}$ and $\Omega_2 = U_2' U_2^{-1}$. The additional complex phase factors can be interpreted as two independent $U(1)$ artificial gauge fields, which manifest themselves as parallel transporters connecting nearest-neighbor points in the discrete group space. We see that the dynamics governed by the single-cross electric Hamiltonian admits a simple analogy in terms of the motion of a charged particle in the presence of a $U(1)\times U(1)$ gauge field. By the successive hopping in the discrete group space $G\times G$ this abstract particle accumulates a phase. Note that this picture is similar to the one proposed in Ref.\ \cite{Olesen:2015baa} for the doubled theory with Abelian gauge group.

Both phase factors, $e^{i\varphi}$ and $e^{i\widetilde{\varphi}}$, satisfy the group composition law and define a one-cocycle. E.g., on the direct link we have
\begin{equation}
  e^{i \varphi(\Omega, U)} e^{i \varphi(\Omega', U)} = e^{i \varphi(\Omega\Omega', U)} ,
\end{equation}
with $\Omega$, $\Omega'$, and $U\in G$. However, the individual phase factors have no physical significance; the physically relevant quantity is an Aharonov-Bohm phase $\Phi(\Omega_1,\Omega_2)$ \cite{Aharonov:1959fk}, which is accumulated for a sequence of nontrivial group transformations $\Omega_1$ and $\Omega_2$, i.e.,\footnote{Note that such a phase cannot be accumulated on the direct (or dual) lattice alone, since the following identity holds
\begin{equation}
  e^{i \varphi(\Omega_1, U)} e^{i \varphi(\Omega_2, U)} \cdots e^{i \varphi(\Omega_n, U)} = e^{i \varphi\left(\prod_{l = 1}^n \Omega_l, U\right)} = 1 ,
\end{equation}
for group elements $\Omega_l\neq\1$, $l = 1,2, \ldots, n$, with $\prod_{l = 1}^n \Omega_l = \1$. The last equality in the above equation follows from the group property, which states that $\varphi(\1, U) = 0 \!\!\mod(2\pi)$.}
\begin{equation}
  \label{Eq:AhoronovBohmPhase}
  e^{i \varphi(\Omega_1, U_2)} e^{i \widetilde{\varphi}(\Omega_2, \Omega_1 U_1)}  e^{i \varphi(\Omega_1^{-1},\Omega_2 U_2)} e^{i \widetilde{\varphi}(\Omega_2^{-1},U_1)} = e^{i \Phi(\Omega_1,\Omega_2)} .
\end{equation}
We recognize that the factor $e^{i\Phi}$ can be identified with the $W$ factors introduced in the previous section, which are similarly defined by a product of one-cocycles. Thus, any Hamiltonian \eqref{Eq:HamiltonianOneCocycle} on a single cross is distinguished by an Aharonov-Bohm phase $\Phi$, which may be classified by determining the admissible one-cocycles. Listing all possible $W$ factors for a given group $G$, we obtain the complete set of distinct doubled theories, either of YM or CS type. This outlines our program for the remaining sections of this work.

It is always possible to choose a gauge, so that the accumulated phases $\Phi$ are expressed in terms of a single one-cocycle. In the following we impose the consistency condition
\begin{equation}
  e^{i \varphi(\Omega, U)} e^{i \varphi(\Omega, U')} = e^{i \varphi(\Omega, U U')} ,
\end{equation}
for which the above expression \eqref{Eq:AhoronovBohmPhase} becomes independent of the state of the cross, i.e.,
\begin{equation}
  e^{i \widetilde{\varphi}(\Omega_2, \Omega_1)}  e^{i \varphi(\Omega_1^{-1},\Omega_2)} = e^{i \Phi(\Omega_1,\Omega_2)} .
\end{equation}
Using $\varphi(\Omega_1^{-1},\Omega_2) = -\varphi(\Omega_1,\Omega_2) \!\!\mod(2\pi)$, which follows from the group property, we recognize that the condition $\varphi(\Omega_1, \Omega_2) = \widetilde{\varphi}(\Omega_2, \Omega_1)\!\!\mod (2\pi)$ implies that $\Phi(\Omega_1,\Omega_2) = 0 \!\!\mod (2\pi)$ and therefore does not constitute a gauge choice. Instead, we employ the asymmetric gauge with
\begin{subequations}
  \begin{IEEEeqnarray}{RCl}
    \varphi(\Omega_1, \Omega_2) &=& 0 \!\!\!\mod(2\pi), \label{Eq:AsymmetricGauge1} \\
    \Phi(\Omega_1,\Omega_2) &=& \widetilde{\varphi}(\Omega_2, \Omega_1 ) \!\!\!\!\mod (2\pi) , \label{Eq:AsymmetricGauge2}
  \end{IEEEeqnarray}
\end{subequations}
for all $\Omega_1, \Omega_2$ in $G$, which leaves no residual gauge degrees of freedom. Of course, the gauge choice $\widetilde{\varphi}(\Omega_2, \Omega_1)=0 \!\!\mod(2\pi)$ is equally valid and there is no reason to prefer one over the other. In the following the former will always be assumed. 

An important consequence of the gauge fixing is that it also constrains the action of the left- and right-multiplication operators. While Eqs.\ \eqref{Eq:AsymmetricGauge1} and \eqref{Eq:AsymmetricGauge2} imply that a nontrivial one-cocycle is introduced on the dual link in the Hamiltonian \eqref{Eq:HamiltonianOneCocycle}, the requirement that the Hamiltonian should be invariant under gauge transformations
\begin{equation}
  [L(\Omega) R(\Omega'), H_E^{\textrm{single-cross}}] = [\widetilde{L}(\Omega) \widetilde{R}(\Omega'), H_E^{\textrm{single-cross}}] = 0, 
\end{equation}
completely determines the action of these group multiplication operators. In particular, we find that for given $e^{i\widetilde{\varphi}}$ (and $e^{i\varphi} = 1$) the one-cocycles associated with left- and right-multiplications read
\begin{equation}
  \omega_L(\Omega,U) = \omega_R(\Omega,U)^{\ast} = e^{i\widetilde{\varphi}(\Omega,U)} ,
\end{equation}
which yields $\omega_L(\Omega,U)\omega_R(\Omega',U) = \omega_L(\Omega\Omega'^{-1},U) = \omega_R(\Omega^{-1}\Omega',U)$, and
\begin{equation}
  \omega_{\widetilde{L}}(\Omega,U) = \omega_{\widetilde{R}}(\Omega,U)^{\ast} = 1 .
\end{equation}
Finally, note that these relations imply the following identities among the $W$ factors of the operator algebra \eqref{Eq:OperatorAlgebra2}:
\begin{equation}
  W_{L\widetilde{L}}(\Omega,\Omega') = W_{R\widetilde{L}}(\Omega,\Omega')^{\ast} = W_{L\widetilde{R}}(\Omega,\Omega') =  W_{R\widetilde{R}}(\Omega,\Omega')^{\ast} = e^{-i \widetilde{\varphi}(\Omega,\Omega')} ,
\end{equation}
In summary, we can say that any choice of group multiplication operator algebra (consistent with the above gauge fixing) determines a corresponding Hamiltonian (and vice versa). The admissible theories are distinguished by a single independent $W$ factor (or, equivalently, a single one-cocycle $\omega_L = \omega_R^{\ast} = e^{i \widetilde{\varphi}}$).

In the following sections, we determine the spectrum of the Hamiltonian \eqref{Eq:HamiltonianOneCocycle}, in the asymmetric gauge \eqref{Eq:AsymmetricGauge1} and \eqref{Eq:AsymmetricGauge2}, and in the strong-coupling regime, when the magnetic contribution to the Hamiltonian can be neglected. In this limit, the dynamics is fully characterized by the electric part of the Hamiltonian, which describes a hopping between group elements. For the CS-type theories the hopping amplitude is complex in general, which arises through the coupling to an abstract $U(1)\times U(1)$ gauge field \cite{Olesen:2015baa}. Distinct theories are defined in terms of the corresponding Aharonov-Bohm phase, which is accumulated by a nontrivial set of group transformations on a single link and its dual. We provide the spectrum of these theories on a single cross, given by the energies $E_n$, $n = 0, 1, \ldots$ and characterize the properties of the ground state and first excited states, with spectral gap $\Delta = E_1 - E_0$. Note that the energy gap becomes infinite in the strong-coupling limit $(e a)^2\rightarrow \infty$.

\subsection{Doubled $\Z(k)$ lattice gauge theory}

The allowed one-cocycles for the doubled $\Z(k)$ lattice gauge theory are derived in \ref{SubSec:One-cocycles and W factors Zk}. Introducing a trivial cocycle, one obtains a doubled theory whose spectrum simply results from the independent combination of the single-link spectra [cf.\ Eq.\ \eqref{Eq:SpectrkumZk}]. A nontrivial doubled CSYM theory is obtained when one applies one of the $k-1$ nontrivial cocycles. While each of these theories is associated with a different Aharonov-Bohm phase, they might not necessarily yield a different spectrum.

\vspace{-4pt}

\subsubsection{Trivial one-cocycle}

In the trivially doubled theory, the unique ground state $E_0 / (e^2 a) = 0$ of the single-cross Hamiltonian $H_E^{\textrm{single-cross}}$ is given by 
\begin{equation}
  \label{Eq:GroundStateZkTrivialCocycle}
  |E_0\rangle =\frac{1}{\sqrt{k}}\sum_{n=0}^{k-1} \hspace{1pt}|\Gamma_1\rangle\otimes|z_n\rangle ,
\end{equation}
in the flux-dual-link basis,  while the first excited state is always four-fold degenerate (with the single exception of the $\Z(2)$ theory, in which case the first excited state has a two-fold degeneracy).

\vspace{-2.5pt}

\subsubsection{Nontrivial one-cocycles}

Introducing a nontrivial one-cocycle in the doubled $\Z(k)$ gauge theory, the lowest-energy state becomes degenerate. The properties of these states depend on the employed cocycle and whether the order of the group $k$ is prime. If $k$ is prime, then the ground state of all CS-type theories are $k$-fold degenerate. Here, we provide only the example of the gauge group $\Z(2)$, for which the two degenerate ground states are given by
\begin{equation}
  |E_0, a\rangle = \frac{1}{\sqrt{2\big(2 + \sqrt{2}\big)}}\left[\big(1+\sqrt{2}\big) |\Gamma_1\rangle\otimes|z_{[a]_2}\rangle + |\Gamma_2\rangle\otimes|z_{[a+1]_2}\rangle \right] ,
  \label{Eq:GroundStatesZ2NontrivialCocycle}
\end{equation}
and $a = 1,2$. Similarly, the first excited states are modified in the presence of a nontrivial one-cocycle. However, since their form depends on the particular case considered, we do not list them here and refer to \cite{Olesen:2015baa} instead for further details of the doubled $\Z(k)$ lattice gauge theories of CS type.

\vspace{-2.5pt}

\subsection{Doubled $S_3$ lattice gauge theory}
\label{SubSec:Doubled S3 lattice gauge theory}

The possible one-cocycles for the doubled $S_3$ lattice gauge theory are summarized in \ref{SubSection:One-cocycles and W factors S3}. Introducing a trivial cocycle yields a naively doubled YM theory. For the group $S_3$ there is only one doubled CS-type theory, corresponding to the single admissible one-cocycle, which is nontrivial.

\vspace{-2.5pt}

\subsubsection{Trivial one-cocycle}

The spectrum of the trivially doubled YM theory is summarized in Tab.\ \ref{Tab:DoubledYMGroupS3}.

\begin{table}[!h]
  \begin{center}
    \begin{tabular}{|c||C{40pt}|C{40pt}|C{40pt}|C{40pt}|C{40pt}|}
      \hline
      & $n = 0$ & $n = 1$ & $n = 2$ & $n = 3$ & $n = 4$ \\
      \hline
      \hline
      $E_n / (e^2a)$ & $0$ & $3/2$ & $3$ & $9/2$ & $6$ \\
      \hline
      $g_n$ & 
      $1$ & $8$ & $18$ & $8$ & $1$ \\
      \hline
    \end{tabular}
    \caption{\label{Tab:DoubledYMGroupS3}Eigenvalues $E_n$ and their degeneracies $g_n$, $n = 0,1,\ldots$, of the single-cross electric Hamiltonian $H_E^{\textrm{single-cross}}$ for the doubled $S_3$ lattice YM theory (with $g_0 = 1$ ground-state degeneracy).}
  \end{center}
\end{table}

Its unique ground state, with energy $E_0 / (e^2 a) = 0$, is given by the symmetric superposition in the flux-dual-link basis
\begin{equation}
  |E_0\rangle =\frac{1}{\sqrt{6}}\sum_{U\in S_3}|\Gamma_1,11\rangle\otimes|U\rangle .
\end{equation}
As it should, this state is invariant under group multiplication on the direct and dual link, respectively. In addition, we provide the first excited states, with energy gap $\Delta / (e^2 a) = 3/2$, that are modified by the introduction of a nontrivial one-cocycle [cf.\ Sec.\ \ref{SubSec:Nontrivial one-cocycle S3}]. In the naive doubled YM theory, these eigenstates are given by
\begin{equation}
  |E_1, ab\rangle = \frac{1}{\sqrt{6}}\sum_{U\in S_3}|\Gamma_3,ab\rangle\otimes|U\rangle ,
\end{equation}
which we label by the indices $a,b = 1,2$, corresponding to the two-dimensional, irreducible $\Gamma_3$ representation. Applying the left- and right-multiplication operators on these states, we obtain
\begin{subequations}
  \begin{IEEEeqnarray}{RCl}
    L(\Omega)R(\Omega')|E_1, ab\rangle &=& \sum_{cd} \Gamma_3(\Omega)_{ca}^{\ast} \hspace{1pt}|E_1, cd\rangle \hspace{0.5pt}\Gamma_3(\Omega')_{db}, \\
    \widetilde{L}(\Omega)\widetilde{R}(\Omega')|E_1, ab\rangle &=& |E_1, ab\rangle ,
  \end{IEEEeqnarray}  
\end{subequations}
cf.\ Eqs.\ \eqref{Eq:FluxDualLinkNontrivialOneCocycleA} and \eqref{Eq:FluxDualLinkNontrivialOneCocycleB}.

\subsubsection{Nontrivial one-cocycle}
\label{SubSec:Nontrivial one-cocycle S3}

The spectrum of the nontrivial doubled CSYM theory is summarized in Tab.\ \ref{Tab:DoubledCSYMGroupS3}.

\begin{table}[!h]
  \begin{center}
    \begin{tabular}{|c||C{60pt}|C{40pt}|C{40pt}|C{40pt}|C{60pt}|}
      \hline
      & $n = 0$ & $n = 1$ & $n = 2$ & $n = 3$ & $n = 4$ \\
      \hline\hline
      $E_n / (e^2a)$ & $3/2\hspace{1pt}(2-\sqrt{2})$ & $3/2$ & $3$ & $9/2$ & $3/2\hspace{1pt}(2+\sqrt{2})$ \\
      \hline
      $g_n$ & $2$ & $8$ & $16$ & $8$ & $2$ \\
      \hline
    \end{tabular}
  \end{center}
  \caption{\label{Tab:DoubledCSYMGroupS3}Eigenvalues $E_n$ and their degeneracies $g_n$, $n = 0,1,\ldots$, of the single-cross electric Hamiltonian $H_E^{\textrm{single-cross}}$ for the doubled $S_3$ lattice CSYM theory (with $g_0 = 2$ ground-state degeneracy).}
\end{table}

In the presence of a nontrivial one-cocycle, cf.\ \ref{SubSection:One-cocycles and W factors S3},
\begin{equation}
  \omega_L(\Omega,U) = \omega_R(\Omega,U)^{\ast} = \omega_2(\Omega,U) ,
\end{equation}
the ground state energy is shifted to $E_0 / (e^2 a) = 3/2\hspace{1pt}(2 - \sqrt{2})\approx 0.879$ and picks up a two-fold degeneracy. The corresponding orthonormal eigenstates are given by
\begin{subequations}
  \begin{IEEEeqnarray}{RCl}
    |E_0, 1\rangle &=& \frac{1}{\sqrt{6\left(2 + \sqrt{2}\right)}}\left[\left(1+\sqrt{2}\right) \sum_{U\in\mathcal{C}_2} |\Gamma_1,11\rangle\otimes|U\rangle + \sum_{U\in\mathcal{C}_{13}}|\Gamma_2,11\rangle\otimes |U\rangle\right] , \\
    |E_0, 2\rangle &=& \frac{1}{\sqrt{6\left(2 + \sqrt{2}\right)}}\left[\left(1+\sqrt{2}\right) \sum_{U\in\mathcal{C}_{13}}|\Gamma_1,11\rangle\otimes |U\rangle + \sum_{U\in\mathcal{C}_2} |\Gamma_2,11\rangle\otimes|U\rangle\right] , 
  \end{IEEEeqnarray}
\end{subequations}
where the summation $\sum_{U\in\mathcal{C}}$ runs over distinct elements of the conjugacy class $\mathcal{C}$. It is useful to compare these states to those obtained in the CS-type $\Z(2)$ theory \eqref{Eq:GroundStatesZ2NontrivialCocycle}, which show a similar structure.

In the CS-type doubled theory, the action of the left- and right-multiplication operators on these states is nontrivial. While the operators $L(\Omega)$, $\widetilde{L}(\Omega)$, etc.\ with $\Omega\in\mathcal{C}_{13}$ leave each of the ground states invariant, those operators that induce a shift $\Omega\in\mathcal{C}_2$ in group space act as follows
\newpage
\begin{subequations}
  \begin{IEEEeqnarray}{RCl}
    L(\mathcal{C}_2) |E_0,a\rangle &=& R(\mathcal{C}_2) |E_0,a\rangle = (-1)^{a} |E_0,a\rangle , \\
    \widetilde{L}(\mathcal{C}_2) |E_0, a\rangle &=& \widetilde{R}(\mathcal{C}_2) |E_0, a\rangle = |E_0, [\hspace{1pt}a\hspace{0.5pt}]_2 + 1\rangle ,
  \end{IEEEeqnarray}
\end{subequations}
where $a = 1,2$, and 
\begin{equation}
  L(\mathcal{C}_2) \widetilde{L}(\mathcal{C}_2) = -\widetilde{L}(\mathcal{C}_2) L(\mathcal{C}_2) .
\end{equation}
Here, it is understood that $L(\mathcal{C}_2)$, $\widetilde{L}(\mathcal{C}_2)$, etc.\ correspond to any operator $L(\Omega)$, $\widetilde{L}(\Omega)$ with the group element $\Omega$ in the set $\mathcal{C}_2$. We observe that the noncommuting algebra of group multiplication operators yields the $\Z(2)$ toric code if restricted to the set of ground states. Thus, the non-Abelian $S_3$ lattice gauge theory of CS type effectively Abelianizes when it is projected to the lowest level.

The cocycle-deformed first excited states are given by
\begin{subequations}
  \begin{IEEEeqnarray}{RCl}
    |E_1, 11\rangle &=& \frac{1}{\sqrt{6}} \left(\sum_{U\in\mathcal{C}_{13}} |\Gamma_3,11\rangle\otimes|U\rangle + \sum_{U\in\mathcal{C}_2}|\Gamma_3,22\rangle\otimes |U\rangle\right) , \\
    |E_1, 12\rangle &=& \frac{1}{\sqrt{6}} \left(\sum_{U\in\mathcal{C}_{13}} |\Gamma_3,12\rangle\otimes|U\rangle - \sum_{U\in\mathcal{C}_2}|\Gamma_3,21\rangle\otimes |U\rangle\right) , \\
    |E_1, 21\rangle &=& \frac{1}{\sqrt{6}} \left(\sum_{U\in\mathcal{C}_{13}} |\Gamma_3,21\rangle\otimes|U\rangle - \sum_{U\in\mathcal{C}_2}|\Gamma_3,12\rangle\otimes |U\rangle\right) , \\
    |E_1, 22\rangle &=& \frac{1}{\sqrt{6}} \left(\sum_{U\in\mathcal{C}_{13}} |\Gamma_3,22\rangle\otimes|U\rangle + \sum_{U\in\mathcal{C}_2}|\Gamma_3,11\rangle\otimes |U\rangle\right) ,
  \end{IEEEeqnarray}
\end{subequations}
with energy gap $\Delta / (e^2 a) = 3/2(\sqrt{2}-1)\approx 0.621$. These states of course still carry the same irreducible representation of $S_3$ as in the case of the trivial one-cocycle, i.e.,
\begin{equation}
  L(\Omega)R(\Omega') \hspace{1pt}|E_1,ab\rangle = \sum_{cd} \Gamma_3(\Omega)_{ca}^{\ast} \hspace{1pt}|E_1,cd\rangle\hspace{0.5pt}\Gamma_3(\Omega')_{db} .
\end{equation}
However, we observe that the operators $\widetilde{L}(\mathcal{C}_2)$ and $\widetilde{R}(\mathcal{C}_2)$ now satisfy
\begin{equation}
  \widetilde{L}(\mathcal{C}_2)|E_1,ab\rangle = \widetilde{R}(\mathcal{C}_2)|E_1,ab\rangle = (-1)^{a+b} |E_1,([\hspace{1pt}a\hspace{0.5pt}]_2+1)([\hspace{1pt}b\hspace{0.5pt}]_2+1)\rangle ,
\end{equation}
where it is understood that $a,b = 1,2$ and (as before) the operators $\widetilde{L}(\mathcal{C}_{13})$ and $\widetilde{R}(\mathcal{C}_{13})$ act as identity operators. In contrast to the lowest level, these states carry a two-dimensional (non-Abelian) representation $\Gamma_3$ and the action of the left- and right-multiplication operators are clearly different. 

Finally, we remark that the other four first excited states as well as the higher-lying states, which involve a $\Gamma_3$ representation, are not changed by the one-cocycle and are therefore not listed explicitly.

\subsection{Doubled $\bar{D}_2$ lattice gauge theory}
\label{SubSec:Doubled D2bar lattice gauge theory}

The possible one-cocycles for the doubled $\bar{D}_2$ lattice gauge theory are summarized in \ref{SubSection:One-cocycles and W factors D2bar}. Introducing a trivial cocycle yields a naively doubled YM theory. For the group $\bar{D}_2$ there are $15$ CS-type theories (corresponding to the number of inequivalent nontrivial one-cocycles), which can be divided in two sets with distinct spectra. Note that these theories similarly Abelianize when they are projected to the lowest energy eigenstate. Therefore, we do not discuss the properties of the low-energy states at length, but only list the level structure and the degeneracies of the different levels.

\subsubsection{Trivial one-cocycle}

As in the previous examples, the unique ground state of the trivially doubled lattice YM theory with energy $E_0/(e^2 a) = 0$ is given by the totally symmetric superposition involving the $\Gamma_1$ representation. There are eight excited states above the ground state with energy gap $\Delta/ (e^2 a) = 3$. The full spectrum of the single-cross Hamiltonian is summarized in \mbox{Tab.\ \ref{Tab:DoubledYMGroupD2bar}}.

\begin{table}[!h]
  \begin{center}
    \begin{tabular}{|c||C{50pt}|C{40pt}|C{40pt}|C{40pt}|C{50pt}|C{40pt}|}
      \hline
      & $n = 0$ & $n = 1$ & $n = 2$ & $n = 3$ & $n = 4$ & $n = 5$ \\
      \hline
      \hline
      $E_n / (e^2a)$ & $0$ & $3$ & $4$ & $6$ & $7$ & $8$ \\
      \hline
      $g_n$ & 
      $1$ & $8$ & $6$ & $16$ & $24$ & $9$ \\
      \hline
    \end{tabular}
    \caption{\label{Tab:DoubledYMGroupD2bar}Eigenvalues $E_n$ and their degeneracies $g_n$, $n = 0,1,\ldots$, of the single-cross electric Hamiltonian $H_E^{\textrm{single-cross}}$ for the doubled $\bar{D}_2$ lattice YM theory (with $g_0 = 1$ ground-state degeneracy).}
  \end{center}
\end{table}

\subsubsection{Nontrivial one-cocycles}

For the nontrivial one-cocycles
\begin{equation}
  \omega_L(\Omega,U) = \omega_R(\Omega,U)^{\ast} \in \left\{\omega_i(\Omega,U) \hspace{1pt}|\hspace{2.5pt} i = 2,3,\ldots,10 \right\} ,
\end{equation}
[cf.\ \ref{SubSection:One-cocycles and W factors D2bar}], we obtain a two-fold degenerate ground state with an energy gap $\Delta/(e^2 a) = 2\sqrt{2} - 1\approx 1.828$ to the set of first excited states. The full spectrum of the corresponding theory is summarized in Tab.\ \ref{Tab:DoubledCSYMGroupD2barA}. Note, while the spectrum of the theory is identical for each of the above cocycles, the eigenstates may still differ.

\begin{table}[!h]
  \begin{center}
    \begin{tabular}{|c||C{50pt}|C{35pt}|C{35pt}|C{35pt}|C{50pt}|C{35pt}|C{35pt}|}
      \hline
      & $n = 0$ & $n = 1$ & $n = 2$ & $n = 3$ & $n = 4$ & $n = 5$ & $n = 6$ \\
      \hline
      \hline
      $E_n / (e^2a)$ & $2(2-\sqrt{2})$ & $3$ & $4$ & $6$ & $2(2+\sqrt{2})$ & $7$ & $8$ \\
      \hline
      $g_n$ & $2$ & $8$ & $4$ & $16$ & $2$ & $24$ & $8$ \\
      \hline
    \end{tabular}
    \caption{\label{Tab:DoubledCSYMGroupD2barA}Eigenvalues $E_n$ and their degeneracies $g_n$, $n = 0,1,\ldots$, of the single-cross electric Hamiltonian $H_E^{\textrm{single-cross}}$ for the doubled $\bar{D}_2$ lattice CSYM theory (with $g_0 = 2$ ground-state degeneracy).}
  \end{center}
\end{table}

In the case of the nontrivial one-cocycles
\begin{equation}
  \omega_L(\Omega,U) = \omega_R(\Omega,U)^{\ast} \in \left\{\omega_i(\Omega,U) \hspace{1pt}|\hspace{2.5pt} i=11,12,\ldots, 16\right\} ,
\end{equation}
the ground state is four-fold degenerate and the energy gap to the eight-fold degenerate first excited state is $\Delta/(e^2 a) = 1$. The full spectrum of this theory is summarized in \mbox{Tab.\ \ref{Tab:DoubledCSYMGroupD2barB}}.

\begin{table}[!h]
  \begin{center}
    \begin{tabular}{|c||C{50pt}|C{40pt}|C{40pt}|C{40pt}|C{50pt}|}
      \hline
      & $n = 0$ & $n = 1$ & $n = 2$ & $n = 3$ & $n = 4$ \\
      \hline
      \hline
      $E_n / (e^2a)$ & $2$ & $3$ & $6$ & $7$ & $8$ \\
      \hline
      $g_n$ & 
      $4$ & $8$ & $20$ & $24$ & $8$ \\
      \hline
    \end{tabular} 
    \caption{\label{Tab:DoubledCSYMGroupD2barB}Eigenvalues $E_n$ and their degeneracies $g_n$, $n = 0,1,\ldots$, of the single-cross electric Hamiltonian $H_E^{\textrm{single-cross}}$ for the doubled $\bar{D}_2$ lattice CSYM theory (with $g_0 = 4$ ground-state degeneracy).}
  \end{center}
\end{table}

\subsection{Doubled $\Delta(27)$ lattice gauge theory}
\label{SubSec:Doubled Delta27 lattice gauge theory}

The possible one-cocycles for the doubled $\Delta(27)$ lattice gauge theory are summarized in \ref{SubSection:One-cocycles and W factors Delta27}. Introducing a trivial cocycle yields a naively doubled YM theory. For the group $\Delta(27)$ there are $80$ doubled CS-type theories (corresponding to the number of different nontrivial one-cocycles) that define two distinct spectra. Again, we restrict ourselves to list the spectra of the single-cross Hamiltonian and the structure of the ground states. 

\subsubsection{Trivial one-cocycle}

The unique ground state of the trivially doubled lattice YM theory with energy $E_0/(e^2 a) = 0$ is given by the totally symmetric superposition involving the $\Gamma_1$ representation. There are $36$ excited states above the ground state with energy gap $\Delta/ (e^2 a) = 12$. The full spectrum of the single-cross Hamiltonian is summarized in Tab.\ \ref{Tab:DoubledYMGroupDelta27}.

\begin{table}[!h]
  \begin{center}
    \begin{tabular}{|c||C{40pt}|C{40pt}|C{40pt}|C{40pt}|C{40pt}|C{40pt}|}
      \hline
      & $n = 0$ & $n = 1$ & $n = 2$ & $n = 3$ & $n = 4$ & $n = 5$ \\
      \hline
      \hline
      $E_n / (e^2a)$ & $0$ & $12$ & $27/2$ & $24$ & $51/2$ & $27$ \\
      \hline
      $g_n$ & $1$ & $36$ & $16$ & $324$ & $288$ & $64$ \\
      \hline
    \end{tabular}
    \caption{\label{Tab:DoubledYMGroupDelta27}Eigenvalues $E_n$ and their degeneracies $g_n$, $n = 0,1,\ldots$, of the single-cross electric Hamiltonian $H_E^{\textrm{single-cross}}$ for the doubled $\Delta(27)$ lattice YM theory (with $g_0 = 1$ ground-state degeneracy).}
  \end{center}
\end{table}

\subsubsection{Nontrivial one-cocycles}

For the set of nontrivial one-cocycles
\begin{equation}
  \omega_L(\Omega,U) = \omega_R(\Omega,U)^{\ast} \in \left\{ \omega_i(\Omega,U) \hspace{1pt}|\hspace{2.5pt} i=2,3,\ldots,9 \right\} ,
\end{equation}
the ground state is three-fold degenerate and the energy gap to the $36$-fold degenerate first excited state is $\Delta/(e^2 a) = \sqrt{3}/2(9 - \sqrt{3})\approx 6.294$. The full spectrum of this theory is summarized in Tab.\ \ref{Tab:DoubledCSYMGroupDelta27Degeneracy3}.

\begin{table}[!h]
  \begin{center}
    \begin{tabular}{|c||C{55pt}|C{35pt}|C{35pt}|C{55pt}|C{35pt}|C{35pt}|C{35pt}|}
      \hline
      & $n = 0$ & $n = 1$ & $n = 2$ & $n = 3$ & $n = 4$ & $n = 5$ & $n = 6$ \\
      \hline
      \hline
      $E_n / (e^2a)$ & $9/2(3 - \sqrt{3})$ & $12$ & $27/2$ & $9/2(3 + \sqrt{3})$ & $24$ & $51/2$ & $27$ \\
      \hline
      $g_n$ & $3$ & $36$ & $12$ & $3$ & $324$ & $288$ & $63$ \\
      \hline
    \end{tabular}
    \caption{\label{Tab:DoubledCSYMGroupDelta27Degeneracy3}Eigenvalues $E_n$ and their degeneracies $g_n$, $n = 0,1,\ldots$, of the single-cross electric Hamiltonian $H_E^{\textrm{single-cross}}$ for the doubled $\Delta(27)$ lattice CSYM theory (with $g_0 = 3$ ground-state degeneracy).}
  \end{center}
\end{table}

On the other hand, for
\begin{equation}
  \omega_L(\Omega,U) = \omega_R(\Omega,U)^{\ast} \in \left\{ \omega_i(\Omega,U) \hspace{1pt}|\hspace{2.5pt} i=10,11,\ldots 27\right\} ,
\end{equation}
the ground state is nine-fold degenerate and the energy gap to the set of first excited states is $\Delta/(e^2 a) = 3$. The full spectrum of this theory is summarized in Tab.\ \ref{Tab:DoubledCSYMGroupDelta27Degeneracy9}.

\begin{table}[!h]
  \begin{center}
    \begin{tabular}{|c||C{40pt}|C{40pt}|C{40pt}|C{40pt}|C{40pt}|C{40pt}|}
      \hline
      & $n = 0$ & $n = 1$ & $n = 2$ & $n = 3$ & $n = 4$  & $n = 5$ \\
      \hline
      \hline
      $E_n / (e^2a)$ & $9$ & $12$ & $18$ & $24$ & $51/2$ & $27$ \\
      \hline
      $g_n$ & $9$ & $36$ & $9$ & $324$ & $288$ & $63$ \\
      \hline
    \end{tabular}
    \caption{\label{Tab:DoubledCSYMGroupDelta27Degeneracy9}Eigenvalues $E_n$ and their degeneracies $g_n$, $n = 0,1,\ldots$, of the single-cross electric Hamiltonian $H_E^{\textrm{single-cross}}$ for the doubled $\Delta(27)$ lattice CSYM theory (with $g_0 = 9$ ground-state degeneracy).}
  \end{center}
\end{table}

\subsection{Center symmetry and confinement}
\label{SubSec:Center symmetry and confinement}

The considerations of the previous sections provide us with a picture of the admissible charge configurations in the strong-coupling regime, when the magnetic part of the Hamiltonian can be neglected. In particular, we observe that in contrast to the discrete non-Abelian group $S_3$, the discrete groups $\bar{D}_2$ and $\Delta(27)$ exhibit charge confinement. This is illustrated in Figs.\ \ref{Fig:NoConfinementS3} -- \ref{Fig:ChargedExcitationsDelta27} and we discuss each of the different non-Abelian groups in the following.

\subsubsection{Charged states in the doubled $S_3$ lattice gauge theory}

\begin{figure}[!h]
  \centering
  \includegraphics[width=0.2\textwidth]{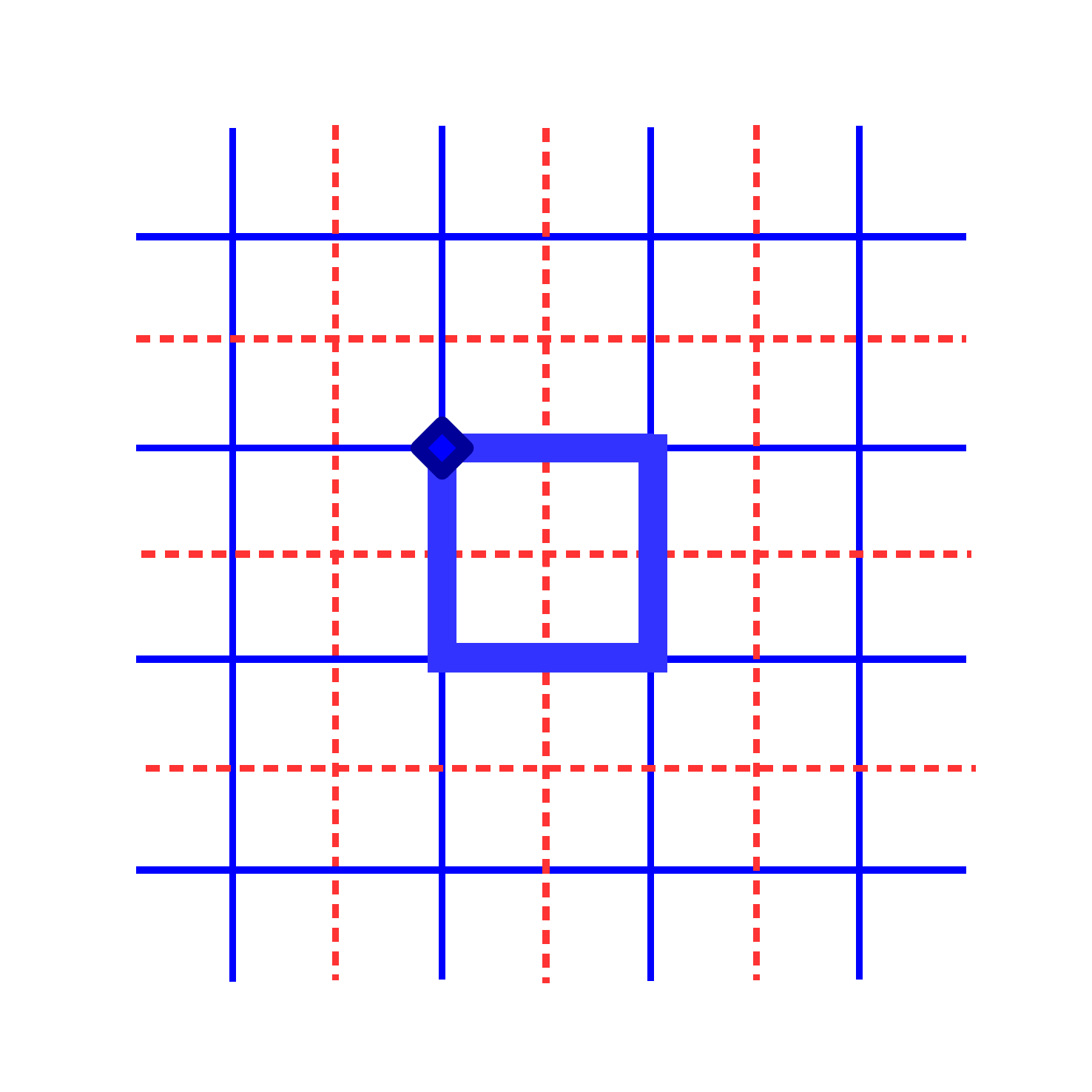} \qquad
  \includegraphics[width=0.2\textwidth]{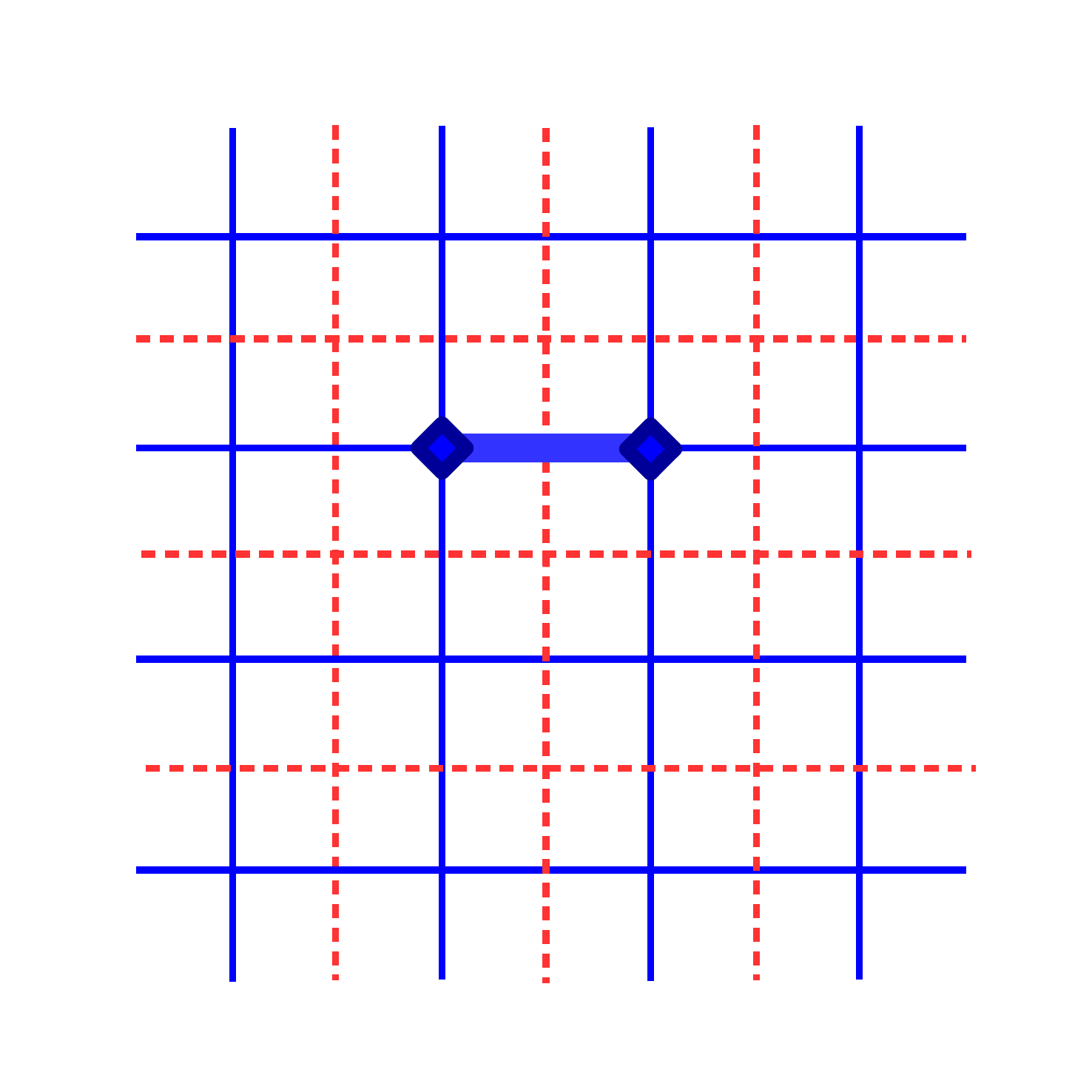}
  \caption{\label{Fig:NoConfinementS3}(Left) A single $\Gamma_3$ charge is not confined because $S_3$ has a trivial center. The energetically favorable electric-flux configuration for a single $\Gamma_3$ charge consists of four units of $\Gamma_3$ electric flux along an elementary plaquette. (Right) Two non-Abelian $\Gamma_3$ charges connected by a $\Gamma_3$ flux string on a single link.}
\end{figure}

The doubled lattice gauge theory with gauge group $S_3$ has an Abelian and a non-Abelian charge, in the $\Gamma_2$ and $\Gamma_3$ representation, respectively, as well as $\Gamma_1$ representation which is associated to vanishing charge. A physical state in the theory is uniquely specified by the corresponding charge assignments at each point of the direct (or dual) lattice. In the nontrivial CS-type theory states with only Abelian charges $\Gamma_2$ carry the same energy as the vacuum -- at least in the absence of the magnetic part of the Hamiltonian. States containing non-Abelian charges $\Gamma_3$ are separated by multiples of the energy gap $\Delta$ [cf.\ \mbox{Sec.\ \ref{SubSec:Doubled S3 lattice gauge theory}}]. Since the gap becomes infinite in the strong-coupling limit, we inquire only about the lowest-energy states in a given charge sector. The group $S_3$ has trivial center and therefore the theory is not confining (in the sense of a linearly rising potential or unbreakable string). That is, we may insert a single charge on the lattice. The lowest-lying state in this sector consists of a closed loop of $\Gamma_3$ electric flux (see Fig.\ \ref{Fig:NoConfinementS3}). Note that the trivially doubled YM theory does not confine either.

\begin{figure}[!h]
  \centering
  \includegraphics[width=0.2\textwidth]{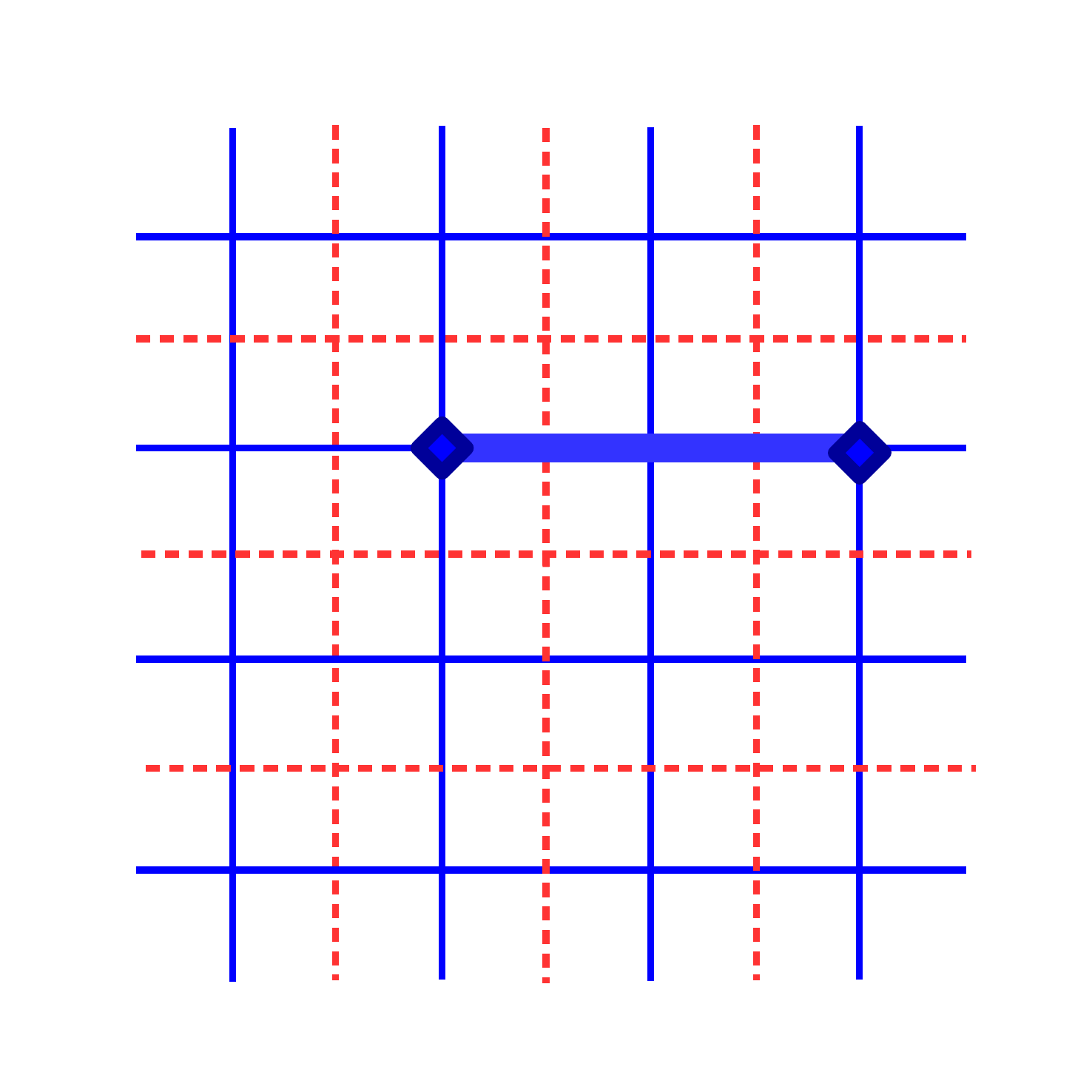}
  \includegraphics[width=0.2\textwidth]{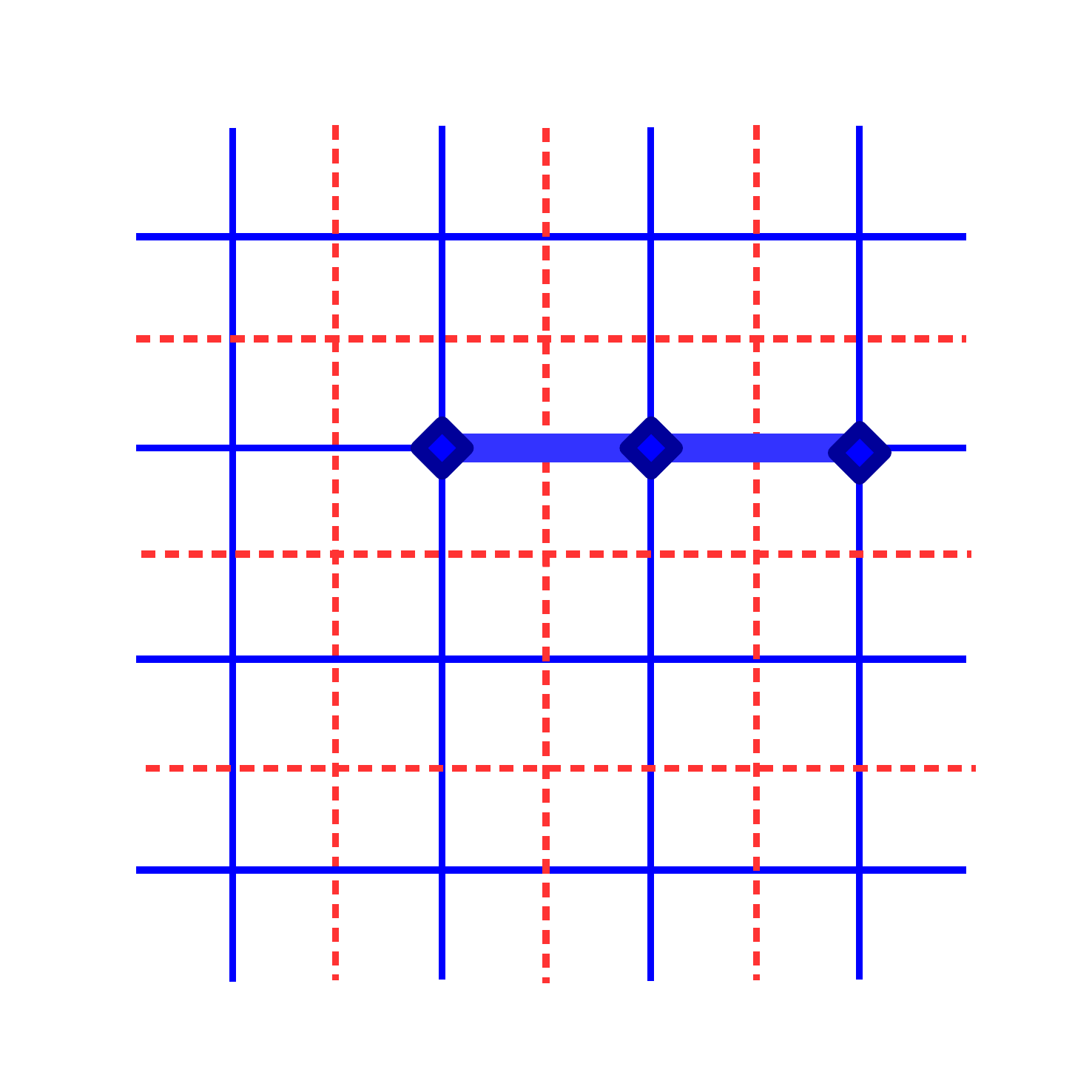}
  \includegraphics[width=0.2\textwidth]{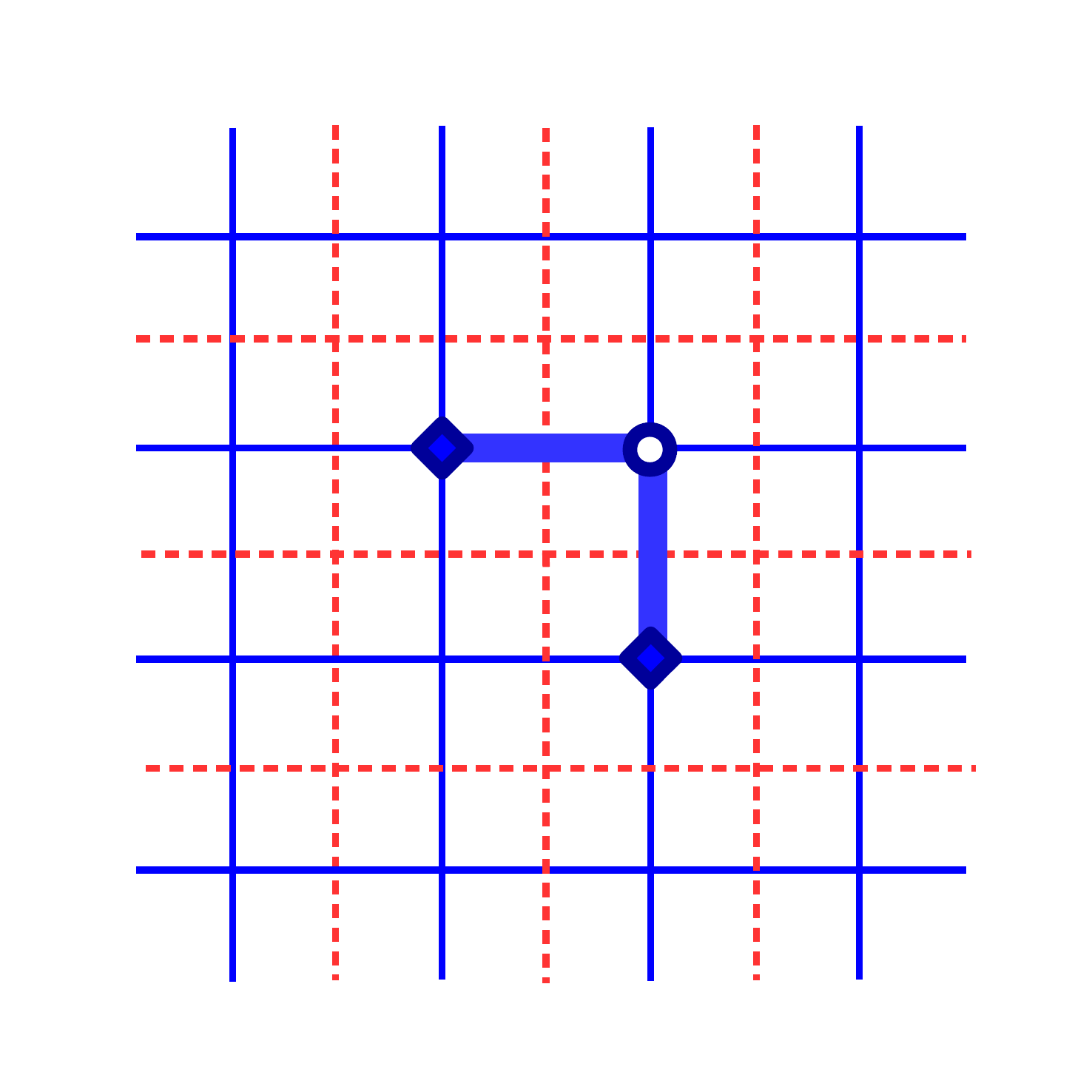}
  \caption{\label{Fig:ChargedExcitationsS3}Charged excitations for a CS-type theory with gauge group $S_3$. While the configuration shown on the left connects two non-Abelian charges $\Gamma_3$ by two units of $\Gamma_3$ electric flux, the other two shown configurations include an additional $\Gamma_3$ or $\Gamma_2$ charge. All of these states carry the same energy.} 
\end{figure}

In the sector consisting of two $\Gamma_3$ charges, the lowest-energy state is given by the particle/flux configuration illustrated in Fig.\ \ref{Fig:NoConfinementS3}. The required energy to separate these charges grows linearly with the number of connecting $\Gamma_3$ electric flux links, until the flux string breaks and forms two single charge excitations in the $\Gamma_3$ representation. Note, however, that there might be multiple charged states with a fixed number of $\Gamma_3$ electric-flux links. An example is illustrated in Fig.\ \ref{Fig:ChargedExcitationsS3}, where a single $\Gamma_3$ flux string along two links connects either two $\Gamma_3$ charges, or three charges (of which only two need to lie in the $\Gamma_3$ representation). Note, while in the CS theory the insertion of a single $\Gamma_2$ charge comes at no additional cost, this does not hold true in the trivially doubled YM theory.

\subsubsection{Charged states in the doubled $\bar{D}_2$ lattice gauge theory}

\begin{figure}[!h]
  \centering
  \includegraphics[width=0.2\textwidth]{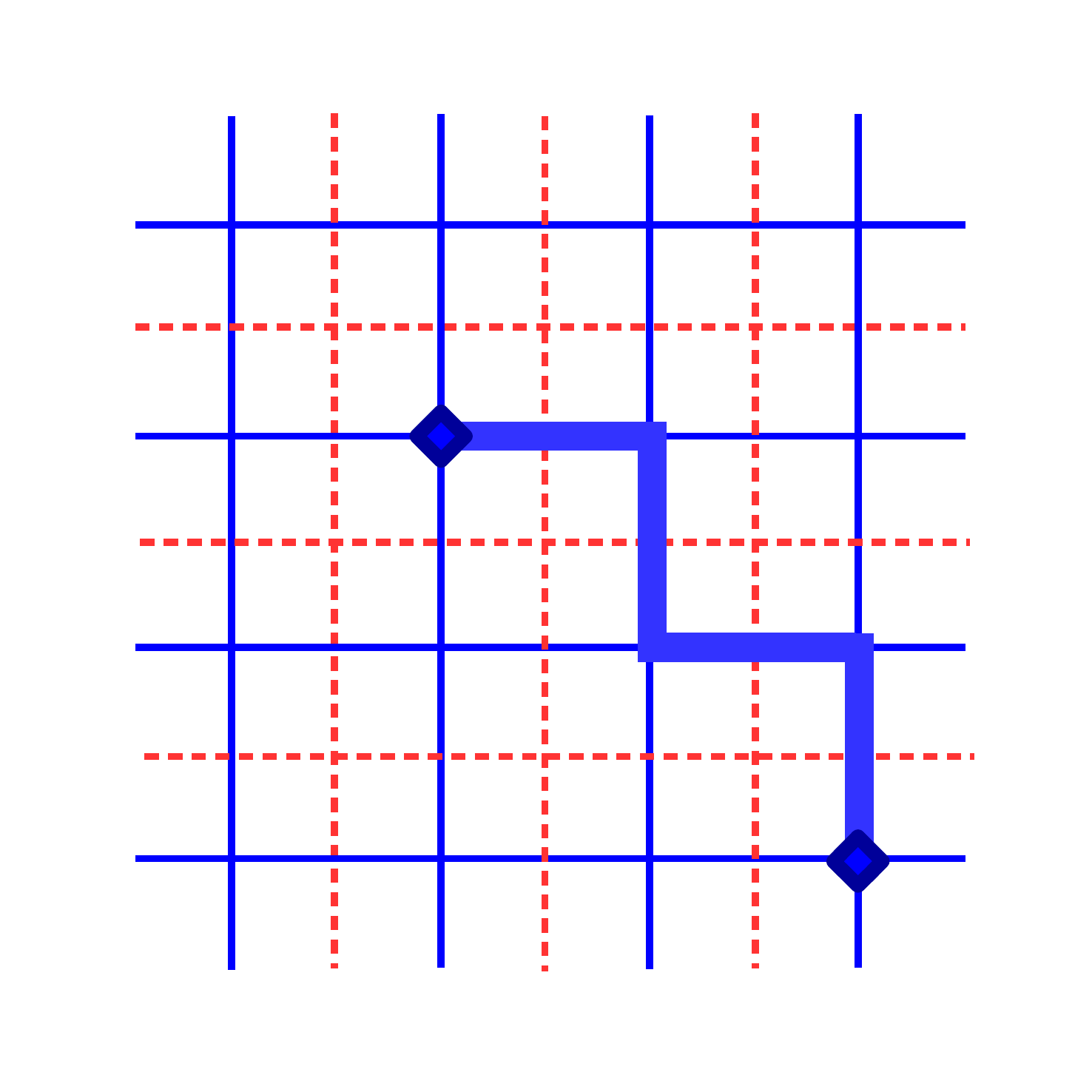}
  \caption{\label{ChargedExcitationsD2bar}Charged excitation for gauge group $\bar{D}_2$, consisting of two $\Gamma_5$ charges connected by a confining electric flux string.}
\end{figure}

The doubled $\bar{D}_2$ gauge theory admits three types of Abelian charges, with one-dimensional representations $\Gamma_2$, $\Gamma_3$, and $\Gamma_4$, and a non-Abelian charge in the two-dimensional $\Gamma_5$ representation, which transforms nontrivially under the $\Z(2)$ center. Thus, there are no single $\Gamma_5$-charge states in the theory. A state consisting of two $\Gamma_5$ charges is connected by a single flux string with nonzero string tension (see Fig.\ \ref{ChargedExcitationsD2bar}). In contrast to the $S_3$ theory, the electric flux string cannot break -- the theory is confining.

\subsubsection{Charged states in the doubled $\Delta(27)$ lattice gauge theory}

\begin{figure}[!h]
  \centering
  \includegraphics[width=0.2\textwidth]{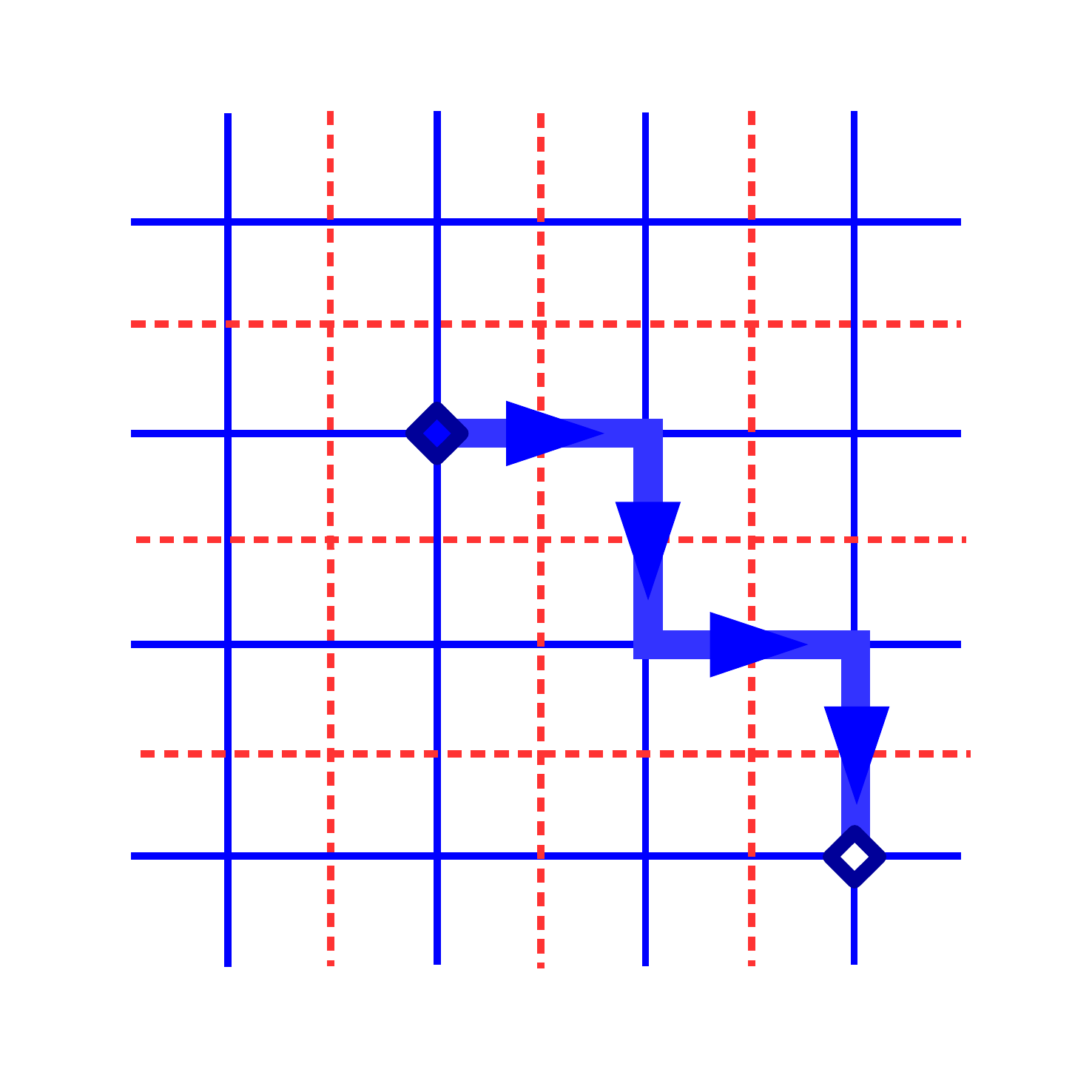}\hspace{20pt}
  \includegraphics[width=0.2\textwidth]{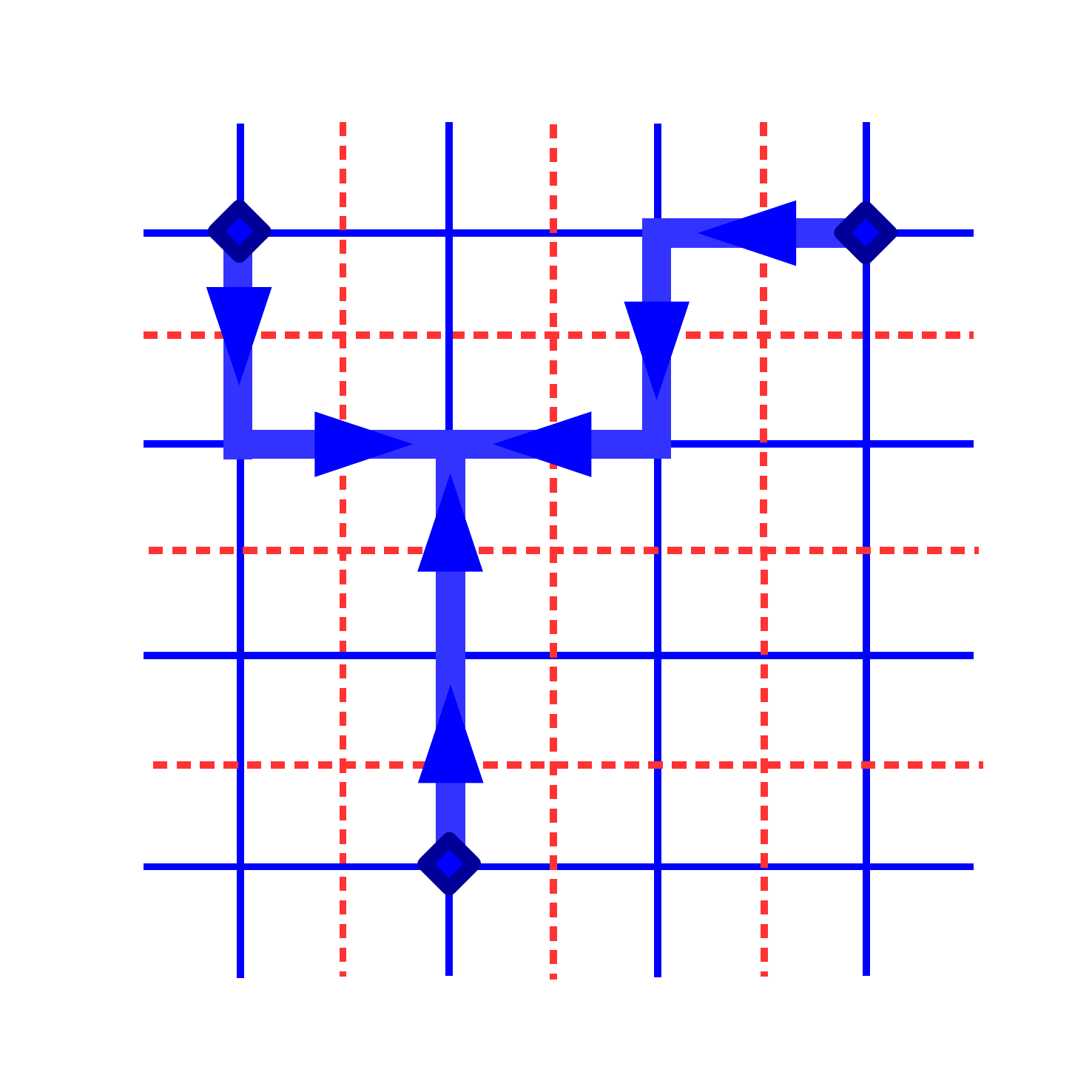}
  \caption{\label{Fig:ChargedExcitationsDelta27}Charged excitations for gauge group $\Delta(27)$. The presence of the nontrivial center $Z(\Delta(27))\cong \Z(3)$ implies that charges with nontrivial representation under center transformations (e.g., the irreducible representations $\Gamma_{10}$ and $\Gamma_{11}$) are bound into two-particle, or three-particle states, where the individual components are confined by an electric flux string.}
\end{figure}

The doubled $\Delta(27)$ lattice gauge theory admits eight Abelian charges characterized by the one-dimensional representations $\Gamma_2$, $\Gamma_3$, \ldots, $\Gamma_9$, and two non-Abelian charges in the three-dimensional representations $\Gamma_{10}$ and $\Gamma_{11}$. Similar to the $\bar{D}_2$ theory, the nontrivial center prohibits a single $\Gamma_{10}$ or $\Gamma_{11}$ charge excitation in the theory. The lowest-energy excitations consist of a pair of $\Gamma_{10}$ and $\Gamma_{11} = \overline{\Gamma}_{10}$ charge particles, connected by a confining string of electric flux. Since these representations are complex this leads to a directed flux string [cf.\ Fig.\ \ref{Fig:ChargedExcitationsDelta27}]. The center of $\Delta(27)$ is $\Z(3)$ and therefore one may envision other bound states consisting of either three $\Gamma_{10}$ or $\Gamma_{11}$ charge particles, connected by single connected string of electric flux. However, since such states must involve at least three links carrying $\Gamma_{10}$ (or $\Gamma_{11}$) flux, they are energetically less favorable than the two-particle states. These particle/flux configurations are illustrated in Fig.\ \ref{Fig:ChargedExcitationsDelta27}.

\section{Conclusions}
\label{Sec:Conclusions}

In this work, we have addressed the properties of doubled lattice CSYM theories with discrete non-Abelian gauge group, thereby extending previous work for Abelian \mbox{groups \cite{Olesen:2015baa}}. We have shown how to realize Hamiltonians for non-Abelian lattice gauge theories that are compatible with a noncommuting algebra of group multiplication operators and therefore allow for theories of the CS type. This construction relies on the identification of one-cocycles that represent the action of gauge transformations on quantum states. They are in one-to-one correspondence with an Aharonov-Bohm phase, which is introduced in the group space on a single pair of links (on the direct lattice and its dual), and define the possible theories within the considered class of Hamiltonians.

We have constructed different Hamiltonians explicitly for the discrete group $\Z(k)\subset U(1)$, which is Abelian, and the (finite) non-Abelian groups $S_3\subset O(2)$, $\bar{D}_2\subset SU(2)$, and $\Delta(27)\subset SU(3)$. Their characteristics were considered in the limit of strong coupling, in which case the theory can be solved exactly. In particular, we have shown that the doubled lattice theories of CS type -- when projected on the lowest-energy eigenstate -- map to a (generalized) toric code. At first, it might sound rather surprising, why a discrete non-Abelian gauge theory may reduce to an Abelian theory. However, one can make sense of this behavior, by noticing that the single-cross electric Hamiltonian describes a hopping that does not distinguish between left- or right-multiplication in group space. Thus, effectively, the considered theory Abelianizes and yields a ground state that conforms with that of a lattice gauge theory with Abelian gauge group (in general, a product of cyclic groups). We emphasize that the Abelianization only affects the ground state (i.e., zero-charge sector) of the theory -- higher-lying states still carry a non-Abelian representation of the gauge group. Of course, one might ask whether these properties are tied to the particular way, by which these theories were constructed. We refer the reader to Ref.\ \cite{Mesterhazy:2016} where such questions will be addressed in detail.

For the investigated lattice CSYM theories with gauge groups $\bar{D}_2$ and $\Delta(27)$ the presence of a nontrivial center implies the confinement of charges, which is a generic phenomenon at strong coupling. One might wonder whether this may turn out to be a useful feature to mitigate errors at finite temperature \cite{Dennis:2001nw,Alicki:2007,Castelnovo:2007,Chesi:2010,Brell:2013wsa,Brown:2014,Freeman:2014}. In particular, the confining string connecting non-Abelian charges restricts their uncontrolled movement on the spatial lattice. This might prevent computational errors that typically arise when a pair of non-Abelian anyons created from the vacuum winds around any one of the nontrivial cycles of the base manifold and annihilate.

So far, we have not investigated the statistics of charged excitations and their braiding properties, which are of immediate interest to applications in topological quantum computing. The presence and nature of excitations with fractional statistics will be addressed in future work.

\section*{Acknowledgments}

We like to thank I.\ Cirac and M.\ L\"uscher for illuminating discussions. The research leading to these results has received funding from the Schweizerischer Na\-tio\-nal\-fonds and from the European Research Council under the European Union's Seventh Framework Programme (FP7/2007-2013), ERC grant agreement 339220.

\begin{appendix}
  
\section{Basics of the Theory of Discrete Groups}
\label{Sec:Basics of the Theory of Discrete Groups}

In this appendix we review some important results from the theory of discrete groups \cite{Serre:1977}. We summarize in particular the properties of the cyclic group $\Z(k)$, the symmetric group $S_3$ (the permutation group of three objects), the binary dihedral (i.e., quaternion) group $\bar{D}_2$, and finally $\Delta(27)$, which belongs to the $\Delta(3 n^2)$ series of discrete groups (see, e.g., Refs.\ \cite{Fairbairn:1964sga,Fairbairn:1982jx,Ludl:2011gn}).

\subsection{Group multiplication and center}
\label{SubSec:Group multiplication and center}

A discrete group $G$ is a set of $d_G$ group elements $\Omega\in G$ endowed with a multiplication rule 
\begin{equation}
\Omega \Omega' = \Omega'' ,
\end{equation}
which satisfies associativity, i.e.,
\begin{equation} 
(\Omega \Omega') \Omega'' = \Omega (\Omega' \Omega'') .
\end{equation}
Furthermore, there is a unit element $\1\in G$, which obeys
\begin{equation}
\Omega \1 = \1 \Omega = \Omega ,
\end{equation}
and each element $\Omega$ has a unique inverse $\Omega^{-1}\in G$, such that
\begin{equation} 
\Omega^{-1} \Omega = \Omega \Omega^{-1} = \1 .
\end{equation}
For an Abelian group the group multiplication is commutative, i.e., $\Omega \Omega' = \Omega' \Omega$, for all $\Omega, \Omega' \in G$. The center $Z(G)$ of a group $G$ consists of those elements $z\in G$ that commute with all group elements, i.e., $z G = G z$. Obviously, if $G$ is Abelian, then $G = Z(G)$.

\subsection{Conjugacy classes}
\label{SubSec:Conjugacy classes}

Conjugacy classes, which we denote by $\mathcal{C}_p$, with index $p = 1, 2, \ldots$, are equivalence classes of group elements. Two elements $\Omega'$ and $\Omega''$ belong to the same conjugacy class if there exists a group element $\Omega$ such that
\begin{equation}
  \Omega'' = \Omega \Omega' \Omega^{-1} .
\end{equation}
The unit element always forms its own conjugacy class, $\mathcal{C}_1 = \{\1\}$, since
\begin{equation}
  \Omega \1 \Omega^{-1} = \Omega \Omega^{-1} = \1 .
\end{equation}
The same is true for all center elements $z$, because
\begin{equation}
  \Omega z\Omega^{-1} = z\Omega \Omega^{-1} = z .
\end{equation}
In an Abelian group every element forms its own conjugacy class, while non-Abelian groups possess conjugacy classes that consist of more than just a single element.

\subsection{Irreducible representations and characters}
\label{SubSec:Irreducible representations and characters}

A unitary representation $\Gamma$ of dimension $d_{\Gamma}\in\N$ associates a $d_\Gamma\times d_\Gamma$ unitary matrix $\Gamma(\Omega)$ with each group element $\Omega$, such that the group multiplication rule is realized by the multiplication of the corresponding matrices, i.e.,
\begin{equation}
  \Gamma(\Omega) \Gamma(\Omega') = \Gamma(\Omega \Omega'). 
\end{equation}
Two representations $\Gamma$ and $\Gamma'$ are unitarily equivalent if there exists a single unitary matrix $V$, such that for all elements $\Omega \in G$
\begin{equation} 
  \Gamma'(\Omega) = V\Gamma(\Omega)V^{\dagger} ,
\end{equation}
where $V^{\dagger}$ denotes the conjugate transpose of $V$. A $d_\Gamma$-dimensional representation is irreducible if there is no unitary transformation that simultaneously block-diagonalizes all $\Gamma(\Omega)$ matrices to blocks of size smaller than $d_\Gamma$. The number of (unitarily inequivalent) irreducible representations $\Gamma_p$, $p = 1,2, \ldots$, equals the number of conjugacy classes. In addition, the dimensions $d_{\Gamma_p}$ of the irreducible representations obey the sum rule
\begin{equation}
  \sum_p d_{\Gamma_p}^2 = d_G .
\end{equation}
Thus, an Abelian group has $d_G$ different one-dimensional representations, while non-Abelian groups also have higher-dimensional irreducible representations.

The conjugate $\overline{\Gamma}$ of a representation $\Gamma$ is realized by complex conjugation
\begin{equation}
  \overline{\Gamma}(\Omega) = \Gamma(\Omega)^{\ast} ,
\end{equation}
which obeys the group multiplication rule
\begin{equation}
  \overline{\Gamma}(\Omega) \overline{\Gamma}(\Omega') = \Gamma(\Omega)^{\ast} \Gamma(\Omega')^{\ast} = \Gamma(\Omega \Omega')^{\ast} = \overline{\Gamma}(\Omega \Omega') .
\end{equation}
A representation $\Gamma$ is real if $\overline{\Gamma} = \Gamma$. It is pseudo-real if $\Gamma$ and $\overline{\Gamma}$ are unitarily equivalent, i.e., if there exists a single unitary matrix $V$ such that
\begin{equation} 
  \overline{\Gamma}(\Omega) = V\Gamma(\Omega)V^{\dagger} ,
\end{equation}
for all $\Omega \in G$. The representation $\Gamma$ is complex, if its conjugate representation is not unitarily equivalent to $\Gamma$.

The trace of the representation matrix $\Gamma(\Omega)$ is known as the character
\begin{equation}
  \chi_\Gamma(\Omega) = \Tr \Gamma(\Omega) .
\end{equation}
From the cyclicity of the trace it follows that the character $\chi_{{}_{\Gamma}}$ is a class function (an invariant of the conjugacy class), i.e., $\chi_{{}_{\Gamma}}(\Omega' \Omega {\Omega'}^{-1}) = \chi_{{}_{\Gamma}}(\Omega)$. The characters of irreducible representations $\Gamma_p$ and $\Gamma_q$ obey the orthogonality relation
\begin{equation}
  \frac{1}{d_G}\sum_{\Omega \in G} \chi_{{}_{\Gamma_p}}(\Omega)^{\ast} \chi_{{}_{\Gamma_q}}(\Omega) = \delta_{p,q} .
\end{equation}
The tensor product of two irreducible representations $\Gamma_p$ and $\Gamma_q$ can be reduced into a direct sum of irreducible representations $\Gamma_r$, i.e.,
\begin{equation}
\Gamma_p \otimes \Gamma_q = \bigoplus_r m_{p,q}(\Gamma_r) \hspace{1pt}\Gamma_r ,
\end{equation}
where the multiplicities $m_{p,q}(\Gamma_r)\in \N$ are determined from the relation
\begin{equation}
  \chi_{{}_{\Gamma_p}}(\Omega) \chi_{{}_{\Gamma_q}}(\Omega) = \sum_r m_{p,q}(\Gamma_r) \chi_{{}_{\Gamma_r}}(\Omega), \quad \Omega \in G .
\end{equation}

\subsection{Distance between group elements in $G$}
\label{SubSec:Distance between group elements in G}

We may define a distance between two group elements $\Omega$ and $\Omega'$
\begin{equation}
  \label{Eq:Distance}
  \mu(\Omega,\Omega') = 1 - \frac{1}{d_\Gamma} \Re \Tr\left[\Gamma(\Omega') \Gamma(\Omega)^{\dagger}\right] ,
\end{equation}
where $\Gamma$ corresponds to a representation of $G$. Depending on the properties of the representation $\Gamma$, which might be complex (or real), Eq.\ \eqref{Eq:Distance} effectively embeds the group $G$ into the group of unitary (orthogonal) matrices for which a concept of a distance is defined. Note that the distance function satisfies the axioms of a pseudometric. While $\mu(\Omega,\Omega) = 0$ always holds, $\mu(\Omega,\Omega') = 0$ does not necessarily imply that $\Omega = \Omega'$. 

\section{The Cyclic Group $\Z(k)\subset U(1)$}
\label{Sec:The Cyclic Group Zk}

Here, we summarize the theory of the Abelian group $\Z(k)$ and construct the possible one-cocycles that define the noncommuting algebra of group multiplication operators in the doubled theory. We refer to Ref.\ \cite{Olesen:2015baa} for an in depth discussion of the doubled $\Z(k)$ lattice gauge theories.
 
\subsection{Group multiplication and center}

The group $\Z(k)$ consists of the $k$-th complex roots of unity, i.e.,
\begin{equation}
  \Z(k) = \{z_n = e^{2 \pi i n/k} \hspace{1pt}|\hspace{1.5pt} n = 0, 1, \ldots, k-1 \} ,
\end{equation}
which form a group under multiplication
\begin{equation}
  z_m z_n = z_{[m+n]_k} ,
\end{equation}
where $[m+n]_k\equiv (m+n) \,(\!\!\!\!\mod k)$. The unit element is given by $z_0 = 1$, and the inverse of $z_n$ is
\begin{equation}
  z_n^{-1} = z_n^{\ast} = e^{- 2 \pi i n/k} = z_{[-n]_k} .
\end{equation}
The center of $\Z(k)$ equals the group itself.

\subsection{Conjugacy classes and irreducible representations}

Due to the fact that $\Z(k)$ is Abelian, its conjugacy classes consist of individual group elements $\mathcal{C}_p$, $p = 1, 2, \ldots, k$, which follows immediately from $z_m z_n z_m^{-1} = z_n$. All its irreducible representations $\Gamma_p$ are one-dimensional. We have
\begin{equation}
\Gamma_p(z_n) = z_n^{p-1} = e^{2 \pi i (p-1)n/k} = z_{[(p-1) n]_k} ,
\end{equation}
which is indeed a representation
\begin{equation}
\Gamma_p(z_m) \Gamma_p(z_n) = z_m^{p-1} z_n^{p-1} = (z_m z_n)^{p-1} = \Gamma_p(z_m z_n) .
\end{equation}

\subsection{Distance between group elements}
\label{SubSec:Distance between group elements Zk}

The group $\Z(k)$ is naturally embedded in $U(1)$. Using the distance between group elements defined in Eq.\ \eqref{Eq:Distance} defined by the representation $\Gamma_2$ carrying unit charge, we obtain
\begin{equation}
  \mu(z_m,z_n) = 1 - \cos(2\pi (m-n) / k) . 
\end{equation}
Thus, we identify nearest-neighbor group elements $z_m$ and $z_n$ as those elements for which $[m-n]_k = [\pm 1]_k$.

\subsection{One-cocycles and $W$ factors}
\label{SubSec:One-cocycles and W factors Zk}

A one-cocycle $\omega$ assigns a one-dimensional representation $\Gamma_p$, $p = 1,2,\ldots, k$, to every group element in $\Z(k)$, i.e.,
\begin{equation}
  \label{Eq:OneCocycleRepresentationZk}
  \omega(z_m,z_n) = \Gamma_{p(n)}(z_m) = z_m^{p(n)-1} ,
\end{equation}
where $p(n)\equiv p(z_n)$. There are $k^k$ such functions $p$ on the group $\Z(k)$, which yield an equal number of candidate one-cocycles. These cocycles should satisfy the consistency condition [cf.\ Eq.\ \eqref{Eq:OneCocycleGroupMultiplication2}]
\begin{equation}
  \label{Eq:ConsistencyConditionOneCocycleZk}
  \omega(z_m,z_{n_1}) \omega(z_m,z_{n_2}) = \omega(z_m,z_{n_1}z_{n_2}) ,
\end{equation}
and this reduces to the following condition on the function $p$:
\begin{equation}
  \left[\hspace{0.5pt}p\left(\left[n_1+n_2\right]_k\right) - p(n_1) - p(n_2) + 1\right]_k = 0 .
\end{equation}
It is solved by the set of linear polynomials modulo $k$, i.e.,
\begin{equation}
  p(n) = \left[a n \right]_k + 1 ,
\end{equation}
with nonnegative integer coefficient $a = 0,1,2,\ldots,k-1$. We therefore find $k$ solutions that satisfy the consistency condition \eqref{Eq:ConsistencyConditionOneCocycleZk}, which are labeled by the index $i = 1,2,\ldots, k$, in the following, i.e.,
\begin{equation}
  \label{Eq:OneCocycleRepresentationZk2}
  \omega_i(z_m,z_n) = z_m^{\left[a_i n\right]_k} = e^{2\pi i a_i m n / k} ,
\end{equation}
and $a_i\neq a_j$, if and only if $i\neq j$. The $W$ factors for the group $\Z(k)$ are constructed from the combination of two (possibly distinct) one-cocycles, as specified by the index pair $(i,j)$. Using Eq.\ \eqref{Eq:OneCocycleRepresentationZk2} and substituting this expression into Eq.\ \eqref{Eq:WFactorOneCocycles}, we get\footnote{Here (and in the following) we simply drop the $L$ and $\widetilde{L}$ indices and identify $W\equiv W_{L\widetilde{L}}$.}
\begin{equation}
  \label{Eq:WFactorZk}
  W(z_m,z_n) = \omega_i(z_n,z_m) \omega_j(z_m,z_n)^{-1} = e^{2 \pi i \hspace{0.5pt} a_{ij} m n/k} ,
\end{equation}
which depends only on the difference $a_{ij} = [a_i-a_j]_k$. Note that both coefficients $a_{ij}$ and $a_i$ take their values in the set $\{0, 1, \ldots, k-1\}$. That is, to obtain all possible $W$ factors, we may simply apply the following gauge choice, $a_j = 1$, i.e., 
\begin{equation}
  \omega_j(\Omega,U) = \Gamma_1(\Omega) = 1 ,
\end{equation}
to reduce the redundancy in \eqref{Eq:WFactorZk}. Thereby, we finally obtain
\begin{equation}
  W(z_m,z_n) = \omega_i(z_n,z_m) = e^{2 \pi i \hspace{0.5pt} a_i m n/k} ,
\end{equation}
by which we see that the one-cocycles are in one-to-one correspondence to the inequivalent doubled lattice gauge theories (as defined by $W$). For the gauge group $\Z(k)$ we obtain $k$ distinct theories.

\section{The Symmetric Group $S_3\subset O(2)$}
\label{Sec:The Symmetric Group S3}

In this appendix we investigate the six-element non-Abelian permutation group of three objects, which has a trivial center and only real representations.

\subsection{Group multiplication and center}

The group $S_3$ consists of the unit element $\1$, the pair permutations $P_{12}$, $P_{23}$, and $P_{31}$, as well as the cyclic permutations $P_{231}$ and $P_{312}$, i.e.,
\begin{equation}
  S_3 = \{ \1, P_{12}, P_{23}, P_{31}, P_{231}, P_{312} \}.
\end{equation}
Tab.\ \ref{Tab:GroupMultiplicationS3} summarizes the group multiplication properties. Only the unit element commutes with all group elements, and therefore the center of $S_3$ is trivial, i.e., $Z(S_3) = \{\1\}$.

\begin{table}[!h]
  \begin{center}
    \begin{tabular}{|c||c|c|c|c|c|c|}
      \hline 
      & $\1$   & $P_{12}$ &  $P_{23}$ &  $P_{31}$ & $P_{231}$ & $P_{312}$ \\
      \hline
      \hline
      $\1$     & $\1$    & $P_{12}$  & $P_{23}$  & $P_{31}$  & $P_{231}$ & $P_{312}$ \\
      \hline
      $P_{12}$  & $P_{12}$ & $\1$     & $P_{312}$ & $P_{231}$ & $P_{31}$  & $P_{23}$  \\
      \hline
      $P_{23}$  & $P_{23}$ & $P_{231}$ &    $\1$  & $P_{312}$ & $P_{12}$  & $P_{31}$  \\
      \hline
      $P_{31}$  & $P_{31}$ & $P_{312}$ & $P_{231}$ &    $\1$  & $P_{23}$  & $P_{12}$  \\
      \hline
      $P_{231}$ & $P_{231}$ & $P_{23}$  & $P_{31}$ & $P_{12}$  & $P_{312}$ &    $\1$  \\
      \hline
      $P_{312}$ & $P_{312}$ & $P_{31}$  & $P_{12}$  & $P_{23}$ &     $\1$ & $P_{231}$ \\
      \hline
    \end{tabular}
  \end{center}
  \caption{\label{Tab:GroupMultiplicationS3}Multiplication table for the symmetric group $S_3$.}
  
\end{table}

\subsection{Conjugacy classes and irreducible representations}
\label{SubSec:Conjugacy classes and irreducible representations}

The conjugacy classes follow readily from Tab.\ \ref{Tab:GroupMultiplicationS3}:
\begin{equation}
  \mathcal{C}_1 = \{\1\}, \quad
  \mathcal{C}_2 = \{P_{12}, P_{23}, P_{31}\}, \quad
  \mathcal{C}_3 = \{P_{231}, P_{312}\} .
\end{equation}
The set
\begin{equation}
\mathcal{C}_{13} = \mathcal{C}_1 \cup \mathcal{C}_3 = \{\1, P_{231}, P_{312}\},
\end{equation}
corresponds to a normal subgroup of $S_3$, which is isomorphic to $\Z(3)$. Identifying all group elements that are related by $\Z(3)$ transformations, we may compactify the group multiplication table as shown below in Tab.\ \ref{Tab:GroupMultiplicationCompactifiedS3}. This reflects the $S_3\hspace{1pt}/\hspace{1.5pt}\Z(3)\cong \Z(2)$ multiplication structure associated with the signature of the permutations (even for elements in $\mathcal{C}_{13}$ and odd for elements in $\mathcal{C}_2$). The non-Abelian group $S_3$ itself is a semidirect product of $\Z(2)$ and $\Z(3)$, i.e., $S_3\cong \Z(3) \rtimes \Z(2)$.

\begin{table}[!h]
  \begin{center}
    \begin{tabular}{|c||c|c|}
      \hline 
      & $\mathcal{C}_{13}$ & $\mathcal{C}_2$ \\
      \hline
      \hline
      $\mathcal{C}_{13}$ & $\mathcal{C}_{13}$ & $\mathcal{C}_2$ \\
      \hline
      $\mathcal{C}_2$   & $\mathcal{C}_2$ & $\mathcal{C}_{13}$ \\
      \hline
    \end{tabular}
  \end{center}
  \caption{\label{Tab:GroupMultiplicationCompactifiedS3}Compactified multiplication table for the symmetric group $S_3$. The conjugacy classes $\mathcal{C}_1 = \{\1\}$ and $\mathcal{C}_3 = \{P_{231}, P_{312}\}$ are combined into the set $\mathcal{C}_{13} = \{\1, P_{231}, P_{312}\}$.}
\end{table}
\begin{table}[!h]
  \begin{center}
    \begin{tabular}{|c||c|c|c|c|c|}
      \hline 
      & $\mathcal{C}_1$ & $\mathcal{C}_2$ & $\mathcal{C}_3$ \\
      \hline
      \hline
      $\chi_{{}_{\Gamma_1}}$ &  $1$    &       $1$      &       $1$      \\
      \hline
      $\chi_{{}_{\Gamma_2}}$ &  $1$    &      $-1$      &       $1$      \\
      \hline
      $\chi_{{}_{\Gamma_3}}$ &  $2$    &       $0$      &      $-1$      \\
      \hline
    \end{tabular}
  \end{center}
  \caption{\label{Tab:CharacterS3}Character table for the symmetric group $S_3$.}
\end{table}

Since $S_3$ has three conjugacy classes, it also has three irreducible representations. Besides the totally symmetric representation $\Gamma_1$ and the totally anti-symmetric representation $\Gamma_2$, which are both one-dimensional, there is a two-dimensional representation $\Gamma_3$ of mixed permutation symmetry. The character table for $S_3$ is shown in Tab.\ \ref{Tab:CharacterS3} from which one can determine the decomposition of products of irreducible representations:
\begin{subequations}
  \begin{IEEEeqnarray}{RCl}
    \Gamma_p\otimes\Gamma_1 &=& \Gamma_p , \quad p = 1,2,3, \\
    \Gamma_2\otimes\Gamma_2 &=& \Gamma_1 , \\
    \Gamma_2\otimes\Gamma_3 &=& \Gamma_3 , \\
    \Gamma_3\otimes\Gamma_3 &=& \Gamma_1\oplus\Gamma_2\oplus\Gamma_3 .
  \end{IEEEeqnarray}
\end{subequations}

\subsection{Distance between group elements in $S_3$}
\label{SubSec:Distance between group elements in S3}

The two-dimensional representation $\Gamma_3$ of the group $S_3$ is real and consists of the orthogonal matrices displayed in Tab.\ \ref{Tab:RepresentationS3}. This representation realizes an embedding of $S_3$ in the orthogonal group $O(2)$, which is non-Abelian (in contrast to the special orthogonal group $SO(2)\cong U(1)$).

\begin{table}[!h]
  \begin{center}
    \begin{tabular}{|c||C{70pt}|C{90pt}|C{90pt}|}
      \hline 
      & $\1$ & $P_{231}$ & $P_{312}$ \\
      \hline
      \hline &&& \\[-9pt]
      $\Gamma_3$ &
      $\left(\begin{array}{cc} 1 & 0 \\ 0 & 1 \end{array}\right)$ &
      $\left(\begin{array}{cc} - \frac{1}{2} & - \frac{\sqrt{3}}{2} \\
        \frac{\sqrt{3}}{2} & - \frac{1}{2} \end{array}\right)$ &
      $\left(\begin{array}{cc} - \frac{1}{2} & \frac{\sqrt{3}}{2} \\
        - \frac{\sqrt{3}}{2} & - \frac{1}{2} \end{array}\right)$ \\[14pt]
      \hline
    \end{tabular}
    \vskip 4pt
    \begin{tabular}{|c||C{70pt}|C{90pt}|C{90pt}|}
      \hline
      & $P_{12}$ & $P_{23}$ & $P_{31}$ \\
      \hline
      \hline &&& \\[-9pt]
      $\Gamma_3$ &
      $\left(\begin{array}{cc} 1 & 0 \\ 0 & -1 \end{array}\right)$ &
      $\left(\begin{array}{cc} - \frac{1}{2} & \frac{\sqrt{3}}{2} \\
        \frac{\sqrt{3}}{2} & \frac{1}{2} \end{array}\right)$ &
      $\left(\begin{array}{cc} - \frac{1}{2} & - \frac{\sqrt{3}}{2} \\
- \frac{\sqrt{3}}{2} & \frac{1}{2} \end{array}\right)$ \\[14pt]
      \hline
    \end{tabular}
  \end{center}
  \caption{\label{Tab:RepresentationS3}Real matrix representation $\Gamma_3$ for group elements in $S_3$.}
\end{table}

Using the representation $\Gamma_3$ to define a distance between group elements $\Omega, \Omega' \in S_3$ [cf.\ Eq.\ \eqref{Eq:Distance}], we obtain
\begin{equation}
  \mu(\Omega,\Omega') = 1 - \frac{1}{2} \Tr \left[\Gamma_3(\Omega') \Gamma_3(\Omega)^{T}\right] ,
\end{equation}
where ${}^{T}$ denotes the matrix transpose. This yields
\begin{subequations}
  \begin{IEEEeqnarray}{RCl}
    \mu(\Omega,\Omega) &=& 0 , \\
    \mu(\1,P_{ij}) &=& \mu(P_{213},P_{ij}) = \mu(P_{312},P_{ij}) = 1 ,
  \end{IEEEeqnarray}
\end{subequations}
where $P_{ij}$ corresponds to an arbitrary pair permutation, while all other elements in the group are separated by a distance of $3/2$. Hence, we observe that different members of the same set $\mathcal{C}_{13}$ or $\mathcal{C}_2$ are separated by a larger distance than elements of $\mathcal{C}_{13}$ from elements in $\mathcal{C}_2$.

\subsection{One-cocycles and $W$ factors}
\label{SubSection:One-cocycles and W factors S3}

A one-cocycle $\omega$ assigns a one-dimensional representation, $\Gamma_1$ or $\Gamma_2$, to each group element in $S_3$. In fact, the cocycle depends not on the group element $U$ itself, but only on the associated conjugacy class $\mathcal{C} = \mathcal{C}(U)$, i.e.,
\begin{equation}
  \omega(\Omega,U) = \Gamma_{p(\mathcal{C})}(\Omega) .
\end{equation}
Since $S_3$ has two one-dimensional representations and three conjugacy classes, there are $2^3 = 8$ candidate one-cocycles. Only two of them, which are listed in Tab.\ \ref{Tab:CocycleS3}, satisfy the consistency condition \eqref{Eq:OneCocycleGroupMultiplication2}. We find that the two allowed one-cocycles $\omega_1$ and $\omega_2$ associate the same representation to the conjugacy classes $\mathcal{C}_1$ and $\mathcal{C}_3$, i.e.,
\begin{equation}
  \omega_i(\Omega, U\in\mathcal{C}_1) = \omega_i(\Omega,U\in\mathcal{C}_3) , \quad i = 1,2 .
\end{equation}
This is a direct consequence of Eq.\ \eqref{Eq:OneCocycleGroupMultiplication2}, which implies that the cocycles descend to a one-dimensional representation of the Abelianization of $S_3$, i.e., $A_{S_3}\cong \Z(2)$ (see Sec.\ \ref{SubSec:One-cocycles and noncommuting operator algebra} and Ref.\ \cite{Mesterhazy:2016} for further details).

\begin{table}[!h]
  \vskip 5pt
  \begin{center}
    \begin{tabular}{|c||c|c|c|c|c|}
      \hline 
      & $\mathcal{C}_1$ & $\mathcal{C}_2$ & $\mathcal{C}_3$ \\
      \hline
      \hline
      $\omega_1$ & $\Gamma_1$ & $\Gamma_1$ & $\Gamma_1$ \\
      \hline
      $\omega_2$ & $\Gamma_1$ & $\Gamma_2$ & $\Gamma_1$ \\
      \hline
    \end{tabular}
  \end{center}
  \caption{\label{Tab:CocycleS3}Allowed one-cocycles $\omega_1$ and $\omega_2$ for the symmetric group $S_3$.}
\end{table}

With these cocycles we may construct the corresponding $W$ factors for the group $S_3$
\begin{equation}
  \label{Eq:WFactorNonAbelian}
  W(\Omega,\Omega') = \omega_i(\Omega',\Omega) \omega_j(\Omega,\Omega')^{-1} .
\end{equation}
Here, we employ the gauge choice $\omega_j(\Omega,\Omega') = \Gamma_1(\Omega) = 1$, ($j = 1$), by which we obtain
\begin{equation}
  W(\Omega,\Omega') = \Gamma_{p_i(\mathcal{C})}(\Omega') , \quad i = 1,2 ,
\end{equation}
where $\mathcal{C} = \mathcal{C}(\Omega)$ is the conjugacy class to which the group element $\Omega$ belongs. Using the two different solutions given in Tab.\ \ref{Tab:CocycleS3}, we may derive the two different types of theories, with distinct $W$ factors, for the symmetric group $S_3$. Inserting the one-cocycle $\omega_1$, we arrive at
\begin{equation}
  W(\Omega,\Omega') = 1 ,
\end{equation}
for all $\Omega,\Omega'\in S_3$, which corresponds to a naively doubled YM theory. On the other hand, we see that the cocycle $\omega_2$ gives rise to a doubled CS-type theory, with
\begin{equation}
  W(\Omega\in\mathcal{C}_2,\Omega'\in\mathcal{C}_2) = -1 ,
\end{equation}
and $W(\Omega,\Omega') = 1$, if either $\Omega\notin\mathcal{C}_2$ or $\Omega'\notin\mathcal{C}_2$.

\section{The Binary Dihedral Group $\bar{D}_2\subset SU(2)$}
\label{Sec:D2bar}

In this appendix we discuss the properties of the eight-element subgroup $\bar{D}_2$ of $SU(2)$. Just as $SU(2)$ itself, this discrete group is non-Abelian, it has the center $\Z(2)$ and a two-dimensional pseudo-real representation.

\subsection{Group multiplication and center}

The group $\bar{D}_2$ can be represented by the following set of group elements
\begin{equation}
  \bar{D}_2 = \{\1, -\1, i\sigma_1, -i\sigma_1, i\sigma_2, -i\sigma_2, i\sigma_3, -i\sigma_3\},
\end{equation}
where $\1$ denotes the $2\times 2$ unit matrix and $\sigma_{\alpha}$, $\alpha = 1,2,3$, are the Pauli matrices. Obviously, the center elements $z\in Z(\bar{D}_2) = \{\pm\1\}$ commute with all group elements and each of them defines its own inverse, i.e., $z^2 = 1$, while $-i\sigma_{\alpha}$ is inverse to $i\sigma_{\alpha}$. The group multiplication rules for $\bar{D}_2$ are given in Tab.\ \ref{Tab:GroupMultiplicationD2bar}.

\begin{table}[!h]
  \begin{center}
    \begin{tabular}{|c||c|c|c|c|c|c|c|c|}
      \hline
      & $\1$ &        $-\1$ &  $i\sigma_1$ &  $-i\sigma_1$ & $i\sigma_2$ & $-i\sigma_2$ & $i\sigma_3$ & $-i\sigma_3$ \\
      \hline
      \hline
      $\1$          &       $\1$ &        $-\1$ &  $i\sigma_1$ &  $-i\sigma_1$ & 
      $i\sigma_2$ & $-i\sigma_2$ & $i\sigma_3$ & $-i\sigma_3$ \\
      \hline
      $-\1$         &      $-\1$ &         $\1$ & $-i\sigma_1$ &  $i\sigma_1$ & 
      $-i\sigma_2$ & $i\sigma_2$ & $-i\sigma_3$ & $i\sigma_3$ \\
      \hline
      $i \sigma_1$  & $i\sigma_1$ & $-i\sigma_1$ &        $-\1$ &        $\1$ & 
      $-i\sigma_3$ & $i\sigma_3$ & $i\sigma_2$ & $-i\sigma_2$ \\
      \hline
      $-i \sigma_1$ & $-i\sigma_1$ & $i\sigma_1$ &        $\1$ &        $-\1$ & 
      $i\sigma_3$ & $-i\sigma_3$ & $-i\sigma_2$ & $i\sigma_2$ \\
      \hline
      $i \sigma_2$  & $i\sigma_2$ & $-i\sigma_2$ & $i\sigma_3$ & $-i\sigma_3$ & 
      $-\1$ &         $\1$ & $-i\sigma_1$ & $i\sigma_1$ \\
      \hline
      $-i \sigma_2$ & $-i\sigma_2$ & $i\sigma_2$ & $-i\sigma_3$ & $i\sigma_3$ & 
      $\1$ &        $-\1$ & $i\sigma_1$ & $-i\sigma_1$ \\
      \hline
      $i \sigma_3$  & $i\sigma_3$ & $-i\sigma_3$ & $-i\sigma_2$ & $i\sigma_2$ & 
      $i\sigma_1$ & $-i\sigma_1$ &        $-\1$ &        $\1$ \\
      \hline
      $-i \sigma_3$ & $-i\sigma_3$ & $i\sigma_3$ & $i\sigma_2$ & $-i\sigma_2$ & 
      $-i\sigma_1$ & $i\sigma_1$ &        $\1$ &        $-\1$ \\
      \hline
    \end{tabular}
  \end{center}
  \caption{\label{Tab:GroupMultiplicationD2bar}Multiplication table for the binary dihedral group $\bar{D}_2$.}
\end{table}

\subsection{Conjugacy classes and irreducible representations}

Based on its multiplication table, one can convince oneself that the group $\bar{D}_2$ has five conjugacy classes
\begin{equation}
  \mathcal{C}_1=\{\1\}, \ \mathcal{C}_2=\{-\1\} , \
  \mathcal{C}_3=\{\pm i\sigma_1\}, \ \mathcal{C}_4=\{\pm i\sigma_2\} , \ 
  \mathcal{C}_5=\{\pm i\sigma_3\} .
\end{equation}
Identifying all group elements that are related to each other by transformations in the center $Z(\bar{D}_2)\cong\Z(2)$, we may combine the conjugacy classes $\mathcal{C}_1$ and $\mathcal{C}_2$ into the set
\begin{equation}
  \mathcal{C}_{12} = \mathcal{C}_1\cup\mathcal{C}_2 = \{\pm \1\},
\end{equation}
which plays the role of a single group element of $\bar{D}_2\hspace{1pt}/\hspace{1.5pt}\Z(2)$. The remaining conjugacy classes $\mathcal{C}_3$, $\mathcal{C}_4$, and $\mathcal{C}_5$ each contain two group elements, which are related by transformations in the center $\Z(2)$. The compactified multiplication table for $\bar{D}_2\hspace{1pt}/\hspace{1pt}\Z(2)$ is shown in Tab.\ \ref{Tab:GroupMultiplicationCompactifiedD2bar}, from which one infers that $\bar{D}_2\hspace{1pt}/\hspace{1.5pt}\Z(2)\cong\Z(2)\times\Z(2)$.

\begin{table}[!h]
  \begin{center}
    \begin{tabular}{|c||c|c|c|c|}
      \hline 
      & $\mathcal{C}_{12}$ & $\mathcal{C}_3$ & $\mathcal{C}_4$ & $\mathcal{C}_5$ \\
      \hline
      \hline
      $\mathcal{C}_{12}$ & $\mathcal{C}_{12}$ & $\mathcal{C}_3$ & $\mathcal{C}_4$ & $\mathcal{C}_5$ \\
      \hline
      $\mathcal{C}_3$   & $\mathcal{C}_3$ & $\mathcal{C}_{12}$ & $\mathcal{C}_5$ & $\mathcal{C}_4$ \\
      \hline
      $\mathcal{C}_4$   & $\mathcal{C}_4$ & $\mathcal{C}_5$ & $\mathcal{C}_{12}$ & $\mathcal{C}_3$ \\
      \hline
      $\mathcal{C}_5$   & $\mathcal{C}_5$ & $\mathcal{C}_4$ & $\mathcal{C}_3$ & $\mathcal{C}_{12}$ \\
      \hline
    \end{tabular}
  \end{center}
  \caption{\label{Tab:GroupMultiplicationCompactifiedD2bar}Compactified multiplication table for the group $\bar{D}_2\hspace{1pt}/\hspace{1.5pt}\Z(2)\cong\Z(2)\times\Z(2)$. The conjugacy classes $\mathcal{C}_1 = \{\1\}$ and $\mathcal{C}_2 = \{-\1\}$, which contain the individual center elements, are combined into the set $\mathcal{C}_{12} = \{\pm \1\}$.}
\end{table}

Since $\bar{D}_2$ has five conjugacy classes, it also has five irreducible representations. The representation $\Gamma_5$, which is defined in terms of the $2\times 2$ matrices $\pm\1$ and $\pm i \sigma_{\alpha}$, $\alpha = 1,2,3$, is pseudo-real. In addition, there are four one-dimensional representations $\Gamma_p$, with $p = 1,2,3,4$. 

\begin{table}[!h]
  \begin{center}
    \begin{tabular}{|c||c|c|c|c|c|}
      \hline 
      & $\mathcal{C}_1$ & $\mathcal{C}_2$ & $\mathcal{C}_3$ & $\mathcal{C}_4$ & $\mathcal{C}_5$ \\
      \hline
      \hline
      $\chi_{{}_{\Gamma_1}}$ &  $1$    &       $1$      &       $1$      &      $1$      &       $1$      \\
      \hline
      $\chi_{{}_{\Gamma_2}}$ &  $1$    &       $1$      &       $1$      &     $-1$      &      $-1$      \\
      \hline
      $\chi_{{}_{\Gamma_3}}$ &  $1$    &       $1$      &      $-1$      &      $1$      &      $-1$      \\
      \hline
      $\chi_{{}_{\Gamma_4}}$ &  $1$    &       $1$      &      $-1$      &     $-1$      &       $1$      \\
      \hline
      $\chi_{{}_{\Gamma_5}}$ &  $2$    &      $-2$      &       $0$      &      $0$      &       $0$      \\
      \hline
    \end{tabular}
  \end{center}
  \caption{\label{Tab:CharacterD2bar}Character table for the binary dihedral group $\bar{D}_2$.}
\end{table}

The character table for $\bar{D}_2$ is shown in Tab.\ \ref{Tab:CharacterD2bar}. While the two-dimensional representation $\Gamma_5$ is pseudo-real and has nontrivial center properties (i.e., nontrivial duality $\mbox{sign}~\Gamma_5(\mathcal{C}_2) = - 1$), the one-dimensional representations $\Gamma_1$, $\Gamma_2$, $\Gamma_3$, and $\Gamma_4$ are real and have trivial duality ($\mbox{sign}~\Gamma_p(\mathcal{C}_2) = 1$, with $p = 1,2,3,4$).

Based on Tab.\ \ref{Tab:CharacterD2bar} one can reduce the products of irreducible representations. The products involving the two-dimensional representation are given by
\begin{subequations}
  \begin{IEEEeqnarray}{RCl}
    \Gamma_p\otimes\Gamma_5 &=& \Gamma_5, \quad p =1,2,3,4 ,  \\
    \Gamma_5\otimes\Gamma_5 &=& \bigoplus_{q=1}^4 \Gamma_q .
  \end{IEEEeqnarray}
\end{subequations}
The products of the one-dimensional representations are listed in Tab.\ \ref{Tab:ProductGammaD2bar}.
\begin{table}[!h]
  \begin{center}
    \begin{tabular}{|c||c|c|c|c|c|c|c|c|c|}
      \hline 
      & $\Gamma_1$ & $\Gamma_2$ & $\Gamma_3$ & $\Gamma_4$ \\
      \hline
      \hline
      $\Gamma_1$ & $\Gamma_1$ & $\Gamma_2$ & $\Gamma_3$ & $\Gamma_4$ \\
      \hline
      $\Gamma_2$ & $\Gamma_2$ & $\Gamma_1$ & $\Gamma_4$ & $\Gamma_3$ \\
      \hline
      $\Gamma_3$ & $\Gamma_3$ & $\Gamma_4$ & $\Gamma_1$ & $\Gamma_2$ \\
      \hline
      $\Gamma_4$ & $\Gamma_4$ & $\Gamma_3$ & $\Gamma_2$ & $\Gamma_1$ \\
      \hline
    \end{tabular}
  \end{center}
\caption{\label{Tab:ProductGammaD2bar}Multiplication table of the one-dimensional representations $\Gamma_p$, $p = 1,2,3,4$, for the binary dihedral group $\bar{D}_2$.}

\end{table}

\subsection{Distance between group elements in $\bar{D}_2$}
\label{SubSec:Distance between group elements in D2bar}

Applying the distance between group elements \eqref{Eq:Distance} defined by the fundamental representation $\Gamma_5$ of $\bar{D}_2$ realizes an embedding in $SU(2)$ and we obtain
\begin{equation}
  \mu(\Omega,\Omega') = 1 - \frac{1}{2} \Re \Tr\left[\Gamma_{5}(\Omega')\Gamma_{5}(\Omega)^{\dagger}\right] .
\end{equation}
This implies that two different members $\Omega$ and $-\Omega$ of the same conjugacy class are separated by a maximal distance $\mu(\Omega,- \Omega) = 2$. The same applies to the center elements, i.e., $\mu(\1,- \1) = 2$. Members of different conjugacy classes (except $\1$ and $-\1$), on the other hand satisfy
\begin{equation}
  \mu(\pm \1,\pm i \sigma_{\alpha}) = \mu(\pm i \sigma_1,\pm i \sigma_2) = \mu(\pm i \sigma_2,\pm i \sigma_3) = \mu(\pm i \sigma_3,\pm i \sigma_1) = 1.
\end{equation}
Thus, members of different conjugacy classes (except $\1$ and $-\1$) are separated by a shorter distance than members of the same class.

\subsection{One-cocycles and $W$ factors}
\label{SubSection:One-cocycles and W factors D2bar}

A one-cocycle $\omega$ assigns a one-dimensional representation $\Gamma_p$, $p = 1,2,3,4$, to each group element in $\bar{D}_2$, i.e., 
\begin{equation}
  \omega(\Omega,U) = \Gamma_{p(\mathcal{C})}(\Omega) ,
\end{equation}
where $\mathcal{C} = \mathcal{C}(U)$. Since the group $\bar{D}_2$ has four one-dimensional representations and five conjugacy classes $\mathcal{C}_q$, $q = 1,2,\ldots, 5$, there are in principle $4^5 = 1024$ one-cocycle candidates. However, only $16$ one-cocycles satisfy the consistency condition \eqref{Eq:OneCocycleGroupMultiplication}, the solutions to which we denote by $\omega_i$, $i = 1,2,\ldots, 16$ [cf.\ Tab.\ \ref{Tab:CocycleD2bar}]. The conjugacy classes $\mathcal{C}_1$ and $\mathcal{C}_2$, which contain the center elements $\pm\1$, are always associated with the same representation, i.e., $p(\mathcal{C}_1) = p(\mathcal{C}_2)$, such that $\mathcal{C}_1$ and $\mathcal{C}_2$ can be combined to $\mathcal{C}_{12} = \mathcal{C}_1\cup\,\mathcal{C}_2$. 

\begin{table}[!h]
  \vskip 5pt
  \begin{center}
    \begin{tabular}{|c||c|c|c|c|c|}
      \hline 
      & $\mathcal{C}_1$ & $\mathcal{C}_2$ & $\mathcal{C}_3$ & $\mathcal{C}_4$ & $\mathcal{C}_5$ \\
      \hline
      \hline
      $\omega_1$ & 
      $\Gamma_1$ & $\Gamma_1$ & $\Gamma_1$ & $\Gamma_1$ & $\Gamma_1$ \\
      \hline
      $\omega_2$ & 
      $\Gamma_1$ & $\Gamma_1$ & $\Gamma_1$ & $\Gamma_2$ & $\Gamma_2$ \\
      \hline
      $\omega_3$ & 
      $\Gamma_1$ & $\Gamma_1$ & $\Gamma_1$ & $\Gamma_3$ & $\Gamma_3$ \\
      \hline
      $\omega_4$ & 
      $\Gamma_1$ & $\Gamma_1$ & $\Gamma_1$ & $\Gamma_4$ & $\Gamma_4$ \\
      \hline
      $\omega_5$ & 
      $\Gamma_1$ & $\Gamma_1$ & $\Gamma_2$ & $\Gamma_1$ & $\Gamma_2$ \\
      \hline
      $\omega_6$ & 
      $\Gamma_1$ & $\Gamma_1$ & $\Gamma_2$ & $\Gamma_2$ & $\Gamma_1$ \\
      \hline
      $\omega_7$ &   
      $\Gamma_1$ & $\Gamma_1$ & $\Gamma_3$ & $\Gamma_1$ & $\Gamma_3$ \\
      \hline
      $\omega_8$ & 
      $\Gamma_1$ & $\Gamma_1$ & $\Gamma_3$ & $\Gamma_3$ & $\Gamma_1$ \\
      \hline
      $\omega_9$ & 
      $\Gamma_1$ & $\Gamma_1$ & $\Gamma_4$ & $\Gamma_1$ & $\Gamma_4$ \\
      \hline
      $\omega_{10}$ & 
      $\Gamma_1$ & $\Gamma_1$ & $\Gamma_4$ & $\Gamma_4$ & $\Gamma_1$ \\
      \hline
      $\omega_{11}$ & 
      $\Gamma_1$ & $\Gamma_1$ & $\Gamma_2$ & $\Gamma_3$ & $\Gamma_4$ \\
      \hline
      $\omega_{12}$ & 
      $\Gamma_1$ & $\Gamma_1$ & $\Gamma_2$ & $\Gamma_4$ & $\Gamma_3$ \\
      \hline
      $\omega_{13}$ & 
      $\Gamma_1$ & $\Gamma_1$ & $\Gamma_3$ & $\Gamma_2$ & $\Gamma_4$ \\
      \hline
      $\omega_{14}$ & 
      $\Gamma_1$ & $\Gamma_1$ & $\Gamma_3$ & $\Gamma_4$ & $\Gamma_2$ \\
      \hline
      $\omega_{15}$ & 
      $\Gamma_1$ & $\Gamma_1$ & $\Gamma_4$ & $\Gamma_2$ & $\Gamma_3$ \\
      \hline
      $\omega_{16}$ & 
      $\Gamma_1$ & $\Gamma_1$ & $\Gamma_4$ & $\Gamma_3$ & $\Gamma_2$ \\
      \hline
    \end{tabular}
    \end{center}
  \caption{\label{Tab:CocycleD2bar}Allowed one-cocycles for the binary dihedral group $\bar{D}_2$.}
\end{table}
With the $16$ allowed one-cocycles we may determine the $W$ factors in the doubled $\bar{D}_2$ theory, i.e.,
\begin{equation}
  W(\Omega,\Omega') = \omega_i(\Omega',\Omega) \omega_j(\Omega,\Omega')^{-1} .
\end{equation}
Proceeding along similar lines as in the case of the $S_3$ theory, we choose the gauge $\omega_j(\Omega,\Omega') = 1$, by which 
\begin{equation}
  W(\Omega,\Omega') = \Gamma_{p_i(\mathcal{C})}(\Omega') , \quad i = 1,2, \ldots,16 ,
\end{equation}
where $\mathcal{C}$ is the conjugacy class associated to $\Omega$. Thus, we obtain $16$ distinct $W$ factors, which correspond to $16$ different doubled $\bar{D}_2$ lattice gauge theories. Inserting the different solutions from Tab.\ \ref{Tab:CocycleD2bar}, we find that only one of them yields a doubled YM theory with trivial $W$ factor, i.e., for $\omega_i = \omega_1$,
\begin{equation}
  W(\Omega,\Omega') = 1 ,
\end{equation}
and $\Omega, \Omega'\in \bar{D}_2$. The remaining one-cocycles $\omega_i$, $i = 2,3, \ldots, 16$, give rise to CS-type theories with $W$ factors
\begin{equation}
  W(\mathcal{C},\mathcal{C}') \in \{\pm 1\} .
\end{equation}
However, note that the single-element conjugacy classes $\mathcal{C}_1$ and $\mathcal{C}_2$ (that contain the center elements $\pm \1$) always yield $W(\mathcal{C},\mathcal{C}') = 1$.

\section{The Discrete Subgroup $\Delta(27)$ of $SU(3)$}
\label{Sec:Delta27}

Here we discuss properties of a 27-element subgroup of $SU(3)$, also known as $\Delta(27)$ \cite{Fairbairn:1964sga,Fairbairn:1982jx,Ludl:2011gn}. This discrete group shares several important features with $SU(3)$: it is non-Abelian, has the center $\Z(3)$, as well as two complex three-dimensional representations.

\subsection{Group multiplication and center}

The group $\Delta(27)$ consists of $27$ group elements, here represented by the following $3\times 3$ matrices:

\begin{subequations}
  \begin{IEEEeqnarray}{RCl}
    D(a,b)&=&\left(\begin{array}{ccc} a & 0 & 0 \\ 0 & b & 0 \\ 0 & 0 & a^{\ast} b^{\ast}
    \end{array}\right) , \\[2pt]
    U(a,b)&=&\left(\begin{array}{ccc} 0 & a & 0 \\ 0 & 0 & b \\ a^{\ast} b^{\ast} & 0 & 0
    \end{array}\right) , \\[2pt]
    L(a,b)&=&\left(\begin{array}{ccc} 0 & 0 & a^{\ast} b^{\ast} \\ a & 0 & 0 \\ 0 & b & 0
    \end{array}\right) ,
  \end{IEEEeqnarray}
\end{subequations}
where the parameters $a$ and $b$ can each take any one of the three values in the set $\{1,e^{2 \pi i/3},e^{-2 \pi i/3}\}$. Hence, there are nine elements in each of the three classes of matrices defined by $D(a,b)$, $U(a,b)$, and $L(a,b)$, such that
\begin{equation}
  \Delta(27) = \left\{D(a,b), U(a,b), L(a,b) \hspace{0pt}\left|\hspace{2pt} a, b\in \{1,e^{2 \pi i/3},e^{-2 \pi i/3}\}\right.\right\}.
\end{equation}
The group multiplication rules are given in Tab.\ \ref{Tab:GroupMultiplicationDelta27} and the inverses of the various group elements are given by their conjugate transpose, i.e., $D(a,b)^{-1} = D(a,b)^{\dagger} = D(a,b)^{\ast}$, $U(a,b)^{-1} = U(a,b)^{\dagger} = L(a^{\ast},b^{\ast}) = L(a,b)^{\ast}$, and  $L(a,b)^{-1} = L(a,b)^{\dagger} = U(a,b)^{\ast}$. The three group elements $D(a,a)=a\1$, where $\1$ denotes the $3\times 3$ unit matrix and $a\in\{1,e^{2 \pi i/3},e^{-2 \pi i/3}\}$, commute with all group elements and therefore the center is $Z(\Delta(27))\cong\Z(3)$.

\begin{table}[!h]
  \begin{center}
    \begin{tabular}{|c||c|c|c|}
      \hline
      & $D(c,d)$ & $U(c,d)$ & $L(c,d)$ \\
      \hline
      \hline &&& \\[-10pt]
      $D(a,b)$ & $D(ac,bd)$ & $U(ac,bd)$ & $L(bc,a^{\ast} b^{\ast} d)$ \\[2pt]
      $U(a,b)$ & $U(ad,bc^{\ast}d^{\ast})$ & $L(bc^{\ast}d^{\ast},a^{\ast}b^{\ast}c)$ & $D(ac,bd)$ \\[2pt]
      $L(a,b)$ & $L(ac,bd)$ & $D(a^{\ast}b^{\ast}c^{\ast}d^{\ast},ac)$ & $U(a^{\ast}b^{\ast}d,ac^{\ast}d^{\ast})$ \\[2pt]
      \hline
    \end{tabular}
  \end{center}
  \caption{\label{Tab:GroupMultiplicationDelta27}Multiplication table for the group $\Delta(27)$.}
\end{table}

\subsection{Conjugacy classes and irreducible representations}

Based on the group multiplication rules one can identify $11$ conjugacy classes, three of which are defined by the center elements
\begin{subequations}
  \begin{IEEEeqnarray}{RCl}
    \mathcal{C}_1 &=& \{\1\} , \\
    \mathcal{C}_2 &=& \{z \1\} = \overline{\mathcal{C}}_3 , \\
    \mathcal{C}_3 &=& \{z^{\ast} \1\} = \overline{\mathcal{C}}_2 ,
  \end{IEEEeqnarray}
\end{subequations}
where $z = e^{2\pi i/3}$ and $\overline{\mathcal{C}}$ is the set of the complex conjugate elements of $\mathcal{C}$, as well as those conjugacy classes consisting of three elements
\begin{subequations}
  \begin{IEEEeqnarray}{RCl}
    \mathcal{C}_4 &=& \left\{D(1,z), \hspace{4pt}D(z,z^{\ast}), \hspace{1pt}D(z^{\ast},1)\right\} = \overline{\mathcal{C}}_5, \\
    \mathcal{C}_5 &=& \left\{D(1,z^{\ast}), D(z^{\ast},z), \hspace{0.5pt}D(z,1)\right\} = \overline{\mathcal{C}}_4, \\
    \mathcal{C}_6 &=& \left\{U(1,1), \hspace{6pt}U(z,z), \hspace{6pt}U(z^{\ast},z^{\ast})\right\} = \overline{\mathcal{C}}_6, \\
    \mathcal{C}_7 &=& \left\{U(1,z), \hspace{5.5pt}U(z,z^{\ast}), \hspace{2pt}U(z^{\ast},1)\right\} = \overline{\mathcal{C}}_8, \\
    \mathcal{C}_8 &=& \left\{U(1,z^{\ast}), \hspace{1pt}U(z^{\ast},z), \hspace{2pt}U(z,1)\right\} = \overline{\mathcal{C}}_7, \\
    \mathcal{C}_9 &=& \left\{L(1,1), \hspace{6pt}L(z,z), \hspace{8pt}L(z^{\ast},z^{\ast})\right\} = \overline{\mathcal{C}}_9, \\
    \mathcal{C}_{10} &=& \left\{L(1,z), \hspace{5.5pt}L(z,z^{\ast}), \hspace{3pt}L(z^{\ast},1)\right\} = \overline{\mathcal{C}}_{11}, \\
    \mathcal{C}_{11} &=& \left\{L(1,z^{\ast}), L(z^{\ast},z), \hspace{3pt}L(z,1)\right\} = \overline{\mathcal{C}}_{10}.
  \end{IEEEeqnarray}
\end{subequations}
When one identifies all group elements that are related to each other by transformations in the center $\Z(3)$, the conjugacy classes $\mathcal{C}_1$, $\mathcal{C}_2$, and $\mathcal{C}_3$ are combined into the set
\begin{equation}
  \mathcal{C}_{123} = \mathcal{C}_1 \cup \mathcal{C}_2 \cup \mathcal{C}_3 = \{\1, z \1, z^{\ast} \1\},
\end{equation}

\begin{table}[!b]
  \begin{center}
    \begin{tabular}{|c||c|c|c|c|c|c|c|c|c|}
      \hline 
      & $\mathcal{C}_{123}$ & $\mathcal{C}_4$ & $\mathcal{C}_5$ & $\mathcal{C}_6$ &
      $\mathcal{C}_7$ & $\mathcal{C}_8$ & $\mathcal{C}_9$ & $\mathcal{C}_{10}$ & $\mathcal{C}_{11}$ \\
      \hline
      \hline
      $\mathcal{C}_{123}$ & $\mathcal{C}_{123}$ & $\mathcal{C}_4$ & $\mathcal{C}_5$ & $\mathcal{C}_6$ &$\mathcal{C}_7$ & $\mathcal{C}_8$ & $\mathcal{C}_9$ & $\mathcal{C}_{10}$ & $\mathcal{C}_{11}$ \\
      \hline
      $\mathcal{C}_4$ & $\mathcal{C}_4$ & $\mathcal{C}_5$ & $\mathcal{C}_{123}$ & $\mathcal{C}_7$ & 
      $\mathcal{C}_8$ & $\mathcal{C}_6$ & $\mathcal{C}_{10}$ & $\mathcal{C}_{11}$ & $\mathcal{C}_9$ \\
      \hline
      $\mathcal{C}_5$ & $\mathcal{C}_5$ & $\mathcal{C}_{123}$ & $\mathcal{C}_4$ & $\mathcal{C}_8$ & 
      $\mathcal{C}_6$ & $\mathcal{C}_7$ & $\mathcal{C}_{11}$ & $\mathcal{C}_9$ & $\mathcal{C}_{10}$ \\
      \hline
      $\mathcal{C}_6$ & $\mathcal{C}_6$ & $\mathcal{C}_7$ & $\mathcal{C}_8$ & $\mathcal{C}_9$ & 
      $\mathcal{C}_{10}$ & $\mathcal{C}_{11}$ & $\mathcal{C}_{123}$ & $\mathcal{C}_4$ & $\mathcal{C}_5$ 
      \\
      \hline
      $\mathcal{C}_7$ & $\mathcal{C}_7$ & $\mathcal{C}_8$ & $\mathcal{C}_6$ & $\mathcal{C}_{10}$ & 
      $\mathcal{C}_{11}$ & $\mathcal{C}_9$ & $\mathcal{C}_4$ & $\mathcal{C}_5$ & $\mathcal{C}_{123}$ \\
      \hline
      $\mathcal{C}_8$ & $\mathcal{C}_8$ & $\mathcal{C}_6$ & $\mathcal{C}_7$ & $\mathcal{C}_{11}$ & 
      $\mathcal{C}_9$ & $\mathcal{C}_{10}$ & $\mathcal{C}_5$ & $\mathcal{C}_{123}$ & $\mathcal{C}_4$ \\
      \hline
      $\mathcal{C}_9$ & $\mathcal{C}_9$ & $\mathcal{C}_{10}$ & $\mathcal{C}_{11}$ & $\mathcal{C}_{123}$&
      $\mathcal{C}_4$ & $\mathcal{C}_5$ & $\mathcal{C}_6$ & $\mathcal{C}_7$ & $\mathcal{C}_8$ \\
      \hline
      $\mathcal{C}_{10}$ & $\mathcal{C}_{10}$ & $\mathcal{C}_{11}$ & $\mathcal{C}_9$ & $\mathcal{C}_4$&
      $\mathcal{C}_5$ & $\mathcal{C}_{123}$ & $\mathcal{C}_7$ & $\mathcal{C}_8$ & $\mathcal{C}_6$ \\
      \hline
      $\mathcal{C}_{11}$ & $\mathcal{C}_{11}$ & $\mathcal{C}_9$ & $\mathcal{C}_{10}$ & $\mathcal{C}_5$&
      $\mathcal{C}_{123}$ & $\mathcal{C}_4$ & $\mathcal{C}_8$ & $\mathcal{C}_6$ & $\mathcal{C}_7$ \\
      \hline
    \end{tabular}
  \end{center}
  \caption{\label{Tab:GroupMultiplicationCompactifiedDelta27}Multiplication table for the sets $\mathcal{C}_{123}$, $\mathcal{C}_4$, $\mathcal{C}_5$, \ldots identified with the group elements of $\Delta(27)\hspace{1pt}/\hspace{1.5pt}\Z(3)\cong\Z(3)\times\Z(3)$. The conjugacy classes $\mathcal{C}_1 = \{\1\}$, $\mathcal{C}_2 = \{z \1\}$, and $\mathcal{C}_3 = \{z^{\ast} \1\}$, with $z = e^{2 \pi i/3}$, which contain the individual center elements are combined to $\mathcal{C}_{123} = \{\1, z \1, z^{\ast} \1\}$.}
\end{table}

\noindent
which plays the role of a single group element of $\Delta(27)\hspace{1pt}/\hspace{1.5pt}\Z(3)$. The other conjugacy classes $\mathcal{C}_4,\mathcal{C}_5,\dots,\mathcal{C}_{11}$ each contain three group elements which are related by transformations in the center. The products of the conjugacy classes, identified with group elements of $\Delta(27)\hspace{1pt}/\hspace{1.5pt}\Z(3)$, are listed in Tab.\ \ref{Tab:GroupMultiplicationCompactifiedDelta27}. From this multiplication table one infers that $\Delta(27)\hspace{1pt}/\hspace{1.5pt}\Z(3)\cong\Z(3)\times\Z(3)$.

Since $\Delta(27)$ has $11$ conjugacy classes it also has $11$ (inequivalent) irreducible representations. The representation that we have used to define $\Delta(27)$ is three-dimensional and complex and corresponds to $\Gamma_{10}$. Its conjugate representation $\Gamma_{11} = \overline{\Gamma}_{10}$ (in which all representation matrices of $\Gamma_{10}$ are complex conjugated) is also three-dimensional. In addition, there are nine one-dimensional representations $\Gamma_1, \Gamma_2, \ldots, \Gamma_9$. The character table of $\Delta(27)$ is shown in Tab.\ \ref{Tab:CharacterDelta27}.

\begin{table}[!h]
  \begin{center}
    \begin{tabular}{|c||c|c|c|c|c|c|c|c|c|c|c|}
      \hline 
      & $\mathcal{C}_1$ & $\mathcal{C}_2$ & $\mathcal{C}_3$ & 
      $\mathcal{C}_4$ & $\mathcal{C}_5$ & $\mathcal{C}_6$ & $\mathcal{C}_7$ & $\mathcal{C}_8$ & 
      $\mathcal{C}_9$ & $\mathcal{C}_{10}$ & $\mathcal{C}_{11}$ \\
      \hline
      \hline
      $\chi_{{}_{\Gamma_1}} = \chi_{{}_{\Gamma_1}}^{\ast}$  
      &  1  &  1  &  1  &  1  &  1  &  1  &  1  &  1  &  1  &  1  &  1  \\
      \hline
      $\chi_{{}_{\Gamma_2}} = \chi_{{}_{\Gamma_3}}^{\ast}$  
      &  1  &  1  &  1  &  1  &  1  & $z$ & $z$ & $z$ &$z^{\ast}$&$z^{\ast}$&$z^{\ast}$\\
      \hline
      $\chi_{{}_{\Gamma_3}} = \chi_{{}_{\Gamma_2}}^{\ast}$  
      &  1  &  1  &  1  &  1  &  1  &$z^{\ast}$&$z^{\ast}$&$z^{\ast}$& $z$ & $z$ & $z$ \\
      \hline
      $\chi_{{}_{\Gamma_4}} = \chi_{{}_{\Gamma_7}}^{\ast}$  
      &  1  &  1  &  1  & $z$ &$z^{\ast}$&  1  & $z$ &$z^{\ast}$&  1  & $z$ &$z^{\ast}$\\
      \hline
      $\chi_{{}_{\Gamma_5}} = \chi_{{}_{\Gamma_9}}^{\ast}$  
      &  1  &  1  &  1  & $z$ &$z^{\ast}$& $z$ &$z^{\ast}$&  1  &$z^{\ast}$&  1  & $z$ \\
      \hline
      $\chi_{{}_{\Gamma_6}} = \chi_{{}_{\Gamma_8}}^{\ast}$  
      &  1  &  1  &  1  & $z$ &$z^{\ast}$&$z^{\ast}$&  1  & $z$ & $z$ &$z^{\ast}$&  1  \\
      \hline
      $\chi_{{}_{\Gamma_7}} = \chi_{{}_{\Gamma_4}}^{\ast}$  
      &  1  &  1  &  1  &$z^{\ast}$& $z$ &  1  &$z^{\ast}$& $z$ &  1  &$z^{\ast}$& $z$ \\
      \hline
      $\chi_{{}_{\Gamma_8}} = \chi_{{}_{\Gamma_6}}^{\ast}$  
      &  1  &  1  &  1  &$z^{\ast}$& $z$ & $z$ &  1  &$z^{\ast}$&$z^{\ast}$& $z$ &  1  \\
      \hline
      $\chi_{{}_{\Gamma_9}} = \chi_{{}_{\Gamma_5}}^{\ast}$  
      &  1  &  1  &  1  &$z^{\ast}$& $z$ &$z^{\ast}$& $z$ &  1  & $z$ &  1  &$z^{\ast}$\\
      \hline
      $\chi_{{}_{\Gamma_{10}}} = \chi_{{}_{\Gamma_{11}}}^{\ast}$
      &  3 & $3z$&$3z^{\ast}$&  0  &  0  &  0  &  0  &  0  &  0  &  0  &  0  \\
      \hline
      $\chi_{{}_{\Gamma_{11}}} = \chi_{{}_{\Gamma_{10}}}^{\ast}$
      &  3 &$3z^{\ast}$& $3z$&  0  &  0  &  0  &  0  &  0  &  0  &  0  &  0  \\
      \hline
    \end{tabular}
  \end{center}
  \caption{\label{Tab:CharacterDelta27}Character table for the group $\Delta(27)$, with $z = e^{2 \pi i/3}$.}
\end{table}

\noindent
All representations except the trivial one $\Gamma_1 = \overline{\Gamma}_1$ are complex. The pairs of conjugate representations are
\begin{equation}
  \overline{\Gamma}_2 = \Gamma_3, \quad \overline{\Gamma}_4 = \Gamma_7, \quad \overline{\Gamma}_5 = \Gamma_9, \quad \overline{\Gamma}_6 = \Gamma_8, \quad \overline{\Gamma}_{10} = \Gamma_{11} .
\end{equation}

Based on the character Tab.\ \ref{Tab:CharacterDelta27} one can deduce the products of irreducible representations. The products involving three-dimensional representations are given by
\begin{subequations}
  \begin{IEEEeqnarray}{RCl}
    \Gamma_p\otimes\Gamma_{10} &=& \Gamma_{10}, \quad p = 1,2, \ldots, 9, \\[2pt]
    \Gamma_{10}\otimes\Gamma_{10} &=& 3\hspace{1pt}\Gamma_{11}, \\
    \Gamma_{11}\otimes\Gamma_{10} &=& \bigoplus_{p = 1}^9 \Gamma_p ,
  \end{IEEEeqnarray}
\end{subequations}
while the remaining reduction formulas can be derived by complex conjugation using $\overline{\Gamma}_{10} = \Gamma_{11}$. The three-dimensional representations $\Gamma_{10}$ and $\Gamma_{11}$ have nontrivial opposite triality, while the one-dimensional representations have trivial triality. The tensor products of the one-dimensional representations are listed in Tab.\ \ref{Tab:ProductGammaDelta27}.

\begin{table}[!h]
  \begin{center}
    \begin{tabular}{|c||c|c|c|c|c|c|c|c|c|}
      \hline 
      & $\Gamma_1$ & $\Gamma_2$ & $\Gamma_3$ & $\Gamma_4$ &
      $\Gamma_5$ & $\Gamma_6$ & $\Gamma_7$ & $\Gamma_8$ & $\Gamma_9$ \\
      \hline
      \hline
      $\Gamma_1$ & $\Gamma_1$ & $\Gamma_2$ & $\Gamma_3$ & 
      $\Gamma_4$ & $\Gamma_5$ & $\Gamma_6$ & $\Gamma_7$ & $\Gamma_8$ & $\Gamma_9$ \\
      \hline
      $\Gamma_2$ & $\Gamma_2$ & $\Gamma_3$ & $\Gamma_1$ & 
      $\Gamma_5$ & $\Gamma_6$ & $\Gamma_4$ & $\Gamma_8$ & $\Gamma_9$ & $\Gamma_7$ \\
      \hline
      $\Gamma_3$ & $\Gamma_3$ & $\Gamma_1$ & $\Gamma_2$ & 
      $\Gamma_6$ & $\Gamma_4$ & $\Gamma_5$ & $\Gamma_9$ & $\Gamma_7$ & $\Gamma_8$ \\
      \hline
      $\Gamma_4$ & $\Gamma_4$ & $\Gamma_5$ & $\Gamma_6$ & 
      $\Gamma_7$ & $\Gamma_8$ & $\Gamma_9$ & $\Gamma_1$ & $\Gamma_2$ & $\Gamma_3$ \\
      \hline
      $\Gamma_5$ & $\Gamma_5$ & $\Gamma_6$ & $\Gamma_4$ & 
      $\Gamma_8$ & $\Gamma_9$ & $\Gamma_7$ & $\Gamma_2$ & $\Gamma_3$ & $\Gamma_1$ \\
      \hline
      $\Gamma_6$ & $\Gamma_6$ & $\Gamma_4$ & $\Gamma_5$ & 
      $\Gamma_9$ & $\Gamma_7$ & $\Gamma_8$ & $\Gamma_3$ & $\Gamma_1$ & $\Gamma_2$ \\
      \hline
      $\Gamma_7$ & $\Gamma_7$ & $\Gamma_8$ & $\Gamma_9$ & 
      $\Gamma_1$ & $\Gamma_2$ & $\Gamma_3$ & $\Gamma_4$ & $\Gamma_5$ & $\Gamma_6$ \\
      \hline
      $\Gamma_8$ & $\Gamma_8$ & $\Gamma_9$ & $\Gamma_7$ & 
      $\Gamma_2$ & $\Gamma_3$ & $\Gamma_1$ & $\Gamma_5$ & $\Gamma_6$ & $\Gamma_4$ \\
      \hline
      $\Gamma_9$ & $\Gamma_9$ & $\Gamma_7$ & $\Gamma_8$ & 
      $\Gamma_3$ & $\Gamma_1$ & $\Gamma_2$ & $\Gamma_6$ & $\Gamma_4$ & $\Gamma_5$ \\
      \hline
    \end{tabular}
  \end{center}
  \caption{\label{Tab:ProductGammaDelta27}Multiplication table for the one-dimensional representations $\Gamma_p$, $p = 1, 2, \ldots, 9$, of the group $\Delta(27)$.}
\end{table}

\subsection{Distance between group elements in $\Delta(27)$}
\label{SubSec:Distance between group elements in Delta27}

Applying the distance between group elements Eq.\ \eqref{Eq:Distance} defined by the three-dimensional fundamental representation $\Gamma_{10}$ (or equivalently $\Gamma_{11}$) of $\Delta(27)$, we obtain
\begin{equation}
  \mu(\Omega,\Omega') = 1 - \frac{1}{3} \Re \Tr\left[\Gamma_{10}(\Omega')\Gamma_{10}(\Omega)^{\dagger}\right] , 
\end{equation}
which realizes an embedding in $SU(3)$. This implies that different members $\Omega$, $z \Omega$, and $z^{\ast} \Omega$ of the same conjugacy class, where $\Omega\in\Delta(27)$ and $z\in Z(\Delta(27))$, are separated by a maximal distance 
\begin{equation}
  \mu(\Omega,z \Omega) =  \mu(z \Omega,z^{\ast} \Omega) = \mu(z^{\ast} \Omega,\Omega) = 3/2 .
\end{equation}
Members of different conjugacy classes (except $\1$, $z \1$, and $z^{\ast}\1$), on the other hand, are separated by the distance $1$. Hence, once again, members of different conjugacy classes (except $\1$, $z \1$, and $z^{\ast} \1$) are separated by a shorter distance than group elements, which belong to the same class.

\subsection{One-cocycles and $W$ factors}
\label{SubSection:One-cocycles and W factors Delta27}

A one-cocycle $\omega$ assigns a one-dimensional representation $\Gamma_p$, $p = 1,2, \ldots, 9$, to each group element in $\Delta(27)$, i.e., 
\begin{equation}
  \label{Eq:OneCocycleDelta27}
  \omega(\Omega,U) = \Gamma_{p(\mathcal{C})}(\Omega) ,
\end{equation}
where $\mathcal{C} = \mathcal{C}(U)$. Since $\Delta(27)$ has nine one-dimensional representations and $11$ conjugacy classes $\mathcal{C}_1$, $\mathcal{C}_2,\dots,\mathcal{C}_{11}$, there are $9^{11}$ possible one-cocycle candidates. The conjugacy classes $\mathcal{C}_1$, $\mathcal{C}_2$, and $\mathcal{C}_3$, which contain the center elements $\1, z \1,$ and $z^{\ast} \1$, are always associated with the same representation, i.e., $p(\mathcal{C}_1) = p(\mathcal{C}_2) = p(\mathcal{C}_3)$, so that $\mathcal{C}_1$, $\mathcal{C}_2$, and $\mathcal{C}_3$ can again be combined into the set $\mathcal{C}_{123}$. We find $81$ allowed cocycles $\omega_i$ by solving the consistency condition Eq.\ \eqref{Eq:OneCocycleGroupMultiplication2}, a subset of which is shown in Tab.\ \ref{Tab:CocycleDelta27}. Note that due to the conjugation symmetry of the representations of $\Delta(27)$, also the one-cocycles \eqref{Eq:OneCocycleDelta27} will be related to each other.

\begin{table}[!h]
  \vskip 10pt
  \begin{center}
    \begin{tabular}{|c||c|c|c|c|c|c|c|c|c|c|c|}
      \hline 
      & 
      $\mathcal{C}_1$ & $\mathcal{C}_2$ & $\mathcal{C}_3$ & $\mathcal{C}_4$ & $\mathcal{C}_5$ & $\mathcal{C}_6$ & $\mathcal{C}_7$ &
      $\mathcal{C}_8$ & $\mathcal{C}_9$ & $\mathcal{C}_{10}$ & $\mathcal{C}_{11}$ \\
      \hline
      \hline
      $\omega_1(\Omega,U) = \omega_1(\Omega,U)^{\ast}$ & 
      $\Gamma_1$ & $\Gamma_1$ & $\Gamma_1$ & $\Gamma_1$ & $\Gamma_1$ & $\Gamma_1$ & $\Gamma_1$ & $\Gamma_1$ &
      $\Gamma_1$ & $\Gamma_1$ & $\Gamma_1$ \\
      \hline
      $\omega_2(\Omega,U) = \omega_3(\Omega,U)^{\ast}$ & 
      $\Gamma_1$ & $\Gamma_1$ & $\Gamma_1$ & $\Gamma_1$ & $\Gamma_1$ & $\Gamma_2$ & $\Gamma_2$ & $\Gamma_2$ &
      $\Gamma_3$ & $\Gamma_3$ & $\Gamma_3$ \\
      \hline
      $\omega_4(\Omega,U) = \omega_7(\Omega,U)^{\ast}$ & 
      $\Gamma_1$ & $\Gamma_1$ & $\Gamma_1$ & $\Gamma_4$ & $\Gamma_7$ & $\Gamma_1$ & $\Gamma_4$ & $\Gamma_7$ &
      $\Gamma_1$ & $\Gamma_4$ & $\Gamma_7$ \\
      \hline
      $\omega_5(\Omega,U) = \omega_9(\Omega,U)^{\ast}$ & 
      $\Gamma_1$ & $\Gamma_1$ & $\Gamma_1$ & $\Gamma_5$ & $\Gamma_9$ & $\Gamma_5$ & $\Gamma_9$ & $\Gamma_1$ &
      $\Gamma_9$ & $\Gamma_1$ & $\Gamma_5$ \\
      \hline
      $\omega_6(\Omega,U) = \omega_8(\Omega,U)^{\ast}$ & 
      $\Gamma_1$ & $\Gamma_1$ & $\Gamma_1$ & $\Gamma_6$ & $\Gamma_8$ & $\Gamma_8$ & $\Gamma_1$ & $\Gamma_6$ &
      $\Gamma_6$ & $\Gamma_8$ & $\Gamma_1$ \\
      \hline
      $\omega_{10}(\Omega,U) = \omega_{11}(\Omega,U)^{\ast}$ & 
      $\Gamma_1$ & $\Gamma_1$ & $\Gamma_1$ & $\Gamma_6$ & $\Gamma_8$ & $\Gamma_9$ & $\Gamma_2$ & $\Gamma_4$ &
      $\Gamma_5$ & $\Gamma_7$ & $\Gamma_3$ \\
      \hline
      $\omega_{12}(\Omega,U) = \omega_{13}(\Omega,U)^{\ast}$ & 
      $\Gamma_1$ & $\Gamma_1$ & $\Gamma_1$ & $\Gamma_2$ & $\Gamma_3$ & $\Gamma_4$ & $\Gamma_5$ & $\Gamma_6$ &
      $\Gamma_7$ & $\Gamma_8$ & $\Gamma_9$ \\
      \hline
      $\omega_{14}(\Omega,U) = \omega_{15}(\Omega,U)^{\ast}$ & 
      $\Gamma_1$ & $\Gamma_1$ & $\Gamma_1$ & $\Gamma_2$ & $\Gamma_3$ & $\Gamma_5$ & $\Gamma_6$ & $\Gamma_4$ &
      $\Gamma_9$ & $\Gamma_7$ & $\Gamma_8$ \\
      \hline
      $\omega_{16}(\Omega,U) = \omega_{17}(\Omega,U)^{\ast}$ & 
      $\Gamma_1$ & $\Gamma_1$ & $\Gamma_1$ & $\Gamma_2$ & $\Gamma_3$ & $\Gamma_6$ & $\Gamma_4$ & $\Gamma_5$ &
      $\Gamma_8$ & $\Gamma_9$ & $\Gamma_7$ \\
      \hline
      $\omega_{18}(\Omega,U) = \omega_{19}(\Omega,U)^{\ast}$ & 
      $\Gamma_1$ & $\Gamma_1$ & $\Gamma_1$ & $\Gamma_4$ & $\Gamma_7$ & $\Gamma_2$ & $\Gamma_5$ & $\Gamma_8$ &
      $\Gamma_3$ & $\Gamma_6$ & $\Gamma_9$ \\
      \hline
      $\omega_{20}(\Omega,U) = \omega_{21}(\Omega,U)^{\ast}$ & 
      $\Gamma_1$ & $\Gamma_1$ & $\Gamma_1$ & $\Gamma_4$ & $\Gamma_7$ & $\Gamma_3$ & $\Gamma_6$ & $\Gamma_9$ &
      $\Gamma_2$ & $\Gamma_5$ & $\Gamma_8$ \\
      \hline
      $\omega_{22}(\Omega,U) = \omega_{23}(\Omega,U)^{\ast}$ & 
      $\Gamma_1$ & $\Gamma_1$ & $\Gamma_1$ & $\Gamma_5$ & $\Gamma_9$ & $\Gamma_4$ & $\Gamma_8$ & $\Gamma_3$ &
      $\Gamma_7$ & $\Gamma_2$ & $\Gamma_6$ \\
      \hline
      $\omega_{24}(\Omega,U) = \omega_{25}(\Omega,U)^{\ast}$ & 
      $\Gamma_1$ & $\Gamma_1$ & $\Gamma_1$ & $\Gamma_5$ & $\Gamma_9$ & $\Gamma_6$ & $\Gamma_7$ & $\Gamma_2$ &
      $\Gamma_8$ & $\Gamma_3$ & $\Gamma_4$ \\
      \hline
      $\omega_{26}(\Omega,U) = \omega_{27}(\Omega,U)^{\ast}$ & 
      $\Gamma_1$ & $\Gamma_1$ & $\Gamma_1$ & $\Gamma_6$ & $\Gamma_8$ & $\Gamma_7$ & $\Gamma_3$ & $\Gamma_5$ &
      $\Gamma_4$ & $\Gamma_9$ & $\Gamma_2$ \\
      \hline
    \end{tabular}
  \end{center}
  \caption{\label{Tab:CocycleDelta27}$27$ among the $81$ allowed one-cocycles for the group $\Delta(27)$, for which we illustrate the conjugation symmetry.}
\end{table}

\newpage
In complete analogy to the groups $S_3$ and $\bar{D}_2$, we observe that the $W$ factors are in one-to-one correspondence to the one-cocycles. Applying a suitable gauge choice (the asymmetric gauge), we find one trivial doubled YM theory, associated to the cocycle $\omega_1$, which yields
\begin{equation}
  W(\Omega,\Omega') = 1 ,
\end{equation}
for all $\Omega, \Omega' = 1$. The remaining one-cocycles $\omega_i$, $i = 2,3,\ldots, 81$ yield (potentially) nontrivial $W$ factors
\begin{equation}
  W(\mathcal{C},\mathcal{C}')\in \{ 1, e^{2\pi i/3}, e^{-2\pi i/3} \} .
\end{equation}

\end{appendix}

\section*{References}

\bibliographystyle{utphys}
\bibliography{references}

\end{document}